\documentclass[a4paper,11pt]{article}

\pdfoutput=1 

\usepackage{jheppub} 

\usepackage{bm}
\usepackage{color}
\usepackage{appendix}
\usepackage{graphicx}
\usepackage{amsmath}
\usepackage{amssymb}
\usepackage{slashed}
\usepackage{tabularx}
\usepackage{wrapfig}
\usepackage{hyperref}
\usepackage{mathabx}

\usepackage[utf8]{inputenc}

\newcommand{\be}{\begin{equation}} \newcommand{\ee}{\end{equation}}
\newcommand{\ba}{\begin{array}{c}} \newcommand{\ea}{\end{array}}
\newcommand{\bea}{\begin{eqnarray}} \newcommand{\eea}{\end{eqnarray}}

\newcommand{\costd}{\cos\theta_{\tau d}^{\rm CM}}

\newcommand\tstrut{\rule{0pt}{2.9ex}}       

\title{\Large The role of right-handed neutrinos in 
 $b\to c \tau\, (\pi \nu_\tau, \rho \nu_\tau, \mu \bar \nu_\mu \nu_\tau) \bar\nu_\tau$  from visible final-state kinematics}

\author[a]{Neus Penalva,}
\author[b]{Eliecer Hern\'andez}
\author[a,1]{and Juan Nieves%
\note{Corresponding author.}}

\affiliation[a]{Instituto de F\'{\i}sica Corpuscular 
(centro mixto CSIC-UV), Institutos de Investigaci\'on de Paterna,
C/Catedr\'atico Jos\'e Beltr\'an 2, E-46980 Paterna, Valencia, Spain}

\affiliation[b]{Departamento de F\'\i sica Fundamental 
  e IUFFyM,\\ Universidad de Salamanca, Plaza de la Merced s/n, E-37008 Salamanca, Spain}
\emailAdd{Neus.Penalva@ific.uv.es}
\emailAdd{gajatee@usal.es}
\emailAdd{jmnieves@ific.uv.es}

\date{\today}
\abstract{In the context of lepton flavor universality violation (LFUV) studies, we fully derive a general tensor formalism
to investigate the role that  left- and right-handed neutrino new-physics 
(NP) terms may have in $b\to c \tau\bar\nu_\tau$ transitions.
We present, for several extensions of the Standard Model (SM), numerical results for the $\Lambda_b\to\Lambda_c\tau\bar\nu_\tau$ semileptonic decay,  which is 
expected to be measured with precision 
at the LHCb. This reaction can be a new source of experimental 
information that can help to confirm, or maybe rule out, LFUV   
presently seen in  $\bar B$ meson decays. The present study analyzes observables that can help 
in distinguishing between different NP scenarios that otherwise provide 
very similar  results for the branching ratios, which are our currently best hints for LFUV.
Since the $\tau$ lepton is very short-lived, we 
consider three subsequent $\tau$-decay modes, two
hadronic $\pi\nu_\tau$ and $\rho\nu_\tau$ and one leptonic $\mu\bar\nu_\mu\nu_\tau$,  
which have been previously
studied  for $\bar B\to D^{(*)}$ decays. Within the tensor formalism that
 we have developed in previous works,
we re-obtain the  expressions for the differential decay width written in terms of 
visible (experimentally accessible) variables of the 
 massive particle created in the $\tau$ decay. There are seven different $\tau$ angular and spin asymmetries
 that are  defined in this way and that can be extracted from experiment. Those 
 asymmetries provide observables that  can help in constraining possible  SM extensions.
}

\begin{document}
\maketitle
\section{Introduction}
\vspace{1cm}
Although there is no single experiment that can still claim the discovery of new physics (NP) beyond
 the Standard Model (SM), there  seems to be however mounting evidence that points 
 in that direction. 
 Lepton flavor universality (LFU), which is inherent to the SM (the exception made of 
 lepton-Higgs couplings), is being challenged in different
  experiments. Thus, the ${\cal R}_{ D^{(*)}}=
  \frac{\Gamma(\bar B\to D^{(*)}\tau\bar\nu_\tau)}
  {\Gamma(\bar B\to D^{(*)}\ell\bar\nu_\ell)}$ ratios, with $\ell=e,\mu
  $, show a  
  $3.1\sigma$ tension~\cite{Amhis:2019ckw} with SM results and the similar
  ${\cal R}_{J/\psi}=\frac{\Gamma(\bar B_c\to J/\psi\tau\bar\nu_\tau)}
  {\Gamma(\bar B_c\to J/\psi\mu\bar\nu_\mu)}$ observable, recently measured by the LHCb Collaboration~\cite{Aaij:2017tyk}, provides also    a  $1.8\,\sigma$ discrepancy with different SM evaluations~\cite{Anisimov:1998uk,Ivanov:2006ni,Hernandez:2006gt,Huang:2007kb,Wang:2008xt,Wen-Fei:2013uea, Watanabe:2017mip, Issadykov:2018myx,Tran:2018kuv,Hu:2018veh,Wang:2018duy,Hu:2019qcn,Leljak:2019eyw,Azizi:2019aaf}.

In the absence of a unique 
possible extension of the SM, one tries to explain
 the discrepancies adopting a phenomenological strategy including, besides 
 NP corrections to the SM vector and axial terms, 
  NP scalar, pseudoscalar and tensor
 $b\to c\tau\bar\nu_\tau$ effective operators that, in principle, 
 affect only  the third quark and lepton generations. Typically, only 
 left-handed neutrinos are considered. The strength of the 
 different NP operators is governed by, complex in general,
 Wilson coefficients that encode the NP  low energy 
 effects and that are fitted to data. 

Further experimental information may come from the analysis of the
$\Lambda_b\to\Lambda_c$ semileptonic decays. In fact, 
the  shape of the $d\Gamma(\Lambda_b\to\Lambda_c
 \mu^-\bar\nu_\mu)/d\omega$ decay width has already been measured by the
 LHCb Collaboration~\cite{Aaij:2017svr}, and there are 
 expectations~\cite{Cerri:2018ypt} that the ${\cal
R}_{\Lambda_c}=\frac{\Gamma(\Lambda_b\to\Lambda_c\tau\bar\nu_\tau)}{
\Gamma(\Lambda_b\to\Lambda_c\mu\bar\nu_\mu)}$ ratio can be obtained
 with a similar precision
 to that reached for  ${\cal R}_{D^{(*)}}$. From the theoretical point 
 of view,  there are 
  precise Lattice
QCD  determinations of the vector and axial form factors~\cite{Detmold:2015aaa}, 
as well as the NP tensor ones~\cite{Datta:2017aue}.  The  
 scalar  and pseudoscalar  form factors, also needed for a full description
 of all NP terms on the low energy Hamiltonian comprising the full set of dimension-6 operators,  can be directly related to vector and axial ones
(see  Eqs.~(2.12) and (2.13) of Ref.~\cite{Datta:2017aue}). This  allows
for a reliable SM determination of  ${\cal
R}_{\Lambda_c}$ ~\cite{Gutsche:2015mxa,Azizi:2018axf, Bernlochner:2018kxh}, as well 
as the evaluation of NP effects~\cite{Dutta:2015ueb,Shivashankara:2015cta,Ray:2018hrx, Li:2016pdv,Datta:2017aue,
Blanke:2018yud,Bernlochner:2018bfn,DiSalvo:2018ngq,Blanke:2019qrx,
Boer:2019zmp,Murgui:2019czp,Ferrillo:2019owd,Mu:2019bin,Colangelo:2020vhu,Hu:2020axt}  that can be compared 
to future experimental determinations.

Although the latest measurements of  the
  ${\cal R}_{D^{(*)}}$ ratios by the Belle Collaboration~\cite{Belle:2019rba}
     constraint the admissible extensions of the 
  SM~\cite{Shi:2019gxi},  disfavoring for instance
  large pure tensor NP scenarios, there is not 
  a unique NP solution to solve the discrepancies (see  Refs.~\cite{Bhattacharya:2018kig,
  Murgui:2019czp, Shi:2019gxi}) and, thus, other observables have
  been proposed as benchmarks to 
 constrain and/or determine the
 most favored NP extension of the SM. These include asymmetries, like the
  $\tau$-forward-backward 
 (${\cal A}_{FB}$) and $\tau$-polarization (${\cal A}_{\lambda_\tau}$) 
 ones, but also different observables related to the four-body
$\bar B\to D^*(D\pi, D\gamma)\tau\bar\nu_\tau$
~\cite{Duraisamy:2013pia,Duraisamy:2014sna,Becirevic:2016hea,
Colangelo:2018cnj} and the full five-body 
$\bar B\to D^*(D Y)\tau(X 
\nu_\tau)\bar\nu_\tau$~\cite{Ligeti:2016npd,Bhattacharya:2020lfm}  
angular distributions. 

The transverse components of the 
 $\tau$ polarization vector ${\cal P}^\mu$ are also  different sources 
 of information. For instance, the $\tau$ polarization vector component perpendicular
  to the plane defined   by the
$\tau$ and final hadron three-momenta, $P_{TT}$, is nonzero only for complex Wilson coefficients. 
Its measurement will not only be an indication of NP beyond the SM but also of CP violation. 
The search for NP in different $\tau$-polarization 
 observables for $\bar B \to D^{(*)}$ decays was explored already twenty five years
  ago in the context of SM extensions with charged Higgs bosons~\cite{Tanaka:1994ay}.
 More recent works~\cite{Nierste:2008qe,Tanaka:2012nw, Ivanov:2017mrj,Alonso:2016gym,
 Alonso:2017ktd,Blanke:2018yud,Asadi:2020fdo} 
have developed this idea. In Ref.~\cite{Penalva:2021gef}, within the formalism previously developed 
in Refs.~\cite{Penalva:2019rgt,
Penalva:2020xup},
we have evaluated the different  components of the tau polarization 
vector (${\cal P}^\mu$)
for the $\Lambda_b\to\Lambda_c$, $\bar B_c\to\eta_c,J/\psi$ and $\bar B\to D^{(*)}$
semileptonic decays, for extensions of the SM involving only left-handed neutrino operators. We have described NP effects in 
the complete two-dimensional space, corresponding to the two independent 
kinematical variables
on which ${\cal P}^\mu$ depends, finding that
its detailed study has indeed a great
potential to discriminate between different NP scenarios for $0^-\to 0^-$ 
decays and also for the $\Lambda_b\to\Lambda_c$ transition.

A caveat in  some of the above $\tau$ polarization-vector analyses is that 
to experimentally measure some of the observables one needs  to be able to establish
 the $\tau$ three-momenta. This is however extremely 
difficult due to the $\tau$ lepton being  very short-lived
 and the fact that the decay products contain neutrinos which escape
detection. A way out of this problem is to concentrate on what is termed as visible kinematics. 
This is achieved 
by considering the subsequent $\tau$ decay and integrating out all variables 
that can not be directly measured, either neutrino-related ones or
 variables defined with respect to the
$\tau$ three-momentum. The price to pay is that one can only access averages of the
full polarization vector components and that all the information on $P_{TT}$ 
is lost after the integration process.
This is for instance what was done, for the $\bar B\to D^{(*)}$ reaction 
in  Ref.~\cite{Alonso:2017ktd}, where, the authors concentrated in the
two subsequent  $\tau\to\pi\nu_\tau$ and $\tau\to\rho\nu_\tau$ hadronic decay modes. Further, 
in Ref.~\cite{Asadi:2020fdo}, it was shown how to extract a total of seven $\tau$
 angular and spin asymmetries from a full analysis of the final-state visible kinematics. A similar 
 study, also for the $\bar B\to D^{(*)}$ reaction but considering in this case 
  the purely leptonic $\tau\to\mu\bar\nu_\mu\nu_\tau$ decay mode, was carried out in 
 Ref.~\cite{Alonso:2016gym}.

In this work we shall present an analysis  parallel to what was done 
in Refs.~\cite{Alonso:2016gym,Alonso:2017ktd,Asadi:2020fdo}, but centered in this case in the
$\Lambda_b\to \Lambda_c\tau\bar\nu_\tau$ semileptonic decay. We only know of one
  analysis of this reaction in terms of visible kinematical variables done in Ref.~\cite{Hu:2020axt}. There, the authors
 construct a measurable angular distribution for the full five body decay
$\Lambda_b\to\Lambda_c(\Lambda \pi)\tau(\pi\nu_\tau)\bar\nu_\tau$ in terms of ten angular observables and they provide results within the SM and different NP models with left-handed neutrinos. 
Three of these observables can be written as linear combinations of the three
 $F_{0,1,2}^\pi$ functions that we introduce in Eq.~\eqref{eq:visible-distr}. Note however that, following Refs.~\cite{Alonso:2016gym,Alonso:2017ktd,Asadi:2020fdo}, we decompose the latter functions in a total of 
 seven angular and spin asymmetries (see Eq.~\eqref{eq:coeffs-1} for the pion $\tau$-decay mode) that, together with the overall normalization, can give separate information on NP and that, hence, we analyze separately.
The rest of the  angular observables analyzed in Ref.~\cite{Hu:2020axt} can not be accessed in our work 
since they require to consider the further $\Lambda_c\to\Lambda\pi$ decay. In our case 
 we  shall also focus on 
NP extensions that  include  right-handed neutrino terms.
The latter  have been suggested 
\cite{Greljo:2018ogz,Asadi:2018wea,Robinson:2018gza, Azatov:2018kzb, Mandal:2020htr}
 as a way
to evade present constraints on NP effective operators 
with only left-handed neutrinos. 
Since interference with the dominant SM left-handed 
terms cancels for massless neutrinos, the contributions from right-handed operators 
are quadratic in the corresponding Wilson coefficients. This means that larger
values of the Wilson 
coefficients may be needed for a purely right-handed NP explanation of the 
discrepancies between SM results and experimental data, which has to be balanced 
with the fact
 that large values of 
the corresponding right-handed NP  Wilson 
coefficients  are more in tension with other low-energy observables
 or collider searches~\cite{Alonso:2016oyd,Akeroyd:2017mhr,Greljo:2018tzh}.
 Here we shall use three different models that include right-handed neutrino NP terms and
 that we take from Ref.~\cite{Mandal:2020htr}. 
 The results obtained within those fits will be compared, not only with SM results, but also with
 the ones obtained from Fit 7 of Ref.~\cite{Murgui:2019czp}, which contains pure left-handed
 neutrino NP operators.
 
The calculations will be done within the tensor formalism that we developed in 
Refs.~\cite{Penalva:2019rgt,Penalva:2020xup} for left-handed neutrino NP operators, which is extended in the present work to account  also for  NP terms constructed out of light right-handed neutrino fields. It is based on the   use of hadron tensors and it provides
  a general description of  any semileptonic decay process where
all hadron polarizations are summed/averaged, being in those cases a useful 
alternative to the commonly used helicity  amplitude approach.
 Within the tensor formalism, 
 we have previously analyzed the $\tau$ polarization 
 vector~\cite{Penalva:2021gef}, but also
 the   role   that  different contributions to the
differential decay widths $d^2\Gamma/(d\omega d\cos\theta_\tau)$ and 
$d^2\Gamma/(d\omega dE_\tau)$, both in the unpolarized and tau helicity-polarized 
 cases, could play in the search of  NP~\cite{Penalva:2019rgt,Penalva:2020xup,
 Penalva:2020ftd}. In the above, $\omega$ stands for
   the product of
the initial and final hadron four-velocities, $\theta_\tau$ is the angle
 made by the three-momenta of the tau 
and final hadron in the center of mass of the final 
two-lepton pair (CM), and $E_\tau$ is the  energy of the tau lepton 
in the  frame where the initial hadron is at rest (LAB). Our studies showed that
the helicity-polarized distributions in the   LAB 
 frame provide  information on NP contributions that cannot be 
 accessed  from the study of the CM differential decay width, the one that is
 commonly analyzed in  the literature. Besides, we have found that  $0^-\to 0^-$ 
 and $1/2^+\to 1/2^+$ decays  seem to better discriminate between different left-handed neutrino NP
  than 
   $0^-\to 1^-$ reactions.

 The present work is organized as follows: In section~\ref{sec:hadron-tensor-formalism}, together with
 appendices~\ref{app:CW},\,\ref{sec:appL},\,\ref{sec:appH} and \ref{app:coeff}, we 
 review  our tensor formalism, and extend it to include
 right-handed neutrino NP terms. 
 We want to stress here that although 
we always refer to $b \to c $ transitions, the hadron and lepton tensors, together with
the expressions for 
the semileptonic differential distributions derived in this work, in the 
presence of  both  left- and right-handed neutrino NP terms,  are valid for 
any   $q \to q '\ell \bar \nu_\ell $ charged-current decay. In  
section~\ref{sec:sdm} we
 collect some of the main theoretical expressions obtained in 
Ref.~\cite{Penalva:2021gef}  concerning
 the spin density operator and the
 $\tau$ polarization vector, which will be needed in the next section. Besides, an  extension of these results to   the case of 
 a $\bar b\to\bar c$ transition is  presented
 (see also appendix~\ref{sec:app-antiquark} in this latter respect). 
 In section~\ref{sec:sequen}, we make a thorough study of the $H_b\to H_c \tau \bar\nu_\tau$ reaction including the subsequent $\tau$-decay, for which we shall consider
 the two  hadronic decay modes $\tau\to\pi\nu_\tau$ and $\tau\to\rho\nu_\tau$ and the leptonic one
$\tau\to\mu\bar\nu_\mu\nu_\tau$. Although we also provide differential
decay widths with respect to variables defined in the $\tau$ rest frame\footnote{In this system, one has access to 
maximal information from the $H_b\to H_c$  semileptonic decay with 
polarized taus, in particular to the CP-violating $P_{TT}$ component of the $\tau$-polarization vector.   In 
section~\ref{sec:tRF}, we detail how $P_{TT}$ can be obtained from an azimuthal-angular asymmetry,  and show results for the CP-violating contributions in  
the baryon $\Lambda_b\to \Lambda_c \tau \bar\nu_\tau$ reaction (Fig.~\ref{fig:PTT}), within a leptoquark model with two nonzero complex Wilson coefficients.   }, we mainly 
concentrate in  obtaining  the  differential decay width in terms of visible kinematic variables and we 
identify (section~\ref{sec:visible}) the seven $\tau$ angular and spin asymmetries mentioned
 above. Some details on 
the   evaluation of the phase-space integrals, which can be rather involved in the
leptonic decay mode, are presented in appendix~\ref{app:phspin}, while the kinematical coefficients that multiply each of 
the observables are discussed in appendix~\ref{app:coeff2}.
Results for the  $\tau$ asymmetries in the $\Lambda_b\to \Lambda_c$ transition  are presented in 
section~\ref{sec:results}. They are obtained within the SM, three different 
NP  extensions that include
right-handed neutrino fields, and a NP model constructed 
 with left-handed neutrino operators alone.  A short summary of our main findings is given in 
section~\ref{sec:conclusions}.

\section{Hadron and lepton tensors in semileptonic decays including new physics with right-handed neutrinos}
\label{sec:hadron-tensor-formalism}
In Refs.~\cite{Penalva:2019rgt,Penalva:2020xup}, we derived a general framework,  
based on the use of general hadron tensors parameterized 
in terms of Lorentz scalar functions,  for describing any  meson or baryon 
semileptonic decay. 
It is an alternative to the helicity-amplitude scheme for the  description of 
processes where all hadron polarizations are summed up and/or averaged.  In these 
two works, NP  with left-handed neutrinos were considered, and here we extend  the 
formalism to include also right-handed neutrino operators. 

\subsection{Effective Hamiltonian}
\label{sec:eh}
We consider an extension of the SM based on the low-energy Hamiltonian comprising
the full set of dimension-6 semileptonic $b\to c$ operators with left- and right-handed neutrinos~\cite{Mandal:2020htr}
\bea
H_{\rm eff}&=&\frac{4G_F V_{cb}}{\sqrt2}\left[(1+C^V_{LL}){\cal O}^V_{LL}+
C^V_{RL}{\cal O}^V_{RL}+C^S_{LL}{\cal O}^S_{LL}+C^S_{RL}{\cal O}^S_{RL}
+C^T_{LL}{\cal O}^T_{LL}\right.\nonumber \\
&+&\left. C^V_{LR}{\cal O}^V_{LR}+
C^V_{RR}{\cal O}^V_{RR}+C^S_{LR}{\cal O}^S_{LR}+C^S_{RR}{\cal O}^S_{RR}
+C^T_{RR}{\cal O}^T_{RR} \right]+h.c.,
\label{eq:hnp}
\eea
with  left-handed neutrino fermionic operators given by 
\be
{\cal O}^V_{(L,R)L} = (\bar c \gamma^\mu b_{L,R}) 
(\bar \ell \gamma_\mu \nu_{\ell L}), \, {\cal O}^S_{(L,R)L} = 
(\bar c\,  b_{L,R}) (\bar \ell \, \nu_{\ell L}), \, {\cal O}^T_{LL} = 
(\bar c\, \sigma^{\mu\nu} b_{L}) (\bar \ell \sigma_{\mu\nu} \nu_{\ell L})
\label{eq:hnp2}
\ee
and the right-handed neutrino ones
\be
{\cal O}^V_{(L,R)R} = (\bar c \gamma^\mu b_{L,R}) 
(\bar \ell \gamma_\mu \nu_{\ell R}), \, {\cal O}^S_{(L,R)R} = 
(\bar c\,  b_{L,R}) (\bar \ell \, \nu_{\ell R}), \, {\cal O}^T_{RR} = 
(\bar c\, \sigma^{\mu\nu} b_{R}) (\bar \ell \sigma_{\mu\nu} \nu_{\ell R}),
\label{eq:hnp2R}
\ee
and where $\psi_{R,L}= (1 \pm \gamma_5)\psi/2$,  $G_F=1.166\times 10^{-5}$~GeV$^{-2}$  and $V_{cb}$ is the corresponding Cabibbo-Kobayashi-Maskawa matrix element. Note that tensor operators with different lepton and quark chiralities
vanish identically\footnote{It follows from
\be
\sigma^{\mu\nu}(1+\chi \gamma_5)\otimes \sigma_{\mu\nu}(1+\chi' \gamma_5) = 
(1+\chi\chi')\sigma^{\mu\nu}\otimes \sigma_{\mu\nu}- (\chi+\chi')\frac{i}{2} 
\epsilon^{\mu\nu}_{\ \ \,\alpha\beta}\sigma^{\alpha\beta}\otimes \sigma_{\mu\nu},
\ee
where we use the convention $\epsilon_{0123}=+1$.
}.

The 10 Wilson coefficients $C^X_{AB}$ ($X= S, V,T$ and $A,B=L,R$) parametrize possible deviations from the SM, i.e. $C^X_{AB}\big|_{\rm SM}$=0. They 
are lepton and flavour dependent, and   complex in general.

\subsection{Hadron and lepton currents}

The  semileptonic differential decay rate of a bottomed hadron
($H_b$) of mass $M$ into a charmed one ($H_c$) of mass $M'$ and 
$\ell \bar\nu_\ell$, measured in its rest frame,  after averaging (summing) over the initial (final) hadron polarizations, 
reads~\cite{Tanabashi:2018oca},
\begin{equation}
  \frac{d^2\Gamma}{d\omega ds_{13}} = \Gamma_0\, \overline\sum\  |{\cal M}|^2,\qquad \Gamma_0=  \frac{G^2_F|V_{cb}|^2
    M'^2}{(2\pi)^3 M},\label{eq:defsec}
  \end{equation}
where ${\cal M} (k,k',p,q,\,{\rm spins} )$ is the transition 
matrix element\footnote{The Lorentz-invariant matrix element, $T$, introduced in the review on {\it Kinematics} of the PDG~\cite{Tanabashi:2018oca} and ${\cal M}$ used in Eq.~\eqref{eq:defsec} are related by  (up to a global phase) 
\be
T= 2G_F V_{cb}\sqrt{2M}\sqrt{2M'}\times {\cal M}.
\ee
}, with $p$, $k'$, $k=q-k'$ and $p'=p-q$, the decaying 
$H_b$ particle,  outgoing charged-lepton,  neutrino and final hadron 
four-momenta, respectively. In addition, $\omega$ is  the product of the two hadron 
four velocities $\omega=(p\cdot p')/(MM')$, which is related to $q^2=(k+k')^2$  via $q^2 = M^2+M'^2-2MM'\omega$, and $s_{13}=(p-k)^2$. Including both left- and right-handed neutrino NP contributions, we have 
\be
{\cal M} = \left(J_{H}^\alpha J^{L}_\alpha+ J_{H} J^{L}+ 
J_{H}^{\alpha\beta} J^{L}_{\alpha\beta}\right)_{\nu_{\ell L}}+\left(J_{H}^\alpha J^{L}_\alpha+ J_{H} J^{L}+ 
J_{H}^{\alpha\beta} J^{L}_{\alpha\beta}\right)_{\nu_{\ell R}}, \\
\ee
with the polarized lepton currents given by 
($u$ and $v$ dimensionful Dirac spinors)
\bea
J^{L}_{(\alpha\beta)}(k,k';h,h_\chi) &=& \frac{1}{\sqrt{2}} 
\bar u_\ell^S(k';h) \Gamma_{(\alpha\beta)} P_5^{h_\chi} v_{\bar\nu_\ell}(k), 
\nonumber\\ \Gamma_{(\alpha\beta)}&=& 1,\gamma_\alpha,\sigma_{\alpha\beta}, \quad
 P_5^{h_\chi} = \frac{1+h_\chi\gamma_5}{2},
  \label{eq:lepton-current}
 \eea
where $h=\pm 1$ stand for  the two possible  charged-lepton polarizations (covariant spin)
along a certain four vector $S^\alpha$ that we choose to measure in the experiment. This is to 
say, the outgoing charged-lepton is produced in the state $u^S_\ell(k'; h)$ defined 
by the condition
%
%
\be
\gamma_5\slashed{S}\,u^S_\ell(k'; h)=h\,u^S_\ell(k'; h).
\ee
The four vector $S^\alpha$ satisfies the constraints  
$S^{\,2}=-1$ and  $S\cdot k'=0$, and the choice $S^\alpha= (|\vec{k}'|, k^{\prime 0}\hat k')
/m_\ell$, with $\hat k'=\vec{k}'/|\vec{k}'|$ and $m_\ell$ the charged lepton
mass, leads to charged-lepton helicity states.  
For later purposes we define here the projector
\be
P_h = \frac{1+h\gamma_5\slashed{S}}{2}.
\ee
In addition, $h_\chi$ accounts for both neutrino chiralities, $R (h_\chi= 1)$ and $L (h_\chi= -1)$.

The dimensionless hadron currents read 
\be
J_{H rr'\,\chi(=L,R)}^{(\alpha\beta)}(p,p') =  \langle H_c; p',r'| \bar c(0) 
O_{H\chi}^{(\alpha\beta)}b(0) | H_b; p, r\rangle, 
\ee
with $c(x)$ and $b(x)$ Dirac fields 
in coordinate space, hadron states normalized as $\langle \vec{p}\,', r'| \vec{p},
r\rangle= (2\pi)^3(E/M)\delta^3(\vec{p}-\vec{p}\,')\delta_{rr'}$ with $r,r'$  spin indexes, and (we recall $h_{\chi=R}=+1$ and $h_{\chi=L}=-1$)
\be
O_{H\chi}^{(\alpha\beta)}= (C^S_{\chi}+h_\chi C^P_{\chi} \gamma_5),\qquad (C^V_{\chi}\gamma^\alpha+ h_\chi C^A_{\chi} \gamma^\alpha\gamma_5),
 \qquad C^T_{\chi}\sigma_{\alpha\beta} (1+h_\chi\gamma_5). \label{eq:Jh}
 \ee
The Wilson coefficients $C_{\chi=L,R}^{S,P,V,A,T}$ in the above definitions are linear combinations of those introduced in the  effective Hamiltonian of Eq.~\eqref{eq:hnp} and are given in Appendix~\ref{app:CW}. Neglecting the neutrino mass, $m_{\nu_\ell}$, there is no interference between the two neutrino chiralities, and the decay probability becomes an
incoherent sum of $\nu_{\ell L}$ and  $\nu_{\ell R}$ contributions, 
\be
|{\cal M}|^2 = |{\cal M}|_{\nu_{\ell L}}^2 + |{\cal M}|_{\nu_{\ell R}}^2 + 
{\cal O}(m_{\nu_\ell}/E_{\nu_\ell}),
\ee
with $E_{\nu_\ell}$ the neutrino energy. The  diagonal lepton tensors needed to obtain  $|{\cal M}|^2$ are 
readily evaluated and they are collected in Appendix~\ref{sec:appL} for $m_{\nu_\ell}=0$.

After summing over polarizations, the hadron contributions can be expressed 
in terms of  Lorentz scalar  structure functions (SFs), which depend on $q^2$, the hadron masses and the 10 NP 
 Wilson coefficients, $C^X_{AB}$, introduced in the effective Hamiltonian of Eq.\eqref{eq:hnp}.  Lorentz, parity and time-reversal  
transformations of the hadron currents (Eq.~\eqref{eq:Jh}) and states
 \cite{Itzykson:1980rh} limit their number, as  discussed in detail in Ref.~\cite{Penalva:2020xup}. 
 The hadron tensors are expressed as linear combination of independent Lorentz (pseudo-)tensor 
 structures, constructed out of the vectors $p^\mu$, $q^\mu$, 
 the metric $g^{\mu\nu}$ and the Levi-Civita pseudotensor 
 $\epsilon^{\mu\nu\delta\eta}$. The coefficients multiplying   
  the (pseudo-)tensors are the  $\widetilde W^{'s}_{\chi=L,R}$ SFs. They depend on $q^2$,  the hadron masses, the Wilson coefficients for each neutrino chirality ($C^{V,A, S,P,T}_{\chi=L,R}$), and the  genuine  hadronic responses ($W's$).   The latter ones are determined 
 by the matrix elements of the involved hadron operators, which for each 
 particular decay are parametrized in terms of form-factors. Symbolically, we have $\widetilde W_\chi = C_\chi W$. There is a total of  16
 independent SFs ($\widetilde W^{'s}_{\chi}$) for each neutrino-chirality set of Wilson coefficients, as shown in Ref.~\cite{Penalva:2020xup}. However, the consideration of both neutrino chiralities does not modify the number of genuine hadronic responses $W's$, and the number of $\widetilde W$ SFs increases due to the greater number of Wilson coefficients. For the sake of clarity, the definition of the $\widetilde W^{'s}_{\chi}$ SFs are compiled here in Appendix~\ref{sec:appH}.

From the general structure of the lepton and hadron tensors, collected 
in Appendices~\ref{sec:appL} and \ref{sec:appH}, and which are at most quadratic 
in $k,k'$ and $p$, 
one can generally write for the decay with a polarized charged lepton.~\cite{Penalva:2020xup,Penalva:2021gef}
\begin{eqnarray}
\frac{2\,\overline\sum\, |{\cal M}|^2 }{M^2}&\simeq& \frac{2\,\overline\sum\, \left(|{\cal M}|_{\nu_{\ell L}}^2 + |{\cal M}|_{\nu_{\ell R}}^2\right) }{M^2}=
{\cal N}(\omega, p\cdot k) + h\bigg\{ \frac{(p\cdot S)}{M}\,
{\cal N_{H_{\rm 1}}}(\omega, p\cdot k) \nonumber\\
&+&\frac{(q\cdot S)}{M}\,
{\cal N_{H_{\rm 2}}}(\omega, p\cdot k)+\frac{\epsilon^{ S k' qp}}{M^3}\,{\cal N_{H_{\rm 3}}}(\omega, p\cdot k)
 \ \bigg\},\label{eq:pol}
\end{eqnarray}
with $\epsilon^{ S k' qp}=\epsilon^{\alpha\beta\rho\lambda}S_\alpha k'_\beta q_\rho 
p_\lambda$   
and  the ${\cal N}$ and  $\cal N_{H_{\rm 123}}$ scalar functions given by
 \bea
 {\cal N}(\omega, k\cdot p)&=&\frac12\Big[{\cal A}(\omega)
+{\cal B}(\omega) \frac{(k\cdot p)}{M^2}+ {\cal C}(\omega) 
\frac{(k\cdot p)^2}{M^4}\Big],\nonumber\\
{\cal N_{H_{\rm 1}}}(\omega, k\cdot p)&=&{\cal A_H}(\omega)
+ {\cal C_H}(\omega)
 \frac{(k\cdot p)}{M^2},\nonumber\\
{\cal N_{H_{\rm 2}}}(\omega, k\cdot p)&=& {\cal B_H}(\omega)
  + {\cal D_H}(\omega) \frac{(k\cdot p)}{M^2}+ {\cal E_H}(\omega) 
  \frac{(k\cdot p)^2}
  {M^4},\nonumber\\ 
 {\cal  N_{H_{\rm 3}}}(\omega, k\cdot p)&=&{\cal F_H}(\omega)+ 
 {\cal G_H}(\omega)\frac{(k\cdot p)}{M^2}.\label{eq:pol2}
  \eea
%
%
There are three independent functions, ${\cal A}, {\cal B}$,   
and ${\cal C}$ , for the non-polarized case, and seven additional ones, ${\cal A_H}, {\cal B_H}, 
{\cal C_H}, {\cal D_H}$, ${\cal E_H}$, ${\cal F_H}$ and ${\cal G_H}$, to describe the process with a defined polarization ($h=\pm 1$) of the outgoing $\ell$  along the four vector $S^\alpha$. Expressions for all of them in terms of the  $\widetilde W$ SFs are given in  Appendix~\ref{app:coeff}. 
As can be seen there, these functions receive contributions from both neutrino chiralities. For ${\cal A}, {\cal B}$,   
 ${\cal C}$, ${\cal F_H}$ and ${\cal G_H}$, it always appears the combination $(L+R)$, i.e. $(\widetilde W_{iL}+ \widetilde W_{iR})$, while 
 for the other functions (${\cal A_H}, {\cal B_H}, {\cal C_H}, {\cal D_H}$ and ${\cal E_H}$) the structure is ($L-R):$ ($\widetilde W_{iL}- \widetilde W_{iR})$.  
An obvious consequence  is that the NP  $L-$ and $R-$neutrino-chirality  contributions cannot be disentangled using only the non-polarized decay, 
and some information is needed from  charged-lepton polarized distributions. 

As can be seen in Appendices~\ref{sec:appH} and \ref{app:coeff}, the $\widetilde W$ SFs present in ${\cal  N_{H_{\rm 3}}}$ 
are generated from the interference of vector-axial with scalar-pseudoscalar terms ($\widetilde W_{I1\chi}$), 
scalar-pseudoscalar with tensor terms ($\widetilde W_{I3\chi}$), and vector-axial with tensor terms ($\widetilde W_{I4\chi,I5\chi,I6\chi}$). 
Since the vector-axial terms are already present in the SM,  at least one of 
the $C^{S,P,T}_\chi$ coefficients must be non-zero to generate a non-zero ${\cal  N_{H_{\rm 3}}}$ term. Besides, 
${\cal N_{H_{\rm 3}}}$ is proportional to the imaginary part of SFs, which requires complex  Wilson coefficients, thus incorporating
violation of the CP symmetry in the NP effective Hamiltonian. This feature makes the study of such contribution of special relevance.

As expected, the ${\cal N}(\omega, k\cdot p)$ and  ${\cal N_{H_{\rm 123}}}(\omega, k\cdot p)$  scalar functions 
give also the antiquark-driven decay $H_{\bar b}\to H_{\bar c} \ell^+ \nu_\ell$, 
as shown in  Appendix~\ref{sec:app-antiquark}.  Moreover,
Eq.~\eqref{eq:anti-lep-ampl} and the results for $\widetilde W$ SFs collected in this work, for NP 
operators involving both left- and right-handed neutrino fields, can be 
straightforwardly used to describe  quark charged-current transitions giving rise to a final $\ell^+\nu_{\ell}$ lepton pair (e.g. $c\to s \ell^+\nu_{\ell}$).

One can use all the formulae given in \cite{Penalva:2020xup} to obtain the  differential
 decay widths for a final $\tau$ with a well defined  helicity either in the laboratory (LAB) or the center
  of mass (CM) frames, where the initial hadron or the outgoing 
 $(\ell\bar\nu_\ell)$-pair are at rest, respectively.
Namely, the $d^2\Gamma/(d\omega dE_\ell)$ and  $d^2\Gamma/(d\omega d\cos\theta_\ell)$ 
distributions for positive and negative helicities 
of the outgoing charged-lepton $\ell$ and where $E_\ell$ is the  LAB energy of 
the charged lepton  and
$\theta_\ell$  is the angle made  by its three-momentum with  
that of the final hadron in the CM frame.
Note that these distributions do not depend on the CP-symmetry breaking term
 ${\cal  N_{H_{\rm 3}}}$, since for both CM and LAB systems 
 $\epsilon^{ S k' qp}=0$, when helicity states are used, 
 i.e. $S^\alpha= (|\vec{k}'|, k^{\prime 0}\hat k')/m_\ell$. 

The CM distribution can be written as
\begin{eqnarray}
  \frac{d^2\Gamma}{d\omega d\cos\theta_\ell}& =& \frac{\Gamma_0 M^3
    M'}{2}
  \sqrt{\omega^2-1}\left(1-\frac{m_\ell^2}{q^2}\right)^2 \Big\{a_0(\omega,h)+a_1(\omega,h)
 \cos\theta_\ell\nonumber \\
 &+&
 a_2(\omega,h)(\cos\theta_\ell)^2\Big\}, \label{eq:GammaCM}
\end{eqnarray}
where the $a_{0,1,2}(\omega,h)$ coefficients are explicitly given in \cite{Penalva:2020xup} as linear combinations of  ${\cal A}, {\cal B}$, ${\cal C}$, ${\cal A_H}, {\cal B_H}, 
{\cal C_H}, {\cal D_H}$ and ${\cal E_H}$. Analogously, the detailed dependence on $E_\ell$ for the LAB distribution 
\begin{eqnarray}
   \frac{d^2\Gamma}{d\omega dE_\ell}& =& \frac{\Gamma_0 M^3}{2} \left\{c_0(\omega)  + c_1(\omega) \frac{E_\ell}{M} +   c_2(\omega) 
     \frac{E^2_\ell}{M^2}\right.\nonumber \\
     &-&\left.h\frac{M}{p_\ell}
 \left(\widehat{c}_0+
     \left[c_0+\widehat{c}_1\right] \frac{E_\ell}{M}+
     \left[c_1+\widehat{c}_2\right] \frac{E^2_\ell}{M^2}+
    \left[ c_2+\widehat{c}_3\right] \frac{E^3_\ell}{M^3}\right)\right\} 
\end{eqnarray}
is also fully addressed in   \cite{Penalva:2020xup}. 

The scheme is totally general and it can be  applied to any charged current semileptonic decay, involving any quark  flavors or initial and final hadron states. Expressions for the $\widetilde W_{i\chi}$ SFs in terms of the Wilson coefficients ($C^X_{AB}$) and the form-factors, used to parameterize  the genuine  hadronic responses ($W_i$), can be obtained from the Appendices of Refs.~\cite{Penalva:2020xup} and \cite{Penalva:2020ftd}, for any $1/2^+ \to 1/2^+ \ell \bar\nu_\ell$, $0^- \to 0^-\ell \bar\nu_\ell$ or $0^- \to 1^-\ell \bar\nu_\ell$ semileptonic decay, regardless of the involved flavors (see Eq.~\eqref{eq:repl} for details).   

In Refs.~\cite{Penalva:2019rgt,Penalva:2020xup}, we presented results for the 
$\Lambda_b\to\Lambda_c\tau\bar\nu_\tau$ decay and  showed that the 
helicity-polarized distributions in the LAB frame provide additional 
information about the NP contributions, which cannot be accessed  by 
analyzing only the CM differential decay widths, as it is commonly proposed
 in the literature (see also the discussion of Eq.~(4.5) in Ref.~\cite{Penalva:2021gef}). In Ref.~\cite{Penalva:2020ftd} we extended  
the study to $\bar B_c\to\eta_c\tau\bar\nu_\tau$, $\bar B_c\to J/\psi\tau\bar\nu_\tau$ as well as the $\bar B\to D^{(*)}\tau\nu_\tau$ decays.  
What we have found is  that the discriminating power between different NP scenarios was better for $\bar B_c\to\eta_c$, $\bar B\to D $ and $\Lambda_b \to \Lambda_c$ 
decays than for   $\bar B_c\to J/\psi$ and $\bar B\to D^*$  reactions. 

In the works of Refs.~\cite{Penalva:2019rgt,Penalva:2020xup,Penalva:2020ftd} only NP left-handed neutrino terms were considered.

\subsection{Spin density operator and charged-lepton polarization vector}
\label{sec:sdm}
The charged-lepton polarization vector ${\cal P}^\mu(\omega, k\cdot p)$ can be readily obtained 
from Eq.~\eqref{eq:pol},
\be
{\cal P}^\mu(\omega, k\cdot p)=\frac1{{\cal N}(\omega, k\cdot p)}\bigg[\ \frac{p^\mu_\perp}{M}
{\cal  N_{H_{\rm 1}}}(\omega, k\cdot p)+\frac{q^\mu_\perp}
 {M}{\cal  N_{H_{\rm 2}}}(\omega, k\cdot p) 
 +\frac{\epsilon^{\mu k'qp}}{M^3}
 {\cal  N_{H_{\rm 3}}}(\omega, k\cdot p)
 \bigg],
\label{eq:PNP}
\ee
with $l_\perp= [l-(l\cdot k'/m_\tau^2) k^{\prime}],\, l = p,q$, which guaranties $k'\cdot {\cal P}=0$.  We refer the reader to Ref.~\cite{Penalva:2021gef}  
for a detailed discussion on the properties of  ${\cal P}^\mu$ and numerical 
calculations, within the SM and different beyond the SM (BSM) scenarios, for  the  
$\Lambda_b\to\Lambda_c\tau\bar\nu_\tau$, $\bar B_c\to\eta_c\tau\bar\nu_\tau$, 
$\bar B_c\to J/\psi\tau\bar\nu_\tau$  and $\bar B\to D^{(*)}\tau\nu_\tau$ decays. Here, we only collect some relations from Ref.~\cite{Penalva:2021gef}, which will be useful to describe the sequential $H_b\to H_c \tau\, (\pi \nu_\tau, \rho \nu_\tau, \mu \bar \nu_\mu \nu_\tau) \bar\nu_\tau$ decays. The spin density operator, $\bar \rho$, and the polarization vector  are related by
\bea 
{\cal P}^\mu&=&{\rm Tr}[\bar\rho\gamma_5\gamma^\mu]=\frac{{\rm Tr}[(\slashed{k'}
+m_\tau){\cal O}(\slashed{k'}+m_\tau)\gamma_5\gamma^\mu]}{{\rm Tr}
[(\slashed{k'}+m_\tau){\cal O}(\slashed{k'}+m_\tau)]}, \nonumber \\
\bar \rho&=&\frac{(\slashed{k'}+m_\tau){\cal O}(\slashed{k'}+m_\tau)}{{\rm Tr}\,[(\slashed{k'}+m_\tau){\cal O}(\slashed{k'}+m_\tau)]}= \frac{\slashed{k'} +m_\tau}{4m_\tau}
\left[I-\gamma_5\,\slashed{\cal P}\right].
\label{eq:pproperties}
\eea
The operator ${\cal O}$ is defined by its relation with the modulus squared of
the invariant amplitude for the production of a final $\tau-$lepton in a $u^S(k';h)$ state,  for a given
momentum configuration of all the particles involved and when all polarizations except that of the $\tau$ lepton are being averaged or summed up,
\be
\overline\sum\, |{\cal M}|^2 = \bar u^S(k';h){\cal O}u^S(k',h)= \frac12 {\rm Tr}
\left[(\slashed{k'}+m_\tau){\cal O}\right]\left( 1+h\,{\cal P}\cdot S\right).
\ee
From the above equation and Eqs.~\eqref{eq:pol} and \eqref{eq:PNP}, it follows 
\be
{\rm Tr}\left[(\slashed{k'}+m_\tau){\cal O}\right]= M^2 {\cal N}(\omega, k\cdot p). \label{eq:reobtainingGamma}
\ee
Finally, the Dirac matrix ${\cal O}$ can be expressed as
\bea
{\cal O}= \overline{\sum_{r,r'}} {\cal O}_{\rm SL}(r,r')\slashed{k}\gamma^0
{\cal O}^\dagger_{\rm SL}(r,r')\gamma^0, \quad {\cal O}_{\rm SL}(r,r') = 
\frac{1}{\sqrt{2}} \sum_{(\alpha,\beta)}\sum_{\chi} J_{H rr'\chi}^{(\alpha\beta)}
\Gamma_{(\alpha\beta)} P_5^{h_\chi}, \label{eq:O}
\eea
where the neutrino mass has also been  neglected here. The operator ${\cal O}_{\rm SL}(r,r')$ gives the Feynman amplitude for the $H_{b}\to H_{c} \ell \bar\nu_\ell$ vertex,
\be
{\cal M} = \bar u^S(k';h){\cal O}_{\rm SL}(r,r') v_{\bar\nu_\ell}(k), \label{eq:OSL}
\ee
with $r,r'$ hadron spin indexes. In the above equation, the antineutrino polarization is not specified, since ${\cal O}$ is obtained after summing also over this degree of freedom,   resulting in  $\slashed{k}$ in Eq.~\eqref{eq:O}.

From Eq.~\eqref{eq:pol-anti} of Appendix~\ref{sec:app-antiquark}, we conclude that the polarized antiquark-driven semileptonic $H_{\bar b}\to H_{\bar c} \ell^+ \nu_\ell$ decay is also described by the polarization vector ${\cal P}^\mu(\omega, k\cdot p)$ given in  Eq.~\eqref{eq:PNP}. The corresponding spin-density operator ($\bar \rho^{\,\bar b \to \bar c}$) reads in this case,
\be
\bar \rho^{\,\bar b \to \bar c} =-\frac{(\slashed{k'}-m_\tau){\cal O}^{\,\bar b \to \bar c}(\slashed{k'}-m_\tau)}{{\rm Tr}\,[(\slashed{k'}-m_\tau){\cal O}^{\,\bar b \to \bar c}(\slashed{k'}-m_\tau)]}=\frac{\slashed{k'} -m_\tau}{4m_\tau}
\left[I+\gamma_5\,\slashed{\cal P}\right].
\ee
The operator ${\cal O}^{\,\bar b \to \bar c}$ is defined by its relation with the modulus squared of
the invariant amplitude for the production of a final anti-tau lepton in a $v^S(k';h)$ state,  for a given
momentum configuration of all the particles involved and when all polarizations except 
that of the anti-tau are being averaged or summed up. One has 
(see Eq.~\eqref{eq:pol-anti})
\be
\overline\sum\, |{\cal M}|^2 = \bar v^S(k';h){\cal O}^{\,\bar b \to \bar c}v^S(k',h)
= \frac12 {\rm Tr}\left[(\slashed{k'}-m_\tau){\cal O}^{\,\bar b \to \bar c}\right]
\left( 1-h\,{\cal P}\cdot S\right),
\ee
with
\be
{\rm Tr}\left[(\slashed{k'}+m_\tau){\cal O}\right]= {\rm Tr}\left[(\slashed{k'}
-m_\tau){\cal O}^{\,\bar b \to \bar c}\right]= M^2 {\cal N}(\omega, k\cdot p), 
\ee
which guaranties that the unpolarized decay distributions are equal for both 
$H_{\bar b}\to H_{\bar c} \tau^+ \nu_\tau$ and $H_{b}\to H_{c} \tau^-   
\bar\nu_\tau$ reactions. Besides,
with these definitions, the probability $P[v^{S}(k',h)]$ that  in an actual 
measurement the anti-tau is found in the
$v^{S}(k',h)$  state is given by
\bea
P[v^{S}(k',h)] &=& \frac{1}{2m_\tau}\bar v^{S}(k',h)\bar \rho^{\,\bar b \to \bar c}\,v^{S}(k',h)\nonumber\\ &=&\frac12(1-h{\cal P}\cdot S)= \frac1{2m_\tau}\bar u^{S}(k',-h)\bar\rho\, u^{S}(k',-h)= P[u^{S}(k',-h)]
\eea
and it is equal to the probability $P[u^{S}(k',-h)]$ that  the $\tau$ is found in the
$u^{S}(k',-h)$  state in the quark $b\to c$ semileptonic decay. 

\section{Sequential $H_b\to H_c \tau\, (\pi \nu_\tau, \rho \nu_\tau, \mu \bar 
\nu_\mu \nu_\tau) \bar\nu_\tau$ decays}
\label{sec:sequen}
The $\tau$ in the final state poses an experimental challenge, because it does not travel
far enough for a displaced vertex and its decay involves at least one more neutrino. 
The maximal accessible information on the $b \to  c \tau\bar\nu_\tau$ transition is encoded 
in the visible decay products of the $\tau$ lepton, for which the three dominant decay 
modes $\tau \to \pi \nu_\tau ,\, \rho \nu_\tau$ and 
$\ell\bar\nu_\ell\nu_\tau$ ($\ell=e,\mu$)  account 
for more than 70\% of the total $\tau$ width  ($ \Gamma_\tau$). Hence in this section, we 
study subsequent decays of the produced $\tau$, after the $b\to c \tau \bar \nu_\tau$ transition,
\bea
H_b \to H_c &\tau^-& \bar \nu_\tau \nonumber \\
&\,\drsh &  \pi^-\nu_\tau ,\, \rho^-\nu_\tau, \, \mu^-\bar\nu_\mu\nu_\tau, \, e^-\bar\nu_e\nu_\tau 
\eea
in the presence of NP left- and right-handed  neutrino operators. Since the lepton $\tau \to e\bar\nu_e\nu_\tau$ distribution can be obtained from the muon-mode ones, assuming LFU in the light sector and replacing $m_\mu \leftrightarrow m_e$, we will only refer to the latter from now on. 
\subsection{Transition matrix element and the $\tau-$polarization vector}
In all cases the Lorentz-invariant amplitude\footnote{From now on, we follow the PDG conventions.} for the decay chain $H_b \to H_c \tau (d\,\nu_\tau) \bar\nu_\tau $ can be cast as ($d= \pi, \rho, \mu\bar\nu_\mu$)
\be
T_d= K \bar u(p_{\nu_\tau})\Big[ \slashed{d}_{(s)}(1-\gamma_5)\frac{(\slashed{k}'
+m_\tau)}{k^{\prime\, 2}-m^2_\tau+i \sqrt{k^{\prime\, 2}} 
\Gamma_\tau (k^{\prime\, 2}) }  {\cal O}_{\rm SL} (r,r')\Big] v_{\bar\nu_\tau}(k),
\ee
with  ${\cal O}_{\rm SL}$ introduced in Eq.~\eqref{eq:OSL}, the 
virtual $\tau$ four-momentum $k'=q-k=p-p'-k$ and $r$ and $r'$  spin 
indexes for the hadrons. In addition,  $K= 4G_F^2 V_{cb}\sqrt{MM'}$ and 
$d_{(s)}^\mu$ is a four-vector (see below), which depends on the $\tau$ decay mode, 
and finally  $s$ is a  polarization index required to specify the state 
of the produced rho or muon. Now using Eqs.~\eqref{eq:pproperties} and 
\eqref{eq:O}, the modulus squared of the invariant amplitude, after 
averaging and summing over polarizations of the initial and final 
particles, reads 
\bea
\overline\sum\, |T|^2 &=& \frac{2|K|^2{\rm Tr}\,[(\slashed{k'}+m_\tau)
{\cal O}(\slashed{k'}+m_\tau)]}{\left(k^{\prime\, 2}-m^2_\tau\right)^2+ 
k^{\prime\, 2} \Gamma^2_\tau (k^{\prime\, 2})}{\cal R}_d(k',p_{\nu_\tau},
p_d,{\cal P}), \nonumber \\
{\cal R}_d(k',p_{\nu_\tau},p_d,\bar \rho)&=&  
\sum_s d^\alpha_{(s)}d^{*\,\beta}_{(s)}{\rm Tr}\,[\gamma_\beta
\slashed{p}_{\nu_\tau}\gamma_\alpha\bar\rho(1+\gamma_5)],
\eea
where we have neglected the $\tau-$neutrino mass, and $p_d$ stands for 
the pion, rho or muon outgoing four-momenta. Next, we can use 
Eq.~\eqref{eq:pproperties} to obtain ${\cal R}_d$ in terms of the tau 
polarization vector  ${\cal P}$,
\bea
&&\hspace{-.5cm}d^\alpha_{(s)} = f_\pi V^*_{ud}\, p_\pi^\alpha, \quad  {\cal R}_\pi = 
\frac{f^2_\pi |V_{ud}|^2 m_\tau}{2}\left(m_\tau^2-m_\pi^2\right)\left[1+ 
\frac{2m_\tau(p_\pi\cdot{\cal P})}{m_\tau^2-m_\pi^2}\right],  \label{eq:polsum1}\\
&&\hspace{-.5cm}d^\alpha_{(s)} = f_\rho V^*_{ud}\, m_\rho \epsilon_s^{* \alpha},
 \quad  {\cal R}_\rho = f^2_\rho |V_{ud}|^2\frac{2m_\rho^2+m_\tau^2}
 {2m_\tau}\left(m_\tau^2-m_\rho^2\right)\left[1+ 
a_\rho \frac{2m_\tau(p_\rho\cdot{\cal P})}{m_\tau^2-m_\rho^2}\right], \label{eq:polsum2} \\ \nonumber \\
&&\hspace{-.5cm}d^\alpha_{(s)} =\frac{1}{\sqrt2}\bar u(p_\mu,s)\gamma^\alpha(1-\gamma_5)v(p_{\bar\nu_\mu}), \quad  {\cal R}_{\mu\bar\nu_\mu} = 16 (p_\mu\cdot p_{\nu_\tau} )
\left[\frac{(k'\cdot p_{\bar\nu_\mu})}{m_\tau} + (p_{\bar\nu_\mu}\cdot{\cal P}) 
\right], \label{eq:polsum3}
\eea
with $ \epsilon_s^{ \alpha}$  the  $\rho$-meson polarization vector, $a_\rho = (m_\tau^2-2m_\rho^2)/(m_\tau^2+2m_\rho^2)$, $f_\pi \sim 93$ MeV and $f_\rho \sim 150 $ MeV. The meson decay constants and the  CKM matrix element $V_{ud}$ determine
\bea
\Gamma_\pi^\tau &=& \Gamma(\tau\to \pi \nu_\tau) =  \frac{G_F^2 f^2_\pi |V_{ud}|^2m^3_\tau}{8\pi}  \left( 1-\frac{m_\pi^2}{m_\tau^2}\right)^2, \nonumber  \\
\Gamma_\rho^\tau &=& \Gamma(\tau\to \rho \nu_\tau) =  \frac{G_F^2 f^2_\rho 
|V_{ud}|^2m_\tau}{8\pi} (m_\tau^2+2m_\rho^2) \left( 1-\frac{m_\rho^2}{m_\tau^2}
\right)^2,
\eea
while for the lepton mode we have ($y=m_\mu/m_\tau$)~\cite{Tsai:1971vv}
\be
\Gamma_\mu^\tau = \Gamma(\tau\to \mu\bar\nu_\mu \nu_\tau) = 
\frac{G_F^2m_\tau^5}{192\pi^3}f(y), \qquad f(y)= 1-8y^2+8y^6-y^8-24y^4\ln y. \label{eq:phase-space-muon}
\ee

Note that  off-shell effects have been neglected in the derivation of Eqs.~\eqref{eq:polsum1}-\eqref{eq:polsum3}. Actually, we will make use of the approximation
\be
\frac{1}{\left[(q-k)^2-m^2_\tau\right]^2+ (q-k)^2 \Gamma^2_\tau\left[(q-k)^2\right]}
\simeq \frac{\pi\delta\left[(q-k)^2 -m_\tau^2\right]}{m_\tau \Gamma_\tau},\label{eq:narrow}
\ee
which puts the $\tau$ on the mass-shell, and it is extremely accurate since $\Gamma_\tau/m_\tau\sim 10^{-12}$.

\subsection{Integration of the phase-space of the final neutrinos}
The total width for the sequential decay $H_b \to H_c \tau (d\,\nu_\tau) \bar\nu_\tau $ is given in the initial hadron rest frame by
\bea
\Gamma_d &=& \frac{(2\pi)^4}{2M}\int\frac{d^3p'}{ 2\sqrt{M^{\prime\, 2}+\vec{p}^{\,\prime\,2}}(2\pi)^3}\int\frac{d^3k}{2|\vec{k}\,|(2\pi)^3}\int\frac{d^3p_d}{ 2\sqrt{m_d^2+\vec{p}_d^{\,2}}(2\pi)^3}\int\frac{d^3p_{\nu_\tau}}{ 2|\vec{p}_{\nu_\tau}|(2\pi)^3} \nonumber \\
&&\times\delta^4\left(p-p'-k-p_d-p_{\nu_\tau}\right)\overline\sum\, |T|^2,
\eea
where for the muon mode, an additional phase-space integration 
$\int\frac{d^3p_{\bar\nu_\mu}}{2|\vec{p}_{\bar\nu_\mu}|(2\pi)^3}$ for 
the outgoing muon antineutrino  is needed and also to take into account 
its four-momentum  in the delta of conservation. The outgoing 
$\bar\nu_\tau(k)$-$\nu_\tau(p_{\nu_\tau})$ tau antineutrino-neutrino pair, 
together with the muon antineutrino $\bar\nu_\mu$ in the case of lepton 
decay mode, are very difficult to detect and hence it is convenient to  
integrate over their variables. This is easily done using the product of 
Dirac delta functions
$\delta^4(q-k'-k)\delta^4(k'-\tilde p_d)$ of Eq.~\eqref{eq:putintau}, 
which is obtained from the delta of conservation of total four momentum 
and the on-shell approximation of Eq.~\eqref{eq:narrow} for the 
$\tau-$propagator. This procedure introduces an integral over the tau 
phase-space, and using Eq.~\eqref{eq:inte-muon-mode} to perform the 
$\nu_\tau$-$\bar\nu_\mu$ neutrino integrations for the muon decay mode, 
we get 
\bea
\Gamma_d &=& \frac{{\cal B}_d MM'}{\pi} \int d\omega \sqrt{\omega^2-1}\int\frac{d^3k'}{ \sqrt{m_\tau^2+\vec{k}^{\,\prime\,2}}}\frac{\delta\big[q^0-\sqrt{m_\tau^2+\vec{k}^{\,\prime\,2}}-|\vec{q}-\vec{k'}\,|\big]}{|\vec{q}-\vec{k'}\,|} \frac{d^2\Gamma_{\rm SL}}{d\omega ds_{13}} \nonumber \\
&&\times \int\frac{d^3p_d}{ \sqrt{m_d^2+\vec{p}_d^{\,2}}} \left[\eta_d(k',  p_d)
+\chi_d(k',p_d)(p_{d}\cdot{\cal P}) \right], \label{eq:defGamma}
\eea
where ${\cal B}_{d=\pi,\,\rho,\, \mu\bar\nu_\mu}$ are the branching fractions 
of $\tau \to \pi\nu_\tau$, $\tau \to \rho\nu_\tau$ and 
$\tau \to \mu\bar\nu_\mu \nu_\tau$ decays. In addition,  
$d^2\Gamma_{\rm SL}/d\omega ds_{13}$ is the unpolarized semileptonic 
$H_b \to H_c \tau \bar\nu_\tau $ differential distribution introduced 
in Eq.~\eqref{eq:defsec}, which is re-obtained thanks to the relation of  
Eq.~\eqref{eq:reobtainingGamma}, $s_{13}=(p-k)^2=(p'+k')^2$, $p=(M,\vec{0}\,)$ 
and $q=p-p' = (M-M'\omega, M' \sqrt{\omega^2-1}\,\hat{q}_{\rm LAB})$, with  
$\hat{q}_{\rm LAB}$ a unitary vector in an arbitrary direction. Indeed, we 
can always take the plane OXZ as the one formed by the three momenta $\vec{k}'$ 
and $\vec{p}^{\,\prime}$ of the outgoing tau and final hadron and perform two of 
the three $d^3k'$ integrations with the help of the Dirac delta function, 
\be
\frac{d^2\Gamma_d}{d\omega ds_{13}} = {\cal B}_d  \frac{d^2\Gamma_{\rm SL}}
{d\omega ds_{13}} \int\frac{d^3p_d}{ \sqrt{m_d^2+\vec{p}_d^{\,2}}}
 \left[\eta_d(\omega,s_{13},  p_d)+\chi_d(\omega,s_{13},p_d)(p_{d}\cdot{\cal P}
 (\omega, s_{13})) \right], \label{eq:defGamma-fin}
\ee
with $\omega$ varying between 1 and $\omega_{\rm max}= (M^2+M^{\prime\,2}-m^2_\tau)
/(2MM')$ and the limits of $s_{13}$ given by $M^2+M(1-m^2_\tau/q^2)
(M'\omega-M \pm M'\sqrt{\omega^2-1}) $.
The scalar functions $\eta_d$ and $\chi_d$ can only depend on masses and 
the scalar product $(k'\cdot p_d)$, where $k'$ is rebuilt in terms of 
$\omega$ and $s_{13}$. The contribution independent of the tau-polarization vector 
reads
\bea
\eta_{d=\pi,\rho} &=& \frac{m_\tau^2}{m^2_\tau-m^2_d} \frac{\delta\big[
\sqrt{m_\tau^2+\vec{k}^{\,\prime\,2}}-\sqrt{m_d^2+\vec{p}_d^{\,2}}-|\vec{k'}
-\vec{p}_d|\big]}{2\pi\, |\vec{k'}-\vec{p}_d|}, \\
\eta_{d=\mu\bar\nu_\mu} &=& \frac{2H(x-2y)H(1+y^2-x)}{\pi\,  m^2_\tau f(y)}
\left[x(3-2x)-y^2(4-3x)\right], \label{eq:def-eta}
\eea
where $H[...]$ is the step function and $x=2(p_\mu\cdot k')/m^2_\tau$ (muon energy 
in the $\tau$ rest frame, except for a constant) in the lepton mode case.    
Note that $\int\frac{d^3p_d}{ \sqrt{m_d^2+\vec{p}_d^{\,2}}} \eta_d$ is normalized 
to 1, and the integration on $ds_{13}$ reconstructs $d\Gamma_{SL}/d\omega$.  A 
further integration on $d\omega$ will give the expected result $\Gamma_d= 
\Gamma_{\rm SL}\,{\cal B}_d$. The term proportional 
to the polarization vector, which contribution vanishes when one fully integrates 
over  $d^3p_d$, reads  
\bea
\chi_{d=\pi,\rho} &=& a_d\frac{2m_\tau^3}{(m^2_\tau-m^2_d)^2} \frac{\delta
\big[\sqrt{m_\tau^2+\vec{k}^{\,\prime\,2}}-\sqrt{m_d^2+\vec{p}_d^{\,2}}-
|\vec{k'}-\vec{p}_d|\big]}{2\pi\, |\vec{k'}-\vec{p}_d|}, \\
\chi_{d=\mu\bar\nu_\mu} &=& \frac{4H(x-2y)H(1+y^2-x)}{\pi\, m^3_\tau f(y)}
\left[1+3y^2-2x\right], \label{eq:def-xi}
\eea
with $a_{d=\pi}=1$ and $a_{d=\rho}=(m_\tau^2-2m_\rho^2)/(m_\tau^2+2m_\rho^2)$, as defined above. 

The $d^3p_d$ integrals in the expression for the $\Gamma_d$ decay width  in Eq.~\eqref{eq:defGamma-fin} can be further worked out analytically thanks to the invariance
of the integrand under proper Lorentz transformations. There are different choices as to what variables to integrate and in what follows we give the result for two different kinematics of the visible product after the $\tau$-decay.

An analogous calculation for the antiquark-driven $H_{\bar b}\to H_{\bar c} \bar\tau\, (\pi \bar\nu_\tau, \rho \bar \nu_\tau, \bar\mu  \nu_\mu \bar\nu_\tau) \nu_\tau$ decays leads to the same expression as in Eq.~\eqref{eq:defGamma-fin}. This is to say the pion/rho/muon distributions are the same for $b\to c \tau \bar\nu_\tau$ and $\bar b\to \bar c \bar\tau \nu_\tau$ processes. 
\subsection{Pion/rho/muon variables in the $\tau$ rest frame.}
\label{sec:tRF}

If the momentum of the $\tau$ lepton is detected, the direction of the outgoing 
visible particle after its decay can be referred to the plane formed by the 
tau and the final hadron. Taking $\vec{k}'$ and  $(\vec{k}'\times \vec{p}^{\,\prime})
\times\vec{k}'$ in the positive $Z$ and $X$ directions respectively, the 
Lorentz-scalar  $(p_{d}\cdot{\cal P})$ can be evaluated in the $\tau$ rest frame (${\cal P}^{* \mu} = 
\Lambda^\mu_{\  \nu}{\cal P}^\nu$, with $\Lambda$ the boost which takes the tau to rest)
\begin{align}
(p_{d}\cdot{\cal P})&=-|\vec{p}^{\,*}_d|\big[\sin\theta^*_d\big(P^*_X(\omega, s_{13})
\cos\phi^*_d+P^*_{Y}(\omega, s_{13})\sin\phi^*_d\big) + P^*_Z(\omega, s_{13})
,\cos\theta^*_d\big]\\
|\vec{p}^{\,*}_{d=\pi,\rho}|&= \left(\frac{m_\tau^2-m_d^2}{2m_\tau}\right),
 \quad |\vec{p}^{\,*}_{\mu}| = \frac{m_\tau}{2} \sqrt{x^2-4y^2},
\end{align}
with $\theta^*_d$ and $\phi^*_d$, the polar and azimuthal angles of $\vec{p}_d$ in  the $\tau-$rest frame, and where the  scalar functions $P^*_{X,Y,Z}(\omega, s_{13})$ determine  the polarization four-vector  in this system, which is given by ${\cal P}^{* \mu}=(0, P^*_X, P^*_Y, P^*_Z)$. Note that these Cartesian   components are obtained as Lorentz scalar products  ($P^*_Z=-{\cal P}^*\cdot n_L$, $P^*_X=-{\cal P}^*\cdot n_T$ and  $P^*_Y=-{\cal P}^*\cdot n_{TT}$)  of the polarization four-vector with 
spatial unit vectors in the positive Z, X and Y axis, which by construction coincide with the directions of  $\vec{k}'$, $(\vec{k}'\times \vec{p}^{\,\prime})\times\vec{k}'$ and  $(\vec{k}'\times \vec{p}^{\,\prime})$, respectively. These scalar products can be now evaluated in the original system with $\vec{k}'\ne \vec{0}$ by using $\Lambda^{-1}$, boost of velocity $\vec{k}'/k^{\prime 0}$, and we find $P^*_{Z,X,Y}= -({\cal P}^*\cdot n_{L,T,TT}) = -({\cal P}\cdot N_{L,T,TT})$, with 
\bea
N_L^\mu=\Big(\frac{|\vec k\,'|}{m_\tau},\frac{k^{\prime0}\vec k\,'} 
{m_\tau|\vec k\,'|}\Big) ,\quad N_{T}^\mu=\Big(0,\frac{(\vec k\,'\times\vec p\,')\times\vec k\,'}
{|(\vec k\,'\times\vec p\,')\times\vec k\,'|}\Big),\quad
N_{TT}^\mu=\Big(0,\frac{\vec k\,'\times\vec p\,'}
{|\vec k\,'\times\vec p\,'|}\Big)\,, \label{eq:N's}
\eea
which allows to identify the Cartesian components of the polarization vector in 
the $\tau$ rest frame with the usual longitudinal and transverse components of the polarization vector $P^*_Z=P_L$, $P^*_X=P_T$ and $P^*_Y=P_{TT}$ in an arbitrary frame~\cite{Penalva:2021gef}. Now integrating over $d|\vec{p}_d|$, we obtain
\bea
\frac{d^4\Gamma_{d}}{d\omega ds_{13}d\cos\theta^*_dd\phi_d^*} &=& \frac{{\cal B}_{d} }{4\pi}\frac{d^2\Gamma_{\rm SL}}{d\omega ds_{13}}  \, \Big(g^d - g^d_P\big[P_T(\omega, s_{13})\sin\theta^*_d\cos\phi^*_d\nonumber \\
&&+P_{TT}(\omega, s_{13})\sin\theta^*_d\sin\phi^*_d + P_L(\omega, s_{13})\cos\theta^*_d\big]
\Big),  \nonumber\\
g^{\pi, \rho} = 1\,, && g^{\mu\bar\nu_\mu}= \frac{2}{f(y)}\int_{2y}^{1+y^2} \sqrt{x^2-4y^2}\left(x(3-2x)-y^2(4-3x)\right)dx, \nonumber\\
g^{\pi, \rho}_P = a_{\pi, \rho}\,, && g^{\mu\bar\nu_\mu}_P = \frac{2}{f(y)}\int_{2y}^{1+y^2} 
(x^2-4y^2)(1+3y^2-2x) dx. \label{eq:Gamma12}
\eea
After integration,  $g^{\mu\bar\nu_\mu}=1$  and $g^{\mu\bar\nu_\mu}_P=-(1-y)^5\left(1+5y+15y^2+3y^3\right)/\,[3f(y)]$.
\begin{figure*}[!h]
\centering
\includegraphics[scale=0.55]{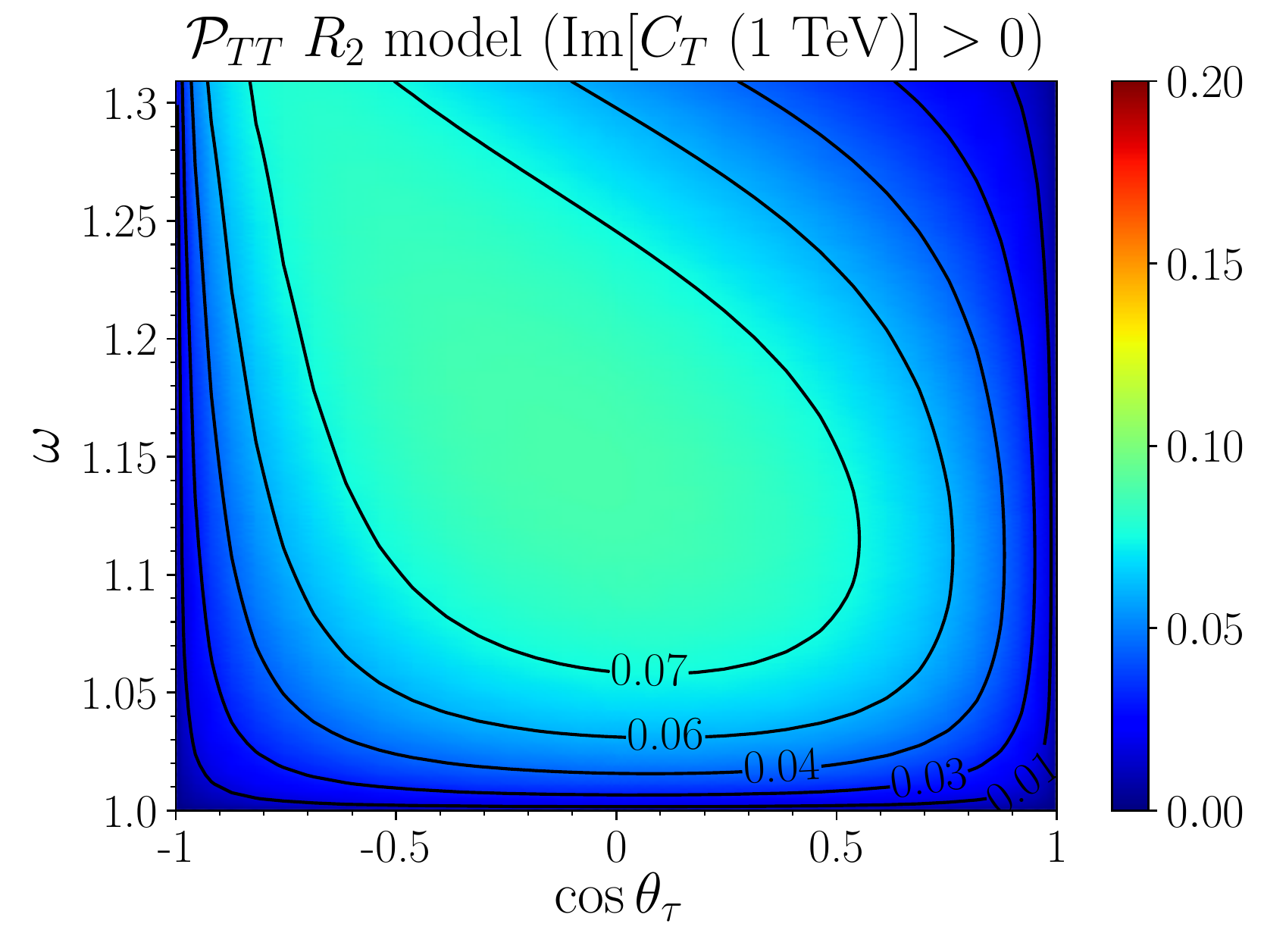}\\
\includegraphics[scale=0.85]{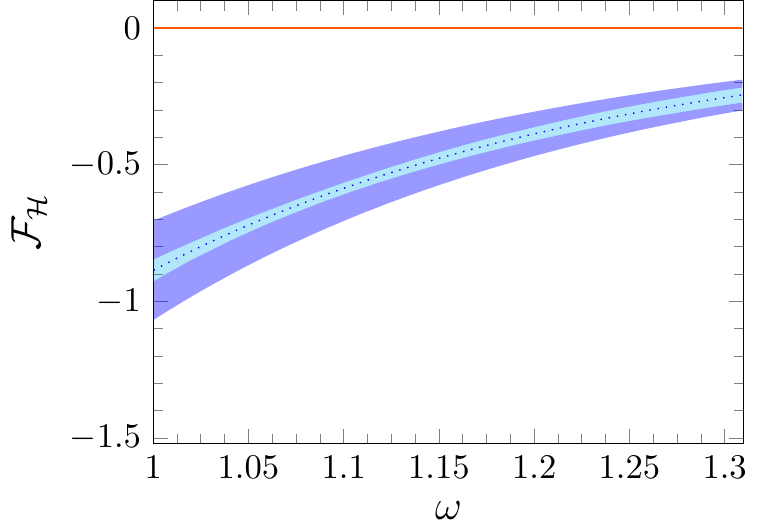}\includegraphics[scale=0.85]{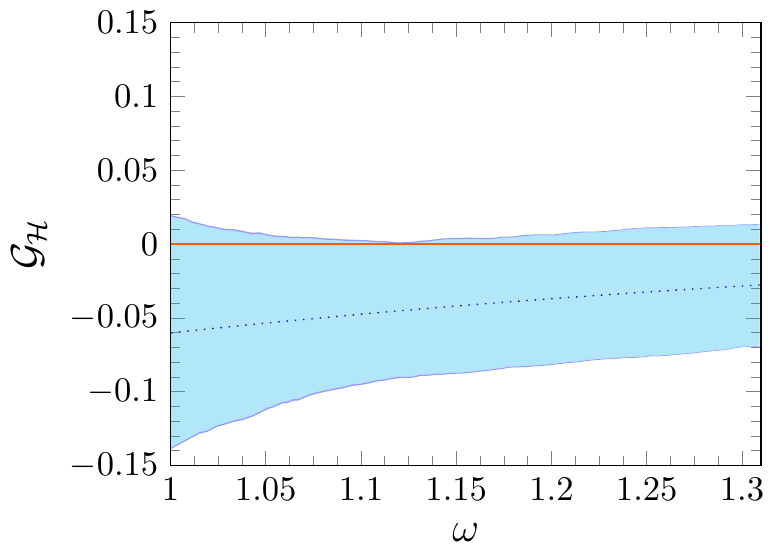}
\caption{ Results for $\Lambda_b\to \Lambda_c\tau\bar\nu_\tau$  evaluated with the $R_2$ leptoquark model of Ref.~\cite{Shi:2019gxi}, for which the two nonzero Wilson coefficients ($C^S_{LL}$ and $C^T_{LL}$) are complex. The unconstrained sign of ${\rm Im\,}[\hat C_T(1\,  {\rm TeV})]$ is taken to be positive. The alternative choice ${\rm Im\,}[\hat C_T(1\,  {\rm TeV})] <0$   would not change the absolute values of the observables shown in this figure, but only their global signs. Top: Two dimensional CM/LAB ${\cal P}_{TT}$ polarization  (Eq.~\eqref{eq:PTTterm}) as a function of  $\omega$ and the CM $\cos\theta_\tau$ variable. The polarization component is obtained using the central values for the form factors (see details in  Sec.~\ref{sec:results} below or in Ref.~\cite{Penalva:2021gef}) and Wilson coefficients. Bottom:  ${\cal F_H}(\omega)$ (left) and ${\cal G_H}(\omega)$ (right) scalar functions entering in the definition of the CP-violating ${\cal N_{H_{\rm 3}}}$ term of  the differential decay width (Eqs.~\eqref{eq:pol} and \eqref{eq:pol2}). The error inherited
 from the form-factor uncertainties is evaluated and propagated  via Monte Carlo, 
 taking into account statistical correlations between the different parameters, and it is depicted as an inner band that accounts for  
 68\% confident-level  intervals. The uncertainty  induced by  the fitted Wilson coefficients is determined 
  using  different $1\sigma$ statistical samples configurations by the authors of Ref.~\cite{Shi:2019gxi}. 
  The two sets of errors are then added in   quadrature  giving rise to the larger uncertainty band. Note that uncertainties on  ${\cal G_H}(\omega)$ are largely dominated by the errors due to the form-factors.}
\label{fig:PTT}
\end{figure*}

Appropriate $\theta_d^*$ and/or $\phi^*_d$ asymmetries can be used to determine the longitudinal and the two transverse components of the tau-polarization four vector, which are two-dimensional functions of the variables $\omega$ and $s_{13}$~\cite{Penalva:2021gef}. The observable $P_{TT}$ is of great interest, since it is given by the CP odd  term ${\cal  N_{H_{\rm 3}}}$ of the ${\cal P}^\mu$ decomposition in Eq.~\eqref{eq:PNP}, which to be different from zero requires the existence of relative complex phases between Wilson coefficients. This component, transverse to the plane formed by the outgoing hadron and tau,  could be obtained integrating over $\cos\theta_d^*$ and looking at the $\phi_d^*$ asymmetry
\bea
P_{TT}(\omega, s_{13}) & = & -\frac{2 g^d}{g^d_P} \, 
\frac{\int_0^\pi d\phi_d^*\frac{d^3\Gamma_d}{d\omega ds_{13} d\phi_d^*}- 
\int_\pi^{2\pi} d\phi_d^*\frac{d^3\Gamma_d}{d\omega ds_{13} d\phi_d^*} }
{\int_0^\pi d\phi_d^*\frac{d^3\Gamma_d}{d\omega ds_{13} d\phi_d^*}+ 
\int_\pi^{2\pi} d\phi_d^*\frac{d^3\Gamma_d}{d\omega ds_{13} d\phi_d^*}}. \label{eq:PTTterm}
\eea
The projection ${\cal P}_{TT}$ is invariant under co-linear boost 
transformations, and thus it is the same in both CM and LAB systems. 
 Measuring a non-zero $ {\cal P}_{TT}$ value in any of these sequential 
 decays  will be a clear   indication of physics BSM and of time reversal 
 (or CP) violation. One can proceed similarly to obtain $P_T$ and $P_L$.

Upon integration on $s_{13}$, the semileptonic $d\Gamma_{\rm SL}/d\omega$ 
differential decay width can be factorized out 
in Eq.~\eqref{eq:Gamma12} by replacing the two-dimensional polarization 
components $P_a(\omega, s_{13})$ by averages on the $s_{13}$ variable weighted 
by the semileptonic distribution
\be
P_a(\omega, s_{13}) \to \langle P_a\rangle(\omega)=
 \left(\frac{d\Gamma_{\rm SL}}{d\omega}\right)^{-1} \int ds_{13}
 \frac{d^2\Gamma_{\rm SL}}{d\omega ds_{13}}P_a(\omega, s_{13}), \quad a=L,T,TT. \label{eq:avg}
\ee
The $d^3\Gamma_{d}/(d\omega d\cos\theta^*_dd\phi_d^*)$ 
distributions thus obtained coincide with the results given in 
Ref.~\cite{Ivanov:2017mrj} for the CM frame (center of mass system of 
the $\tau\bar\nu_\tau$  pair).

Both the two-dimensional and the averaged polarization components in the CM and LAB 
frames  were detailedly studied, in Ref.~\cite{Penalva:2021gef}, in the presence of 
NP involving only left-handed neutrino operators. Results were obtained for the 
$\Lambda_b\to\Lambda_c\tau\bar\nu_\tau$, $\bar B_c\to\eta_c\tau\bar\nu_\tau$, 
$\bar B_c\to J/\psi\tau\bar\nu_\tau$  and $\bar B\to D^{(*)}\tau\nu_\tau$ decays, 
and in the case of the baryon decay, an  special attention to BSM signatures 
derived from complex NP contributions was paid. We complete here the analysis 
of section 4.2.1 of Ref.~\cite{Penalva:2021gef} for the 
$\Lambda_b\to\Lambda_c\tau\bar\nu_\tau$ decay by showing 
in Fig.~\ref{fig:PTT}, the  CP-violating observables ${\cal P}_{TT}(\omega,\cos\theta_\tau)$  
$\left(= -{\cal P}\cdot N_{TT}^{\rm LAB/CM}\right)$, ${\cal F_H}(\omega)$ and 
${\cal G_H}(\omega)$ (see Eqs.~\eqref{eq:pol} and \eqref{eq:pol2}) obtained within 
the leptoquark model~\cite{Shi:2019gxi} employed in that section, and  which 
predicts complex Wilson coefficients. The polarization $\langle P_{TT}\rangle (\omega)$ 
displayed in the left-bottom plot of Fig. 10 of Subsec. 4.2.1 of 
Ref.~\cite{Penalva:2021gef} can be obtained from the average indicated in 
Eq.~\eqref{eq:avg} using the two-dimensional ${\cal P}_{TT}$ shown in the top 
plot of Fig.~\ref{fig:PTT}, and which could be measured by looking at the azimuthal 
asymmetry proposed in  Eq.~\eqref{eq:PTTterm}. This average $\langle P_{TT}\rangle 
(\omega)$ will be a linear combination of the ${\cal F_H}(\omega)$ and
 ${\cal G_H}(\omega)$ scalar functions, also displayed here in  Fig.~\ref{fig:PTT}, 
 which encode the maximal information contained in the CP-violating 
 ${\cal N_{H_{\rm 3}}}$ term of the $\Lambda_b\to\Lambda_c\tau\bar\nu_\tau$ 
 differential decay width. One cannot determine ${\cal F_H}(\omega)$ and 
 ${\cal G_H}(\omega)$ only from $\langle P_{TT}\rangle (\omega)$. Therefore, 
 to extract both of them, it would be necessary  to analyze the dependence of 
 the $d^3\Gamma_d/(d\omega d\cos\theta_\tau d\phi_d^*)$ sequential decay distribution 
 on $\phi^*_d$, which will allow to obtain the  
two-dimensional ${\cal P}_{TT}(\omega,\cos\theta_\tau)$ polarization component.

\subsection{Visible pion/rho/muon variables in the CM frame.}
\label{sec:visible}
 When the  tau momentum cannot be fully reconstructed experimentally, 
the previous expressions are no longer useful, since  the kinematics of the 
decay-product is referred to the $\tau$ direction.   
It is therefore suitable to construct observables directly from final-state 
kinematics of the visible decay particle $\pi,\rho, \mu$, 
without relying on the reconstruction of the tau momentum, which needs to 
be integrated out ($s_{13})$. We take the energy of the charged particle 
in the $\tau-$decay, $E_d$ and its angle $\theta_d$  with the final hadron 
$H_c$, both variables defined in the CM frame ($\vec{q}=0$, $W$ boson at 
rest). This kinematical set up has been extensively used in the literature 
to analyze NP signatures in $\bar B\to D^{(*)} \tau\, (\pi \nu_\tau, 
\rho \nu_\tau, \mu \bar \nu_\mu \nu_\tau) \bar\nu_\tau$ 
decays~\cite{Kiers:1997zt,Nierste:2008qe,Alonso:2016gym,Alonso:2017ktd,
Asadi:2020fdo}, although these studies have not considered BSM right-handed neutrino 
fields. Moreover,  a similar polarimetry analysis  for the 
$\Lambda_b\to\Lambda_c\tau\bar\nu_\tau$ reaction has not be done yet,  
despite  the good prospects that LHCb can measure it in the near future, 
given the large number of $\Lambda_b$ baryons which are produced at LHC.

Following the notation in Ref~\cite{Tanaka:2010se}, we introduce 
\be 
\gamma= \frac{q^2+m^2_\tau}{2m_\tau\sqrt{q^2}}, \quad \beta = 
\frac{\sqrt{\gamma^2-1}}{\gamma}= \frac{q^2-m^2_\tau}{q^2+m^2_\tau}, 
\quad \xi_d= \frac{E_d}{m_\tau \gamma},
\ee
with $\gamma$ and $\beta$ defining the boost 
 from the $\tau$ rest 
frame to the CM one. In addition, the dimensionless variable $\xi_d$ is the CM 
ratio of the energies of the tau-decay massive product and the tau lepton. Let 
us call now $\theta_{\tau d}^{\rm CM}\in [0,\pi[$, the angle formed by the 
$\vec{p}_\tau$ and $\vec{p}_d$ in the CM reference system. We have
\bea
\tau\to (\pi,\rho)\nu_\tau &\Rightarrow& (\costd)(\omega,\xi_d)= 
\frac{2\gamma\xi_d-(1+y^2)/\gamma}{2\beta \sqrt{\gamma^2\xi_d^2-y^2}}, \nonumber \\
\tau\to \mu \bar\nu_\mu\nu_\tau &\Rightarrow& (\costd)(\omega,\xi_d,x )= 
\frac{2\gamma\xi_d-x/\gamma}{2\beta \sqrt{\gamma^2\xi_d^2-y^2}},
\eea
where we have used $y=m_d/m_\tau$ for all decay modes. For the hadronic channels 
$\costd$, obtained in that case from the condition $(k'-p_d)^2=0$, is totally fixed 
by  $\omega$ and the energy of the tau-decay massive product, while for the lepton 
mode it also depends on the additional variable $x=2(p_\mu\cdot k')/m^2_\tau$ 
introduced above. Next, requiring $(\costd)^2 \le 1$, we obtain the allowed region 
for the energy of the pion or rho mesons\footnote{The outgoing $\pi$ or $\rho$ 
hadron could exit at rest only for a single value $\sqrt{q^2}=m^2_\tau/m_d$ of the 
phase-space,  which is likely not accessible, since we expect $m_{\pi,\rho}
< m^2_\tau/(M-M')$.  }
\bea
\tau\to (\pi,\rho)\nu_\tau &\Rightarrow& \frac{1-\beta}{2}+  
\frac{1+\beta}{2}y^2=\xi_1  \le \xi_d \le \xi_2= \frac{1+\beta}{2}+   
\frac{1-\beta}{2}y^2,  \label{eq:ed-range}
\eea
or bounds that the variable $x$ should satisfy in the lepton mode,
\bea
\tau\to \mu \bar\nu_\mu\nu_\tau &\Rightarrow&  x_- \le x \le x_+, 
\quad x_\pm = 2\gamma^2\xi_d \pm 2\gamma\beta  \sqrt{\gamma^2\xi_d^2-y^2}.
\label{eq:x-bounds}
\eea
Also for this latter case, the maximum allowed value of $\xi_d$ is still $\xi_2$, 
which corresponds to $E_d^{\rm max}=(q^2+m_d^2)/(2\sqrt{q^2})$. However, in certain 
circumstances, $\xi_d$ can be as low as $y/\gamma$ for all reachable $q^2$. This 
is to say a kinematics where  the daughter massive lepton is at rest in the CM frame, 
which would be compatible with the energy-momentum conservation thanks to 
the other 
two neutrinos present in the tau-decay final state. In general, one finds
\bea
y \le \sqrt{\frac{1-\beta}{1+\beta}}= \frac{m_\tau}{\sqrt{q^2}} &\Rightarrow & 
y/\gamma  \le \xi_d \le \xi_2, \label{eq:ed-range-lepton} \\
\nonumber \\
y >  \sqrt{\frac{1-\beta}{1+\beta}}= \frac{m_\tau}{\sqrt{q^2}}&\Rightarrow 
&\xi_1  \le \xi_d \le \xi_2. \label{eq:ed-range-lepton-wrong}
\eea
For the $b\to c$ semileptonic decays analyzed here, we have $y=m_\mu/m_\tau$ 
and $\sqrt{q^2} \le (M-M')< m^2_\tau/m_\mu$, which corresponds to the range of 
Eq.~\eqref{eq:ed-range-lepton}. Thus,  the outgoing muon can exit at rest for any 
$q^2$ value, fixing in this way the minimum reachable value for $\xi_d$ to 
$y/\gamma$ ($E_d= m_\mu)$ independently of $q^2$ (or $\omega$). In a hypothetical 
case, for which $y^2 > (1-\beta)/(1+\beta)$ (Eq.~\eqref{eq:ed-range-lepton-wrong}), 
the outgoing massive particle could not exit with zero momentum.  

The bounds of Eq.~\eqref{eq:x-bounds} should be combined with the product of step 
functions $H(x-2y)H(1+y^2-x)$ which appears in the definition of  
$\eta_{d=\mu\bar\nu_\mu}$ and $\chi_{d=\mu\bar\nu_\mu}$ in Eqs.~\eqref{eq:def-eta} 
and \eqref{eq:def-xi}, respectively. While $2y$ is always smaller  than the lower 
bound in Eq.~\eqref{eq:x-bounds}, the combined use of $(1+y^2)$ and the upper 
bound $x_+$ is more subtle  and it leads to the following available phase space
\bea
y \le \sqrt{\frac{1-\beta}{1+\beta}} &\Rightarrow & \varphi(\omega,\xi_d,x)= H(\xi_d-y/\gamma)H(\xi_1-\xi_d)H(x-x_-)H(x_+-x)\nonumber \\
&&+H(\xi_d-\xi_1)H(\xi_2-\xi_d)H(x-x_-)H(1+y^2-x),
\label{eq:phase-space}\\
\nonumber \\
y >  \sqrt{\frac{1-\beta}{1+\beta}}&\Rightarrow & \widetilde\varphi(\omega,\xi_d,x)= 
H(\xi_d-\xi_1)H(\xi_2-\xi_d)H(x-x_-)H(1+y^2-x), \label{eq:phase-space-wrong}
\eea
in agreement  with the results of Ref~\cite{Tanaka:2010se} obtained for $y=0$.  

Taking in this case the outgoing hadron momentum $\vec{p}^{\,\prime}$ as the positive $Z$ 
direction and $\vec p\,'\times \vec k\,'$ as the positive $Y$ direction, one can write $k^{\prime\mu}= m_\tau \gamma (1,\beta\sin\theta_\tau,0, 
\beta\cos\theta_\tau)$ and
\be
\costd=  \sin\theta_\tau\sin\theta_d \cos\phi_d+ \cos\theta_\tau\cos\theta_d
\Rightarrow \cos\phi_d = \frac{\costd-\cos\theta_\tau\cos\theta_d}
{\sin\theta_\tau\sin\theta_d}=z_0, \label{eq:defz0}
\ee
with $\theta_d$ and $\phi_d$ the polar and azimuthal angles of the three-momentum
$\vec{p}_d$ in  the CM frame, $\vec{p}_d= m_\tau \sqrt{\gamma^2\xi_d^2-y^2}
\left(\sin\theta_d\cos\phi_d, \sin\theta_d\sin\phi_d,
 \cos\theta_d\right)$.  The variable $z_0$ above  depends on $\omega$ and 
 $\cos\theta_\tau$ for the hadron modes, while it also depends  on $x$, 
 through $\costd$, in the lepton channel. The condition $(\cos\phi_d)^2 \le 1$ 
 limits the values of the  cosinus of the CM $\tau$-polar angle to the 
 range\footnote{Note that, $\cos(\theta_d-\theta_{\tau d}^{\rm CM})
 -\cos(\theta_d+\theta_{\tau d}^{\rm CM}) = 2\sin\theta_{\tau d}^{\rm CM}
 \sin\theta_d \ge 0$.}
\be
\cos(\theta_d+\theta_{\tau d}^{\rm CM}) \le \cos\theta_\tau \le 
\cos(\theta_d-\theta_{\tau d}^{\rm CM}). \label{eq:theta-limits}
\ee
In addition, we can express the  scalar product $(p_{d}\cdot{\cal P}(\omega, s_{13}))$ in terms of the CM-variables as:
\bea
\frac{(p_{d}\cdot{\cal P})}{m_\tau} &=& \gamma\left(\gamma\beta\xi_d-\sqrt{\gamma^2\xi_d^2-y^2}\costd\right)P_L^{\rm CM}(\omega, \cos\theta_\tau) \nonumber \\
&&+ \sqrt{\gamma^2\xi_d^2-y^2} \Big\{ P_{TT}^{\rm CM}(\omega, \cos\theta_\tau)\sin\theta_d\sin\phi_d \nonumber \\
&&+  \frac{P_{T}^{\rm CM}(\omega, \cos\theta_\tau)}{\sin\theta_\tau}
\left[\cos\theta_\tau\costd-\cos\theta_d  \right] \Big\}, \label{eq:pxP}
\eea
where we have used ${\cal P}^\mu=\left(P_L N_L^\mu +P_T N_T^\mu + P_{TT} 
N_{TT}^\mu\right)_{\rm CM}$, with the vectors  $N_{L,T, TT}^\mu$ computed using  
Eq.~\eqref{eq:N's}, the CM final hadron and tau four momenta and the relations $P_{L,T,TT}=-({\cal P}\cdot N_{L,T,TT})$ 
(see Ref.~\cite{Penalva:2021gef} for further details). We recall that in Eq.~\eqref{eq:pxP}, the definition of $\costd$ involves $\cos\phi_d$ (see Eq.~\eqref{eq:defz0}). 

Taking into account the dependence of $(p\cdot k), (p\cdot N_{L,T}^{\rm CM})$ and
 $(q\cdot N_{L,T}^{\rm CM})$ on $\cos \theta_\tau$, ones finds~\cite{Penalva:2021gef} (the $P_{TT}^{\rm CM}$ term will not contribute after integrating in $\phi_d$)
 \bea
  P_L^{\rm CM}(\omega,\cos\theta_\tau)&=& \frac{\delta a_0(\omega)+ 
  \delta a_1(\omega) \cos\theta_\tau + \delta a_2(\omega) \cos^2\theta_\tau}
  {a_0(\omega) + a_1(\omega) \cos\theta_\tau + a_2(\omega) \cos^2\theta_\tau}, \nonumber \\ 
  \frac{P_T^{\rm CM}(\omega,\cos\theta_\tau)}{\sin\theta_\tau}&=& 
  \frac{p'_0(\omega)+ p'_1(\omega) \cos\theta_\tau }{a_0(\omega) + a_1(\omega) 
  \cos\theta_\tau + a_2(\omega) \cos^2\theta_\tau}, \label{eq:PLPT}
 \eea
where, for $i=1,2,3$, one has
\bea
\delta a_i(\omega)= a_i(\omega, h=-1)- a_i(\omega, h=+1),\nonumber\\
a_i(\omega)= a_i(\omega, h=-1)+ a_i(\omega, h=+1),
\eea
which are functions of ${\cal A_H}, {\cal B_H}, {\cal C_H}, {\cal D_H}$ and 
${\cal E_H}$ and ${\cal A}, {\cal B}$ and ${\cal C}$ respectively 
(see Eqs.~\eqref{eq:pol}, \eqref{eq:pol2} and \eqref{eq:GammaCM}\,). Besides,
\be
p'_0(\omega)= \frac{M^{\prime}\sqrt{\omega^2-1}}{M\sqrt{q^2}}\left(
\frac{2M{\cal A_H}(\omega)}{1-m^2_\tau/q^2}+ M_\omega {\cal C_H}(\omega)\right),
\quad p'_1(\omega)= \frac{M^{\prime\, 2}(\omega^2-1)}{M\sqrt{q^2}}{\cal C_H}(\omega). 
\label{eq:defpprima1}
\ee

Now, we are in conditions to address the integration of the $\phi_d$ and 
$\cos\theta_\tau$  (or $s_{13}$)\footnote{The pair $(\omega,\cos\theta_\tau)$ 
fixes  $s_{13}$ and $d\omega\, ds_{13}= M M' \left(1-m^2_\tau/q^2\right)
\sqrt{\omega^2-1}\,d\omega\, d\cos\theta_\tau$. },
\begin{itemize}
\item For the hadronic modes, the $\eta_{d=\mu\bar\nu_\mu}$ and 
$\chi_{d=\mu\bar\nu_\mu}$  functions contain an energy conservation Dirac delta function, 
which can be used to integrate $d\phi_d$,
\be
  \frac{\delta\big[\sqrt{m_\tau^2+\vec{k}^{\,\prime\,2}}-\sqrt{m_d^2+\vec{p}_d^{\,2}}
  -|\vec{k'}-\vec{p}_d|\big]}{|\vec{k'}-\vec{p}_d|} = 
  \frac{\delta\left(\cos\phi_d-z_0\right)}{|\vec{k'}||\vec{p}_d|
  \sin\theta_\tau\sin\theta_d}, \label{eq:aux}
\ee
with $z_0$ introduced in Eq.~\eqref{eq:defz0}. Now, we use 
\bea
\int_0^{2\pi} d\phi_d \,\delta\left(\cos\phi_d-z_0\right) g\left(\cos\phi_d,\sin\phi_d\right)&=& \int_{-1}^1 dz \frac{\delta(z-z_0)}{\sqrt{1-z^2}}\Big[g\left(z,\sqrt{1-z^2}\right)\nonumber \\
&&+g\left(z,-\sqrt{1-z^2}\right)\Big]
\eea
to carry out the  $\phi_d$ integration. As a result, we find that the $P_{TT}$ contribution vanishes and considering $\sin\theta_\tau\sin\theta_d$ in Eq.~\eqref{eq:aux}, we obtain a common factor 
\be 
\frac{1}{\sin\theta_\tau\sin\theta_d\sqrt{1-z^2_0}} = 
\frac{1}{\sqrt{\left[\cos(\theta_d-\theta_{\tau d}^{\rm CM})-\cos\theta_\tau\right]
\left[\cos\theta_\tau-\cos(\theta_d+\theta_{\tau d}^{\rm CM})\right]}}. \label{eq:common-factor}
\ee
Now, the integration over $\cos\theta_\tau$ can be  easily done using the analytical integrals compiled in Eq.~\eqref{eq:integrals}, 
which appear when  the factor from the above equation, 
the longitudinal and transverse polarization components given in Eq.~\eqref{eq:PLPT}, the expression for the CM  
$d^2\Gamma_{SL}/(d\omega d\cos\theta_\tau)$ in Eq.~\eqref{eq:GammaCM} and the limits of Eq.~\eqref{eq:theta-limits} are considered.
\item In the leptonic mode, the $\eta_{d=\mu\bar\nu_\mu}$ and 
$\chi_{d=\mu\bar\nu_\mu}$  functions do not contain any Dirac delta. However, 
many of the results of the previous case can also be used here. To fix the  OXZ 
plane, it is necessary to detect both the hadron and tau momenta, and given the 
expected experimental difficulties to reconstruct the $\tau$-trajectory, we 
integrate the azimuthal $\phi_d$ angle.  It holds   
\bea
\int_0^{2\pi} d\phi_d\,  g\left(\cos\phi_d,\sin\phi_d\right)&=& 
\int_{-1}^1 dz_0 \frac{g\left(z_0,\sqrt{1-z_0^2}\right)+g\left(z_0,-\sqrt{1-z_0^2}
\right)}{\sqrt{1-z_0^2}}.
\eea
Here, we see  again that the contribution of the CP-violating polarization component 
$P_{TT}$ cancels out, and some kind of $\phi_d$-asymmetry, similar to that of 
Eq.~\eqref{eq:PTTterm}, would be needed to isolate this term. In  addition, 
$z_0=z_0(x)$ and the change of variables 
\be
dz_0 = \frac{dx}{|\partial x/ \partial z_0|} =\frac{dx}{2\gamma\beta \sqrt{\gamma^2\xi_d^2-y^2} \sin\theta_\tau\sin\theta_d}
\ee
completes the reconstruction of the factor of Eq.~\eqref{eq:common-factor}. 
Altogether,  we can integrate $d\cos\theta_\tau$ in a similar way to what  we 
have shown  above in detail 
for the hadronic channel. The main difference is that to obtain  
the $d^3\Gamma_d/(d\omega  d\xi_d d\cos\theta_d)$ distribution of 
visible variables, we  still have to integrate over the $x$ variable, 
taking into account the accessible  phase-space $\varphi(\omega,\xi_d,x)$ given in Eq.~\eqref{eq:phase-space}.
\end{itemize}
Thus,  finally we obtain for $y^2\le m_\tau^2/\sqrt{q^2}$ 
\bea
\frac{d^3\Gamma_d}{d\omega  d\xi_d d\cos\theta_d} & = & {\cal B}_{d}
\frac{d\Gamma_{\rm SL}}{d\omega} \Big\{  F^d_0(\omega,\xi_d)+ F^d_1(\omega,\xi_d) 
\cos\theta_d + F^d_2(\omega,\xi_d)P_2(\cos\theta_d)\Big\}, \label{eq:visible-distr}
\eea
with  $P_2$  the Legendre polynomial of order two. The contributions of the different waves for the hadronic modes read
\bea 
F^{\pi,\rho}_0(\omega,\xi_d)&=& \frac{1}{2\beta(1-y^2)}\,\left(1+\frac{a_{\pi,\rho}\left(1+y^2-2\xi_d\right)}{\beta(1-y^2)}\, \langle P_L^{\rm CM}\rangle(\omega)\right),\nonumber \\
F^{\pi,\rho}_1(\omega,\xi_d)&=& \frac{3}{2\beta(1-y^2)}\,\Bigg\{\frac{a_1\costd}{3a_0+a_2}+\frac{a_{\pi,\rho}}{1-y^2}\Bigg(\frac{1+y^2-2\xi_d}{\beta} \frac{\delta a_1}{3a_0+a_2}\costd\nonumber \\
&&-\frac{8\sqrt{\gamma^2\xi_d^2-y^2}}{3\pi}\left[\sin\theta_{\tau d}^{\rm CM}\right]^2\langle P_T^{\rm CM}\rangle\Bigg)\Bigg\}, \nonumber \\
F^{\pi,\rho}_2(\omega,\xi_d)&=& \frac{1}{\beta(1-y^2)}\,\Bigg\{\frac{a_2}{3a_0+a_2}P_2(\costd)+\frac{a_{\pi,\rho}}{1-y^2}\Bigg(\frac{1+y^2-2\xi_d}{\beta} \frac{\delta a_2}{3a_0+a_2}
P_2(\costd)\nonumber \\
&&-\sqrt{\gamma^2\xi_d^2-y^2}\left[\sin\theta_{\tau d}^{\rm CM}\right]^2\costd 
\frac{3p'_1}{3a_0+a_2}\Bigg)\Bigg\}. \label{eq:coeffs-1}
\eea
 For the lepton channel, for which we remind that $\costd$ depends also on the integration variable $x$, we have 
\bea
F^{\mu\bar\nu_\mu}_0(\omega,\xi_d)&=& \frac{1}{\beta f(y)}\int dx\, \varphi(\omega,\xi_d,x)\Big\{G_1(x,y)+\langle P_L^{\rm CM}\rangle(\omega) G_L(x,y) \Big\},\nonumber \\
F^{\mu\bar\nu_\mu}_1(\omega,\xi_d)&=& \frac{3}{\beta f(y)}\int dx\, \varphi(\omega,\xi_d,x)\Bigg\{\frac{a_1(\omega)G_1(x,y)+\delta a_1(\omega)G_L(x,y)}{3a_0(\omega)+a_2(\omega)}\costd \nonumber \\
&&-\frac{8\sqrt{\gamma^2\xi_d^2-y^2}\,}{3\pi}\langle P_T^{\rm CM}\rangle(\omega)\left[\sin\theta_{\tau d}^{\rm CM}\right]^2G_T(x,y) \Bigg\},\nonumber \\
F^{\mu\bar\nu_\mu}_2(\omega,\xi_d)&=& \frac{2}{\beta f(y)}\int dx\, \varphi(\omega,\xi_d,x)\Bigg\{\frac{a_2(\omega)G_1(x,y)+\delta a_2(\omega)G_L(x,y)}{3a_0(\omega)+a_2(\omega)}P_2(\costd) \nonumber \\
&&-\sqrt{\gamma^2\xi_d^2-y^2} \frac{3p'_1(\omega)}{3a_0(\omega)+a_2(\omega)}\left[\sin\theta_{\tau d}^{\rm CM}\right]^2\costd G_T(x,y) \Bigg\},\label{eq:coeffs-2}
\eea
with the functions $G_1(x,y)= x(3-2x)-y^2(4-3x)$, 
$G_L(x,y)=\left(x-2\xi_d\right)\left(1+3y^2-2x\right)/\beta$ and 
$G_T(x,y)=\left(1+3y^2-2x\right)$. The integrations on the variable $x$ 
are straightforward in all cases since only polynomials are involved. The actual 
expressions,  lengthy ones in some cases,  have been collected  in 
Appendix~\ref{app:coeff2} where we also provide, more visual, two-dimensional graphic 
representations of their $(\omega, \xi_d)$ dependence.

Note that  $G_1$ and $G_L$ provide the overall normalization
\be 
\frac{2}{\beta f(y)}\int d\xi_d\int dx \varphi(\omega,\xi_d,x)G_1(x,y) = 1, 
\quad \int d\xi_d\int dx \varphi(\omega,\xi_d,x)G_L(x,y) = 0,
\ee
where the equivalent ones for the hadronic modes are trivially satisfied.  Upon 
integration on $\cos\theta_d$, and taking the massless limit $y\to 0$, we recover 
the results of Ref.~\cite{Tanaka:2010se} identifying  $2F^d_0(\omega,\xi_d)$ here 
with $f(q^2,\xi)+P_L(q^2)g(q^2,\xi)$ in that reference. 
 Note that, besides  some differences in the notation, there is a sign change 
 in the definition of the polarization terms we provide here with respect   to the ones in
 Refs.~\cite{Tanaka:2010se, Alonso:2016gym}.

In the sequential $\tau$-decay distribution of Eq.~\eqref{eq:visible-distr}, all 
information on the  $b\to c \tau \bar\nu_\tau$ transition is encoded  in the 
$\omega$-dependent functions $a_i, \delta a_i, p'_1$ and 
$\langle P^{\rm CM}_{L,T}\rangle$. As already mentioned, they can be expressed in terms of 
the ${\cal A}, {\cal B}$ and ${\cal C}$ and ${\cal A_H},{\cal B_H},{\cal C_H}, 
{\cal D_H}$ and ${\cal E_H}$ ones introduced here in Eqs.~\eqref{eq:pol} and 
\eqref{eq:pol2}. The first set of three functions (or equivalently $a_{0,1,2}$) 
determine the unpolarized $H_b\to H_c \tau \bar\nu_\tau$ semileptonic 
$d^2\Gamma_{SL}/(d\omega d\cos\theta_\tau)$ distribution\footnote{Note that $a_0, a_1$ 
and $a_2$ could be obtained from the terms in 
Eq.~\eqref{eq:visible-distr} which come from the $\eta_d$ contribution  of 
Eq.~\eqref{eq:defGamma-fin}, since $d\Gamma_{\rm SL}/d\omega\,\propto\, (a_0+a_2/3)$.}. 
The helicity-asymmetry coefficients $\delta a_{i=0,1,2}(\omega)$ involve only the second 
set of functions,  while $p'_1$  only involves  ${\cal C_H}$.
Finally, the angular weighted averages of the longitudinal and transverse components 
of the tau polarization vector are exhaustively discussed  in Ref.~\cite{Penalva:2021gef},
 where (Appendix B) analytical expressions   in terms of ${\cal A_H},{\cal B_H},
 {\cal C_H}, {\cal D_H}$ and ${\cal E_H}$ and the combination $(3a_0+a_2)$, can be 
 found. Note also that $\langle P^{\rm CM}_L\rangle = \frac{3\delta a_0+\delta a_2}
 {3a_0+a_2}$
and $\langle P^{\rm CM}_T\rangle = \frac{3\pi p'_0}{4 (3a_0+a_2)}$.
Thus for fixed $\omega$, the combined $(\xi_d,\cos\theta_d$) analysis of the  
$d^3\Gamma_d/(d\omega  d\xi_d d\cos\theta_d)$ distribution provides, in addition 
to $a_0$, $a_1$ and $a_2$, five independent observables  
$\delta a_{i=0,1,2}(\omega)$, $\langle P^{\rm CM}_{T}\rangle$ and 
$p'_1$, which can be used to fully determine the five 
${\cal A_H},{\cal B_H},{\cal C_H}, {\cal D_H}$ and ${\cal E_H}$ $\omega$-functions, 
and that give the maximal information on NP  in the  $b\to c \tau \bar\nu_\tau$ 
transition, without considering CP-violation. 
 CP non-conserving  contributions,  encoded in the $P_{TT}$ component of the
tau polarization vector, canceled out when we carried out the $\phi_d$ integration. 
As noted above, the measurement of such angle would require to detect the $\tau$-three 
momentum. Hence, the ${\cal F_H}$ and ${\cal G_H}$ functions, which are responsible 
for CP violation, 
 are only accessible by including
additional information. For $\bar B \to D^*$, some CP-odd observables (triple product 
asymmetries), defined using angular distributions involving the kinematics of the 
products of the $D^*$ decay,  have also been  presented~\cite{Duraisamy:2013pia, 
Duraisamy:2014sna, Ligeti:2016npd, Bhattacharya:2020lfm}. These asymmetries  are 
sensitive to the relative phases of the Wilson coefficients, as are the ${\cal F_H}$ and 
${\cal G_H}$ scalar functions.

We note that the expression found here for the visible  distribution in 
Eq.~\eqref{eq:visible-distr} recovers the results presented in 
Refs.~\cite{Alonso:2016gym,Alonso:2017ktd, Asadi:2020fdo}  for $\bar B \to D^{(*)}$ 
transitions, accounting  in the leptonic mode also for effects due to the finite  
mass of the outgoing muon/electron from the tau decay. Thus, there is a correspondence 
between $n(q^2)$ and the asymmetries $A_{FB}, P_L, P_\perp, Z_L, Z_\perp, Z_Q$ and 
$A_Q$ introduced in Eq.~(1.1) of Ref.~\cite{Asadi:2020fdo} for the hadron modes, 
and  $a_{i=0,1,2}$, $\delta a_{i=0,1,2}, p'_1$ and $\langle P^{\rm CM}_{T}\rangle$ 
(or  ${\cal A}, {\cal B}$ and ${\cal C}$, and ${\cal A_H},{\cal B_H},{\cal C_H}, 
{\cal D_H}$ and ${\cal E_H}$) used here.  In fact, the relationships become apparent
 when comparing equations (3.12) and (3.13) of \cite{Asadi:2020fdo} and the 
 Eqs.~\eqref {eq:visible-distr}-\eqref {eq:coeffs-1} in this work,
\bea
n(q^2) &\propto&  \left(3a_0(\omega)+a_2(\omega)\right),\quad A_{FB}(q^2) = \frac{3 a_1(\omega)/2}{3a_0(\omega)+a_2(\omega)},\quad A_{Q}(q^2) = \frac{a_2(\omega)}{3a_0(\omega)+a_2(\omega)}\nonumber \\
P_L(q^2)&=& - \langle P^{\rm CM}_L\rangle, \quad Z_{L}(q^2) = -\frac{3 \delta a_1(\omega)/2}{3a_0(\omega)+a_2(\omega)},\quad Z_{Q}(q^2) = -\frac{\delta a_2(\omega)}{3a_0(\omega)+a_2(\omega)}\nonumber \\
P_\perp(q^2)&=& - \langle P^{\rm CM}_T\rangle, \quad Z_\perp(q^2) =  
- \frac{p'_1(\omega)}{3a_0(\omega)+a_2(\omega)}. \label{eq:observ}
\eea
\begin{table}[t]
\begin{center}
\begin{tabular}{|c|c|c|}
    \hline
Unpolarized $\tau$ &\rule{0pt}{3.5ex} $\mathcal{A},\, \mathcal{B},\,
 \mathcal{C}$ & $ n,\, A_{{FB}},\, A_Q$ \\\hline
Polarized $\tau$ & \rule{0pt}{3.0ex}\hspace{-3mm} $\mathcal{A_H},\, 
\mathcal{B_H},\, \mathcal{C_H},\, \mathcal{D_H},\, \mathcal{E_H}$  & $\langle P_L^{\rm CM}\rangle,\, \langle P_T^{\rm CM}\rangle,\, Z_L,\, Z_Q,\, Z_\perp$\hspace{-1.5mm} \\
Polarized $\tau$,\,complex Wilson coeff. & \rule{0pt}{3.0ex}\hspace{-3mm} $\mathcal{F_H},\, \mathcal{G_H}$ 
& $\langle P_{TT}\rangle,\, \ Z_T$ \\
 \hline
\end{tabular}
    \caption{ For each row, the observables in the second column contain the same 
    physical information as those compiled in the third one. The quantities in the first row determine the decay 
    for unpolarized taus, while the ones in the second and third rows describe 
    the decays for polarized taus. Finally, observables in the third row
     are zero unless the Wilson coefficients are complex.}
   \label{tab:equival}
   \end{center}
\end{table}
In addition, the remaining two asymmetries $P_T$ (related to our $\langle P_{TT}\rangle$) and $Z_T$ mentioned in 
\cite{Asadi:2020fdo} should correspond to linear combinations of the
${\cal F_H}$ and ${\cal G_H}$ scalar functions within the tensor formalism presented 
in Sec.~\ref{sec:hadron-tensor-formalism}. As mentioned above, these CP-violating 
contributions  
cancel out on integration over the azimuthal angle $\phi_d$, which measurement 
would require  detecting both the hadron and tau momenta. 
These relations are  schematically shown in Table~\ref{tab:equival} where, 
for each row, the 
observables in the second and the third columns are equivalent in the sense that they
contain the same physical information. In addition, we show which quantities
determine the decay for unpolarized (first row)
and polarized (second and third rows) final taus, as well as which of them
 require complex Wilson coefficients (third row).

As we have seen, the tensor scheme  
of Sec.~\ref{sec:hadron-tensor-formalism} allows to straightforwardly compute 
the $d^3\Gamma_d/(d\omega  d\xi_d d\cos\theta_d)$ distribution of visible  
variables  for any $H_b\to H_c \tau\, (d \nu_\tau) \bar\nu_\tau$ decay, including 
  left- and/or right-handed NP neutrino operators.

Each of the observables in Eq.~\eqref{eq:observ}, which are embedded in one of the  
$F^d_{0,1,2}(\omega,\xi_d)$ partial waves introduced in Eq.~\eqref {eq:visible-distr}, 
are affected by 
kinematical $C_n, C_{A_{FB}}, \dots ,
C_{Z_\perp}$ coefficients.
Specifically, one can write
\begin{eqnarray}
 F^d_0(\omega,\xi_d) &=& C_n^d(\omega,\xi_d)+C_{P_L}^d(\omega,\xi_d)\,\langle P^{\rm CM}_L\rangle(\omega), \nonumber \\
 F^d_1(\omega,\xi_d) &=& C_{A_{FB}}^d(\omega,\xi_d)A_{FB}(\omega)+C_{Z_L}^d(\omega,\xi_d)Z_L(\omega)
 + C_{P_T}^d(\omega,\xi_d)\,\langle P^{\rm CM}_T\rangle(\omega), \nonumber \\ 
 F^d_2(\omega,\xi_d) &=& C_{A_Q}^d(\omega,\xi_d)A_{Q}(\omega)+
 C_{Z_Q}^d(\omega,\xi_d)Z_Q(\omega)+ C_{Z_\perp}^d(\omega,\xi_d)Z_\perp(\omega).
 \label{eq:coeff}
\end{eqnarray}
Those coefficients are  tau-decay mode dependent and in the case of the
$\pi$ and $\rho$ hadronic ones they can be easily read out from 
Eq.~\eqref{eq:coeffs-1}. The corresponding expressions for the fully 
$\mu\bar\nu_\mu$ leptonic mode are collected in Appendix ~\ref{app:coeff2}.
 There, in Figs.~\ref{fig:coeff1} and \ref{fig:coeff2}, we also provide, for all three tau-decay modes considered 
 in this work, their $(\omega,\xi_d)$-graphic 
representations. What we actually show are the products of each of the
coefficients times the kinematical factor ${\cal K}(\omega)=\sqrt{\omega^2-1}\,(1-m_\tau^2/q^2)^2$
that makes part of the
$d\Gamma_{\rm SL}/d\omega$ semileptonic decay width. 
The visual inspection 
of the different panels in Figs.~\ref{fig:coeff1}  and \ref{fig:coeff2} provides 
immediate information 
on which regions of the available  $(\omega, \xi_d)$ phase-space might result 
more sensitive to (or adequate to extract from) each of the observables of 
Eq.~\eqref{eq:observ}. 
Taking into account the numerical values of the coefficients,  the hadron
 channels, and in particular the pion mode, seem, in general,  to be more  
 convenient to determine the 
 semileptonic quantities  of Eq.~\eqref{eq:observ}. Probably, the best strategy 
 would be to perform a 
multi-parametric fit of the 
$d^3\Gamma_d/(d\omega  d\xi_d d\cos\theta_d)$ experimental data to the theoretical predictions
 of Eqs.~\eqref {eq:visible-distr}-\eqref {eq:coeffs-1}.

\section{Results for the visible pion/rho/muon distributions in the
presence of NP right-handed  neutrino operators}
\label{sec:results}
\begin{table}
\begin{center}
\begin{tabular}{c|ccccc}
                        &  SM  & L Fit 7 \cite{Murgui:2019czp}&  R S3 \cite{Mandal:2020htr}& R S5a \cite{Mandal:2020htr}& R S7a \cite{Mandal:2020htr}
                       \\\hline\tstrut
 $\Gamma_{e(\mu)}$ & ~$2.15\pm 0.08$ & $-$ & $-$\\ \tstrut 
 $\Gamma_\tau$ & ~$0.715\pm 0.015$ 
 &~  $0.89\pm 0.05$&~ $0.81\pm0.04$ &~ $0.81\pm0.04$ &~ $0.81\pm0.06$ \\ \tstrut
 ${\cal R}_{\Lambda_c}$ & ~$0.332 \pm 0.007$   & ~ $0.41\pm 0.02$ &  ~ $0.378\pm0.017$ & ~ $0.378\pm0.017$& ~ $0.38\pm 0.03$\\ \hline
\end{tabular}
\end{center}
\caption{Total decay widths $\Gamma_\tau=\Gamma\left(\Lambda_b\to\Lambda_c\tau\bar\nu_\tau\right)$ 
and  $\Gamma_{e(\mu)}=\Gamma\left(\Lambda_b\to\Lambda_c\, e(\mu)\bar\nu_{e(\mu)}\right)$ [units of  $\left(10\times |V_{cb}|^2 {\rm ps}^{-1}\right)$]
and  ratios ${\cal R}_{\Lambda_c} =\Gamma\left(\Lambda_b\to\Lambda_c\tau\bar\nu_\tau\right)
/\Gamma\left(\Lambda_b\to\Lambda_c\, e(\mu)\bar\nu_{e(\mu)}\right)$ 
obtained in the SM, 
 the NP model Fit 7 of Ref.~\cite{Murgui:2019czp}, which involves 
only left-handed neutrinos, and other three ones taken from 
Ref.~\cite{Mandal:2020htr}, where all included NP operators use right-handed 
neutrino fields (see text for details). Errors induced by the uncertainties in
the form-factors and  Wilson Coefficients are added in quadrature.}
\label{tab:ratios}
\end{table}
 We will consider three different extensions of the SM including right-handed 
neutrino fields, that correspond to  the more promising ones, in terms of the pulls from the 
SM hypothesis, among those discussed in Ref.~\cite{Mandal:2020htr}. We will show 
predictions for the observables collected in Eq.~\eqref{eq:observ}, extracted 
from the visible distributions of the tau-decay massive products, for the baryon 
$\Lambda_b\to \Lambda_c$ reaction. We will compare these NP results with those 
obtained in the SM, and within an extension of the SM determined by Fit 7 of 
Ref.~\cite{Murgui:2019czp} constructed only with left-handed neutrino operators. 
We focus in the baryon decay for the sake of brevity, since some of the observables 
of Eq.~\eqref{eq:observ} with right-handed neutrinos were already shown in  
\cite{Mandal:2020htr} for the meson $\bar B\to D^{(*)}$ semileptonic decays, 
where the extensions considered in this work were fitted. Moreover, 
$\bar B_c \to \eta_c, J/\psi$ transitions, studied in our previous works, 
follow in general a similar pattern to that seen in the analog ones from 
$\bar B$-meson decays. In addition, we will not include results from 
Fit 6 of Ref.~\cite{Murgui:2019czp}, as we did in previous 
studies~\cite{Penalva:2019rgt,Penalva:2020xup,Penalva:2020ftd,Penalva:2021gef}, 
since this NP scenario, which involves only left-handed neutrinos, provides 
polarized tau-distributions more similar to the SM ones than those obtained with 
the model Fit 7 of the same reference.
 
The $\Lambda_b\to \Lambda_c$ form factors used here are directly obtained (see Appendix E
 of  Ref.~\cite{Penalva:2020xup}) from those calculated in the  lattice quantum Chromodynamics (LQCD) simulations of Refs.~\cite{Detmold:2015aaa} (vector and axial ones) and \cite{Datta:2017aue} (tensor NP form factors) using $2+1$ flavors of
 dynamical domain-wall fermions.  The NP scalar and pseudoscalar
  form factors   are directly related to the vector and axial ones and we 
use  Eqs.~(2.12) and (2.13) of Ref.~\cite{Datta:2017aue} to evaluate them. We 
use the errors and  statistical correlation-matrices, provided in the LQCD papers,
 to Monte Carlo transport the form-factor uncertainties to the different 
 observables 
 shown in this work. For the model Fit 7 and the right-handed neutrino 
 scenarios, we shall use  statistical samples of Wilson coefficients 
 selected such that the $\chi^2$-merit function computed in 
 Refs.~\cite{Murgui:2019czp} and \cite{Mandal:2020htr}, respectively, 
 changes at most by one unit from its value at the fit minimum. Both 
 sets of errors are then added in quadrature and displayed in the 
 predictions.

The  analysis carried out in Refs.~\cite{Mandal:2020htr, 
Murgui:2019czp} considers only input from the  $\bar B \to D^{(*)}$ 
meson transitions. Namely, the most recent world-average correlated values of 
${\cal R}_D$ and ${\cal R}_{D^*}$ from the Heavy Flavor Averaging Group 
\cite{HFLAV:2019otj}, the value of the $q^2$-integrated  lepton 
polarization asymmetry [$P_\tau(D^*) = \int dq^2 (d\Gamma_{\rm SL}/dq^2)
P_L(q^2)/\Gamma_{\rm SL}]$ and the longitudinal  $D^*$ polarization, 
$F_L^{D^*}$, measured by Belle~\cite{Hirose:2016wfn, Belle:2019ewo}, 
and the $q^2$ distributions of the $D$ and $D^*$ 
mesons~\cite{BaBar:2013mob,Belle:2015qfa}, together with bounds from 
the leptonic decay $\bar B_c \to \tau \bar\nu_\tau$.
 
The scenario 3 of Ref.~\cite{Mandal:2020htr} induces exclusively 
$b\to c \tau \bar\nu_{\tau R}$  right-handed neutrino NP interactions, 
and particularly the vector boson mediator only contributes to the vector 
Wilson coefficient $C^V_{RR}$. It trivially follows that for any $H_b\to 
H_c$ decay, the  ${\cal A}, {\cal B}$ and ${\cal C}$ [${\cal A_H},
{\cal B_H},{\cal C_H}, {\cal D_H}$ and ${\cal E_H}$] functions will 
take the SM values scaled by a factor $(1+|C^V_{RR}|^2)$ 
[$(1-|C^V_{RR}|^2)$]. Therefore, $n(q^2)$ and consequently the total 
semileptonic width in the tau mode will be enhanced by  $(1+|C^V_{RR}|^2)$ 
with respect to the SM result. No signatures of NP will appear in the  
$A_{FB}(q^2)$ and $A_{Q}(q^2)$ pion/rho/muon angular asymmetries, while 
$ P_L(q^2), Z_{L}(q^2), Z_{Q}(q^2), P_\perp(q^2)$ and  $Z_\perp(q^2)$ 
will be scaled down by the factor $(1-|C^V_{RR}|^2)/(1+|C^V_{RR}|^2)$ as 
compared to the SM predictions.

The presence of a vector leptoquark at the high-energy scale leads to the scenario 5 of Ref.~\cite{Mandal:2020htr}, where  
both left- and right-handed neutrino operators contribute at the $m_b$ scale. 
In  Fit 5a, only  right-handed neutrino fields are considered, which 
give rise to non-vanishing $C^V_{RR}$ and $C^S_{LR}$ Wilson coefficients, 
though the latter one is determined in Ref.~\cite{Mandal:2020htr} with large errors. 
Including also left-handed neutrino operators does not improve the $\chi^2$ and the 
left-handed Wilson coefficients are compatible with zero within one sigma.

A scalar leptoquark is considered in  scenario 7a of  Ref.~\cite{Mandal:2020htr}, 
where  a solution dominated by $C^V_{RR}$, with  an additional Wilson coefficient 
$C^T_{RR}$ compatible with zero within one sigma, and $C^S_{RR} \approx -8 C^T_{RR}$,
 is found. As in the previous case, 
adding the left-handed operators that contribute in the presence of the scalar 
leptoquark leads to a solution
compatible with vanishing left-handed Wilson coefficients.

We note that none of these three possibilities with only right-handed neutrino 
fields can generate values of the longitudinal $D^*$ polarization within its 
current one sigma experimental range. NP models, like Fit 7 of  
Ref.~\cite{Murgui:2019czp}, with a significant contribution from $C^V_{RL}$ 
reduces the tension with the $F_L^{D^*}$ measurement. We should also mention 
that for the right-handed neutrino scenarios 3, 5a and 7a, the Wilson coefficient
 $C^V_{RR}$ is found to be in the range $0.3-0.5$, taking into account uncertainties, 
 and such relatively large values are
challenged by mono-tau searches at LHC~\cite{Greljo:2018tzh}.

\begin{figure*}[!h]
\centering
\includegraphics[scale=0.75]{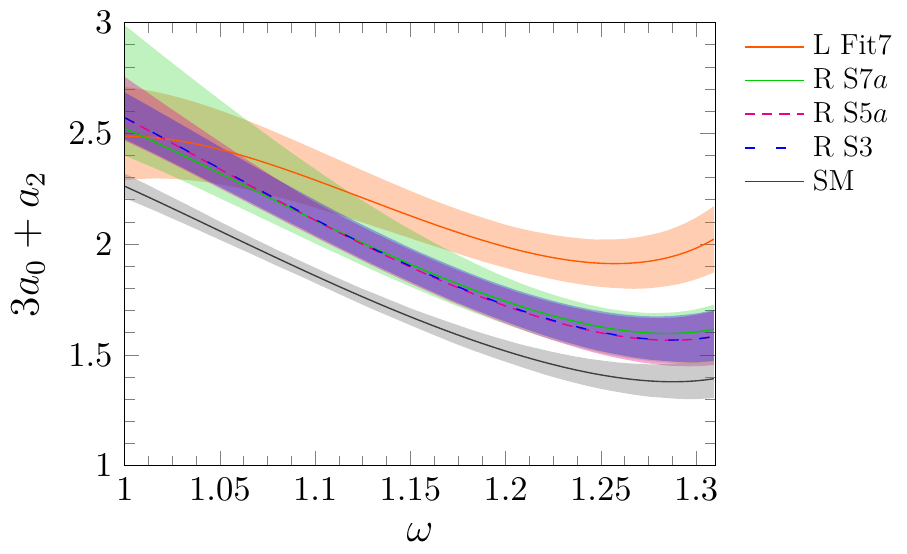}\hspace{0.05cm}
\includegraphics[scale=0.75]{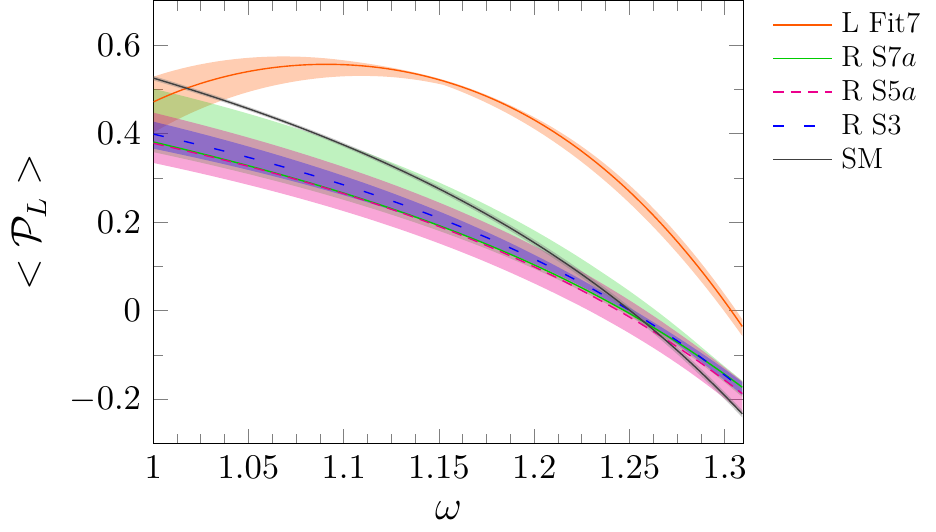}\\
\hspace*{-1.5cm}\includegraphics[scale=0.75]{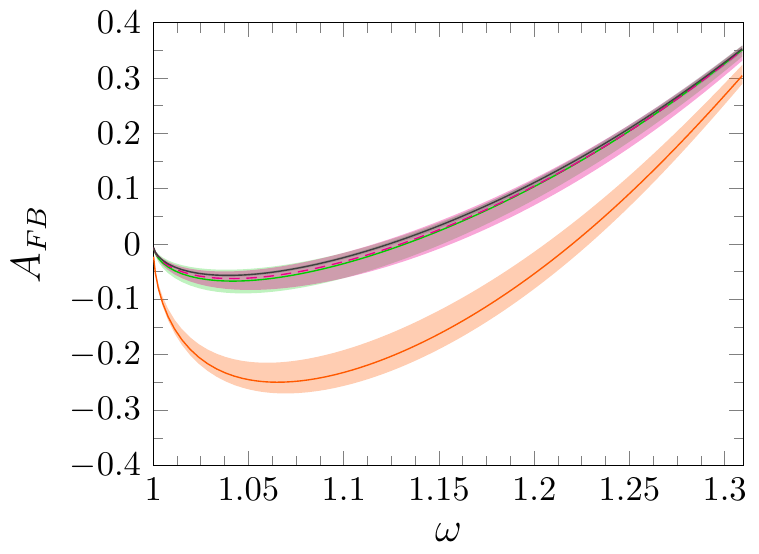}\hspace{1.7cm}\includegraphics[scale=0.75]{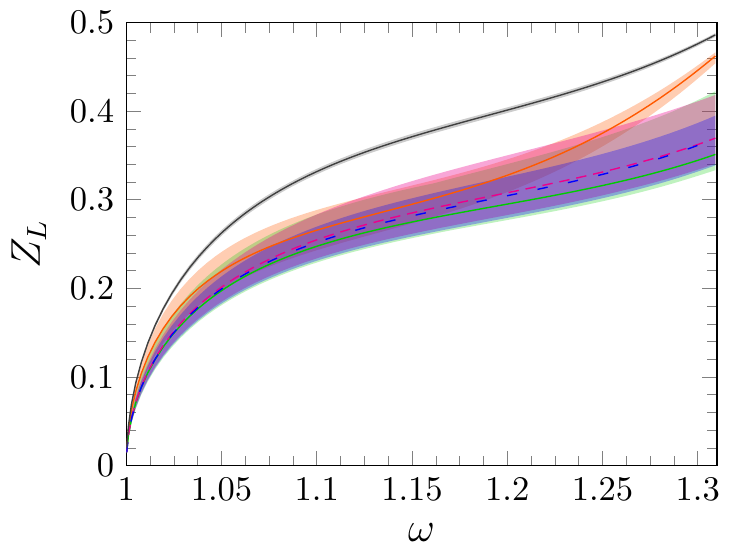}\\
\hspace*{-1.4cm}\includegraphics[scale=0.75]{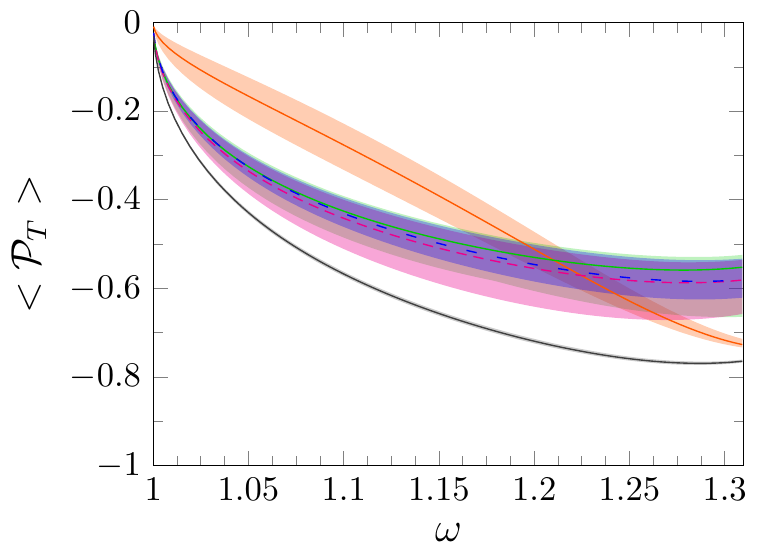}\hspace{1.2cm}
\includegraphics[scale=0.75]{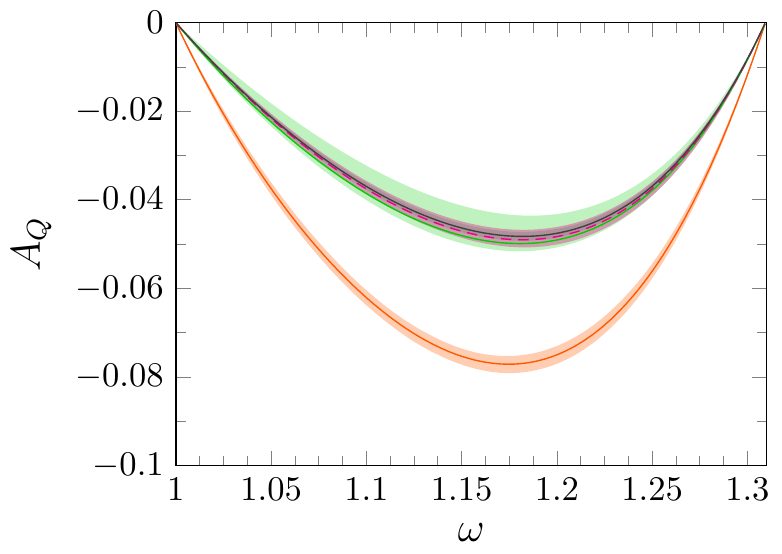}\\
\hspace*{-1.2cm}\includegraphics[scale=0.75]{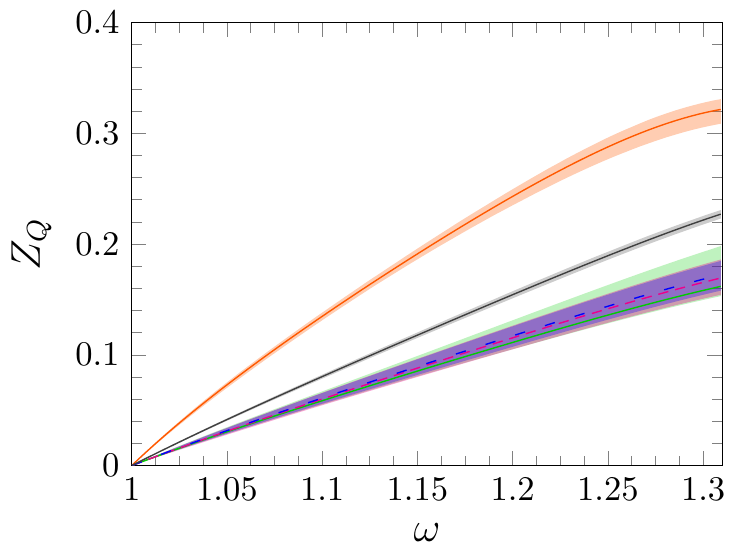}
\hspace{1.6cm}\includegraphics[scale=0.75]{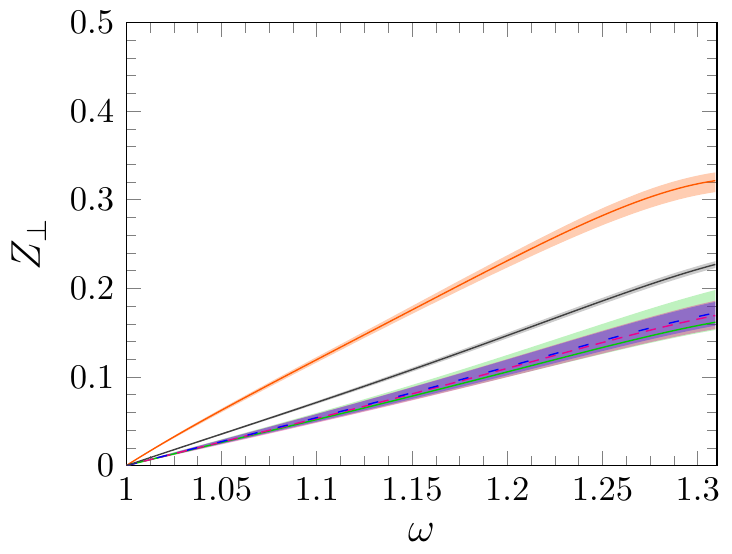}
\caption{Predictions for the semileptonic observables $(3a_0+a_2), \langle P^{\rm CM}_L\rangle,
 A_{FB}, Z_{L}, \langle P^{\rm CM}_T\rangle, A_{Q}, Z_{Q}$ and  $Z_\perp$ introduced in 
 Eq.~\eqref{eq:observ}, as a function of $\omega$, for the  $\Lambda_b\to\Lambda_c\tau
 \bar\nu_\tau$ semileptonic decay. We show results obtained within the SM, the 
 NP model Fit 7 of Ref.~\cite{Murgui:2019czp}, which involves only left-handed 
 neutrinos, and other three ones taken from Ref.~\cite{Mandal:2020htr}, where all 
 included NP operators use right-handed neutrino fields. Error bands account for uncertainties induced by both form-factors and fitted Wilson coefficients (added in quadrature). The right-handed neutrino scenario 3 and the SM lead to the same results for 
the $A_{FB}(q^2)$ and $A_{Q}(q^2)$ angular asymmetries.}
\label{fig:ZsyAes}
\end{figure*}

In Fig.~\ref{fig:ZsyAes}, we show predictions for $(3a_0+a_2), \langle P^{\rm CM}_L\rangle, A_{FB}, 
Z_{L}, \langle P^{\rm CM}_T\rangle, A_{Q}, Z_{Q}$ and  $Z_\perp$ defined in Eq.~\eqref{eq:observ}, 
for the SM, the NP model Fit 7 of Ref.~\cite{Murgui:2019czp}  and for scenarios 3, 5a and 7a 
from Ref.~\cite{Mandal:2020htr}, which incorporate NP operators constructed using 
right-handed neutrino fields. All these quantities can be obtained from the   
$S,P$- and $D$- wave contributions ($F^d_{0,1,2}(\omega,\xi_d)$) to the 
$d^3\Gamma_d/(d\omega  d\xi_d d\cos\theta_d)$ differential distribution, associated 
to any of the $\Lambda_b\to \Lambda_c\tau (\pi \nu_\tau, \rho \nu_\tau, 
\mu \bar \nu_\mu \nu_\tau )\bar\nu_\tau$ sequential decays studied in the previous section. 
As already stressed, in the absence of CP-violation, this set of observables 
provides the maximal information (scalar functions ${\cal A}, {\cal B}, 
{\cal C}$, and ${\cal A_H},{\cal B_H},{\cal C_H}, {\cal D_H}, {\cal E_H}$ 
in Eqs.~\eqref{eq:pol} and \eqref{eq:pol2}) which can be extracted from the 
analysis of the semileptonic $\Lambda_b\to \Lambda_c\tau \bar\nu_\tau$ 
transition, considering the most general polarized state for the final tau (see Table~\ref{tab:equival}).

The $(3a_0+a_2)\, \propto \, d\Gamma_{\rm SL}/d\omega$ distributions displayed in the first 
panel of the figure lead to the results for the integrated widths compiled in 
Table~\ref{tab:ratios}, and it cannot disentangle among the three right-handed 
neutrino 
scenarios examined in this work. However, these distributions are useful to 
efficiently 
separate between the SM and any of its extensions fitted to the violations of LFU 
observed in 
$B$-meson decays. Moreover, for relatively large values of $\omega> 1.15$,  neutrino 
left-handed and right-handed NP models predict significantly different   
$d\Gamma_{\rm SL}/d\omega$ differential decay widths. 

In the other seven panels of Fig.~\ref{fig:ZsyAes}, we show tau angular and polarization asymmetries, as a function of $\omega$. 
Relative errors in these observables are smaller than for  $(3a_0+a_2)$, since 
they are defined as ratios for which the form-factor uncertainties largely cancel 
out. None of these observables are useful in distinguishing between the three 
scenarios with right-handed neutrinos taken from Ref.~\cite{Mandal:2020htr}. 
Furthermore, the angular asymmetries $A_{FB}(q^2)$ and $A_{Q}(q^2)$, and to some 
extent the longitudinal polarization average  $\langle P^{\rm CM}_L\rangle $, 
 do not distinguish   between  SM and these latter NP models either. The predictions 
 from Fit 7 of Ref.~\cite{Murgui:2019czp} are significantly different from those 
 obtained within the SM and the right-handed neutrino models in all cases, except 
 for  $Z_L$, where all the extensions of the SM give similar results. The $D$-wave 
 polarization asymmetries $Z_Q$ and $Z_\perp$ seem quite adequate 
 to distinguish the left-handed Fit 7 and the right-handed neutrino models, since 
 the first type of NP extension produces an increase in the prediction  of the SM, 
 while the latter NP scenarios reduce the results of the SM.

\section{Summary}
\label{sec:conclusions}
We have  given the hadron and lepton tensors and  the semileptonic differential 
distributions in the presence of  both  left- and right-handed neutrino NP terms, 
and  the most general polarization state for the final tau. The formalism  is  valid 
for any  quark $q \to q\, '\ell \bar \nu_\ell $ or antiquark 
$\bar q \to \bar q\, '
\bar \ell\nu_\ell $ charged-current decay, although  we have usually referred to 
$b \to c $ transitions. This framework  is an alternative to the helicity amplitude 
one to  describe processes where all hadron polarizations 
are summed up and/or averaged. The results of the first part of this work complete 
the scheme presented in Ref.~\cite{Penalva:2020xup}, where only left-handed neutrino 
fields were considered.

In section~\ref{sec:tRF}, we have discussed the $d^3\Gamma_d/(d\omega 
d\cos\theta_\tau d\cos\theta_d^* d\phi_d^*)$ sequential decay distribution in the 
$\tau$ rest frame, and how it can be used to extract the LAB or CM two dimensional 
$P_L(\omega,\cos\theta_\tau)$, $P_T(\omega,\cos\theta_\tau)$ and $P_{TT}(\omega,
\cos\theta_\tau)$ components of the $\tau$-polarization vector. These observables, 
together  with the unpolarized  $d^2\Gamma/(d\omega d\cos\theta_\tau)$ distribution,  
provide the maximum information from the $H_b\to H_c$  semileptonic decay with 
polarized taus~\cite{Penalva:2021gef}, including the CP-violating contributions 
driven by the ${\cal F_H}$ and ${\cal G_H}$ scalar functions (Eqs.~\eqref{eq:pol} 
and \eqref{eq:pol2}). These latter functions are non zero only when some of the 
Wilson coefficients are complex, and are extracted from $P_{TT}(\omega,\cos
\theta_\tau)$, the polarization vector component  transverse to the plane formed 
by the outgoing hadron 
and tau. We have detailed how $P_{TT}$ could be obtained integrating over 
$\cos\theta_d^*$, and looking at the $\phi_d^*$ asymmetry defined in 
Eq.~\eqref{eq:PTTterm}. Results for the CP-violating contributions in  
the baryon $\Lambda_b\to \Lambda_c \tau \bar\nu_\tau$ reaction are shown 
in Fig.~\ref{fig:PTT} within the $R_2$ leptoquark model of Ref.~\cite{Shi:2019gxi}, 
for which the two nonzero Wilson coefficients ($C^S_{LL}$ and $C^T_{LL}$) are complex.
If the tau momentum is not determined, the $\tau$ rest frame cannot 
be defined and the former results cannot be experimentally accessed. 

Reconstructing the $\tau$ momentum in the final state poses an experimental 
challenge, because the $\tau$ does not travel
far enough for a displaced vertex and its decay involves at least one more invisible 
neutrino. Direct $\tau$ polarization measurements are even more complicated 
 to  perform. Therefore, the maximal 
accessible information on the $b \to  c \tau\bar\nu_\tau$ transition is  encoded 
in the visible decay products of the $\tau$ lepton. For that  reason, we have 
studied the sequential $H_b\to H_c \tau\, (\pi \nu_\tau, \rho \nu_\tau, \mu \bar
\nu_\mu \nu_\tau) \bar\nu_\tau$ decays.

Without relying on the reconstruction of the tau momentum, we have derived the 
so-called visible decay particle $\pi,\rho, \mu$, 
distributions~\cite{Alonso:2016gym,Alonso:2017ktd, Asadi:2020fdo}, 
valid for any $H_b\to H_c$  
semileptonic decay.
We take as visible kinematical variables the energy $E_d$ (or the variable $\xi_d$, which 
is proportional to the energy) of the charged particle in the $\tau$ decay  and the 
angle $\theta_d$
made by its  three-momentum with that of
the final hadron $H_c$, both variables defined in the CM 
frame ($W$ boson at rest). The scheme allows to account for the full set of dimension-6 
semileptonic $b\to c$ operators with left- and right-handed neutrinos considered in 
Ref.~\cite{Mandal:2020htr}.

In the absence of CP-violation,  the  analysis of the dependence on ($\omega, \xi_d$) 
of the $S,P$- and $D$-wave contributions ($F^d_{0,1,2}(\omega,\xi_d),\, d=\pi 
\nu_\tau, \rho \nu_\tau, \mu \bar \nu_\mu \nu_\tau$) to the $d^3\Gamma_d/(d\omega  
d\xi_d d\cos\theta_d)$ differential distribution provides the maximal information, 
which can be extracted from the analysis of the semileptonic $H_b\to H_c\tau 
\bar\nu_\tau$ transition, considering the most general polarized state for the final 
tau. This exhaustive information (scalar functions ${\cal A}, {\cal B}, {\cal C}$, 
and ${\cal A_H},{\cal B_H},{\cal C_H}, {\cal D_H}, {\cal E_H}$ in Eqs.~\eqref{eq:pol} 
and \eqref{eq:pol2}) can be rewritten in terms of the overall unpolarized  
normalization distribution $d\Gamma_{\rm SL}/d\omega$, and seven angular and spin 
asymmetries [ see Table~\ref{tab:equival} and Eq.~\eqref{eq:observ}] introduced in 
Ref.~\cite{Asadi:2020fdo} for $B$-meson 
decays. We have found that, in  general, the hadronic tau-decay channels, and in 
particular the pion mode, are more  convenient to determine the $H_b\to H_c\tau 
\bar\nu_\tau$ semileptonic observables  than the lepton $\tau \to \mu \bar\nu_\mu 
\nu_\tau$ channel. For this latter mode, we have provided, for the very first time, 
expressions where the muon mass is not set to zero.

We have considered three different extensions of the SM, taken from the
recent study in Ref.~\cite{Mandal:2020htr}, that include
right-handed neutrino fields,  
and we have shown predictions (Fig.~\ref{fig:ZsyAes}) for the semileptonic observables 
defined in Eq.~\eqref{eq:observ}, for the  $\Lambda_b\to \Lambda_c$ decay. 
We have compared these NP results with those obtained in the SM, and within an 
extension of the SM determined by Fit 7 of Ref.~\cite{Murgui:2019czp} constructed 
 with left-handed neutrino operators alone. 

None of the semileptonic decay asymmetries turned out to be useful in 
distinguishing between the three scenarios with right-handed neutrinos. The 
predictions from Fit 7 of Ref.~\cite{Murgui:2019czp} are, however, significantly different 
from those obtained within the SM and the right-handed neutrino models in all cases, 
except for  $Z_L$, where all the extensions of the SM give similar results. 
The $D$-wave polarization asymmetries $Z_Q$ and $Z_\perp$ seem quite adequate  
to distinguish the left-handed Fit 7 and the right-handed neutrino models.

We are aware that the measurement of these  observables  is rather difficult.
At present,  $\Lambda_b$'s are  only  produced at the LHC, where the corresponding
 $\tau$ decay modes are difficult to reconstruct. However the LHCb collaboration
 has already published semileptonic decay results where the $\tau$ has been
 reconstructed through the $\tau\to\mu\nu_\tau\bar\nu_\mu$ decay
 mode~\cite{Aaij:2015yra, Aaij:2017tyk}. It is reasonable to expect and extension of
 this selection strategy to  $\Lambda_b$ semileptonic decays\footnote{A private
 communication with M. Pappagallo (deputy physics coordinator of the LHCb
 experiment) confirms that a measurement   of the  
${\cal B}(\Lambda_b\to\Lambda_c\tau(\mu\nu_\tau\bar\nu_\mu)\bar\nu_\tau)/{\cal
 B}(\Lambda_b\to\Lambda_c\mu\bar\nu_\mu)$  ratio is already
 ongoing.}. The  other two $\tau$ decay modes,  $\pi\nu_\tau$ and $\rho\nu_\tau$,
 analyzed  in this work  have a lower  reconstruction efficiency and are
  not being exploited at the moment.

\section*{Acknowledgements}
We warmly thank C. Murgui, J. Camalich, A. Pe\~nuelas, A. Pich  and M. Artuso for useful
discussions.
This research has been supported  by the Spanish Ministerio de
Econom\'ia y Competitividad (MINECO) and the European Regional
Development Fund (ERDF) under contracts FIS2017-84038-C2-1-P, PID2020-112777GB-I00 and PID2019-105439G-C22, 
the EU STRONG-2020 project under the program H2020-INFRAIA-2018-1, 
grant agreement no. 824093 and by  Generalitat Valenciana under contract PROMETEO/2020/023. 
\appendix

\section{ Wilson coefficients $C_{\chi=L,R}^{S,P,V,A,T}$ }
\label{app:CW}
We compile in this appendix the coefficients that enter into the definition of the hadron operators in Eq.~\eqref{eq:Jh}. For left-handed neutrinos ($\chi=L$),  we have 
\bea
C^V_L&=&(1+C^V_{LL}+C^V_{RL}), \quad C^A_L=(1+C^V_{LL}- C^V_{RL}),  \nonumber \\
C^S_L&=&(C^S_{LL}+ C^S_{RL}),\quad C^P_L = (C^S_{LL}- C^S_{RL}), \quad C^T_L
= C^T_{LL}, \label{eq:WCL}
\eea
while for right-handed neutrinos ($\chi=R$),
\bea
C^V_R&=&(C^V_{LR}+ C^V_{RR}), \quad C^A_R=-(C^V_{LR}- C^V_{RR}), \quad C^S_R=(C^S_{LR}+ C^S_{RR}), \nonumber \\
C^P_R&=&-(C^S_{LR}- C^S_{RR}), \quad C^T_R= C^T_{RR}, \label{eq:WCR}
\eea
where $C^X_{AB}$ ($X= S, V,T$ and $A,B=L,R$) appear in the BSM effective Hamiltonian of Eq.~\eqref{eq:hnp}, taken from Ref.~\cite{Mandal:2020htr}.

\section{Lepton tensors}
\label{sec:appL}

From Eq.~\eqref{eq:lepton-current}, in the limit of massless neutrinos, 
we obtain ($h_\chi= \pm 1, h= \pm 1$)
\be
J^{L}_{(\alpha\beta)}(k,k';h,h_\chi)[J^{L}_{(\rho\lambda)}(k,k';h,h_\chi)]^* = 
\frac12 {\rm Tr}\Big[(\slashed{k}'+ m_{\ell}) 
\Gamma_{(\alpha\beta)} P_5^{h_\chi}\slashed{k}\,\gamma^0 
\Gamma_{(\rho\lambda)}^\dagger \gamma^0P_h\Big]. \label{eq:LT}
\ee
The different $\Gamma_{(\alpha\beta)}$ and $\Gamma_{(\rho\lambda)}$ operators give 
rise to the following lepton tensors (we use the convention $\epsilon_{0123}=+1$ and the short-notation $\epsilon_{\alpha k' k S}=\epsilon_{\alpha\delta\eta\sigma} k^{\prime \delta}k^\eta S^\sigma$, etc.)
\begin{eqnarray}
L(k,k';h,h_\chi) &=& \frac12\left[k\cdot k' - m_\ell h h_\chi (k\cdot S)\right], \label{eq:leptensorSP}\\
L_\alpha(k,k';h,h_\chi)&=&\frac{m_\ell}{2} k_\alpha - \frac{h\,h_\chi}{2}
\left(k'_\alpha\, k\cdot S-S_\alpha\, k\cdot k' - i h_\chi \epsilon_{\alpha k' k S}\right), \label{eq:leptensor-vec}\\
L'_{\rho\lambda}(k,k';h,h_\chi)&=& L'_{\rho\lambda}(k,k';h_\chi)- h h_\chi L'_{\rho\lambda}(k, m_\ell S; h_\chi), 
  \label{eq:leptensor-ST}\\
L_{\alpha\rho}(k,k';h,h_\chi) &=& L_{\alpha\rho}(k,k';h_\chi)+hh_\chi L_{\alpha\rho}(k, m_\ell S;h_\chi), \label{eq:leptensorVA}\\
L_{\alpha\rho\lambda}(k,k';h,h_\chi) &=& \frac{i m_{\ell}}{2}  \left( g_{\alpha\lambda}k_\rho
-g_{\alpha\rho}k_\lambda -i h_\chi\epsilon_{\alpha\rho\lambda k}\right) 
+ \frac{i hh_\chi}{2} \Bigg\{ k'_\alpha (S_\rho k_\lambda - S_\lambda k_\rho)
 \nonumber \\
&+&  k_\alpha (S_\rho k'_\lambda - S_\lambda k'_\rho) +
 S_\alpha (k_\rho k'_\lambda - k_\lambda k'_\rho)+ (k\cdot k') 
(g_{\alpha\rho}S_\lambda -g_{\alpha\lambda}S_\rho ) \nonumber \\
&+& (S\cdot k) 
(g_{\alpha\lambda}k'_\rho -g_{\alpha\rho}k'_\lambda )+ih_\chi \Big [ (k\cdot k') \epsilon_{\alpha \rho\lambda S } 
  + S_\lambda \epsilon_{\alpha\rho k' k} 
- S_\rho \epsilon_{\alpha\lambda k' k}\nonumber \\
&+& k_\alpha 
\epsilon_{\rho\lambda S k'}- k'_\alpha 
\epsilon_{\rho\lambda k S}  \Big]\Bigg\}, \label{eq:lepdif}\\
L_{\alpha\beta\rho\lambda}(k,k';h,h_\chi) & =& \frac12\, L_{\alpha\beta\rho\lambda}(k,k';h_\chi)
- \frac{h h_\chi}{2} L_{\alpha\beta\rho\lambda}(k, m_\ell S; h_\chi ), \label{eq:leptensor}
\end{eqnarray}
which correspond to $(\Gamma_{(\alpha\beta)}, \Gamma_{(\rho\lambda)})=(1, 1), (\gamma_\alpha,1), (1,\sigma_{\rho\lambda}), (\gamma_\alpha,\gamma_\rho)$, $(\gamma_\alpha,\sigma_{\rho\lambda})$ and $(\sigma_{\alpha\beta},\sigma_{\rho\lambda})$, respectively, and in Eqs.~\eqref{eq:leptensorVA}, \eqref{eq:leptensor-ST} and \eqref{eq:leptensor}
\begin{eqnarray}
L_{\alpha\rho}(k,k';h_\chi)&=& \frac12 \left(k'_\alpha k_\rho+k_\alpha k'_\rho -
g_{\alpha\rho} k\cdot k'- ih_\chi\epsilon_{\alpha\rho k'k}\right),\nonumber \\ 
 L'_{\rho\lambda}(k,k';h_\chi)&=& \frac{i}{2 } \left( k_{\rho} k'_{\lambda}-k_{\lambda} k'_{\rho} -i h_\chi\epsilon_{\rho\lambda k'k}\right),\nonumber \\ 
L_{\alpha\beta\rho\lambda}(k,k';h_\chi) &=& g_{\beta \rho} (k_\alpha k'_\lambda+ k_\lambda k'_\alpha)- g_{\beta \lambda} (k_\alpha k'_\rho+ k_\rho k'_\alpha) 
 -g_{\alpha \rho} (k_\beta k'_\lambda+ k_\lambda k'_\beta) \nonumber \\
 &+&  g_{\alpha \lambda} (k_\beta k'_\rho+ k_\rho k'_\beta)+(k\cdot k')(g_{\alpha \rho}g_{\beta \lambda} 
 -g_{\alpha \lambda}g_{\beta \rho}) \nonumber \\
 &-&i h_\chi
 \left(k'_\alpha \epsilon_{\beta\lambda\rho k} - 
 k'_\beta \epsilon_{\alpha\lambda\rho k}  + k_\rho 
 \epsilon_{\alpha\beta\lambda k'} - k_\lambda 
 \epsilon_{\alpha\beta\rho k'}  \right).
\end{eqnarray}

\section{Hadron tensors}
\label{sec:appH}

We collect here the hadron tensors that should be contracted with  the
corresponding lepton ones, compiled in the previous appendix,  to obtain 
$\overline\sum  |{\cal M}|^2_{\nu_{\ell\chi}}$, $\chi=L,R$. 
The tensorial decompositions, for a given set $C_\chi^{S,P,V,A,T}$ of  NP Wilson coefficients (see Eqs.~\eqref{eq:WCL} and \eqref{eq:WCR}), 
are taken from Ref.~\cite{Penalva:2020xup}.

\begin{itemize}

\item The spin-averaged squared of the $O_{H\chi}^\alpha$ operator matrix element leads to
\bea
W^{\alpha\rho}_\chi(p,q, C^V_\chi, C^A_\chi) &=& \overline{\sum_{r,r'}} \langle H_c;  p',r' 
|(C^V_\chi V^\alpha + h_\chi C^A_\chi A^\alpha)  | H_b; p,r\rangle \times \nonumber \\
&\times& \langle H_c; p',r' | (C^V_\chi V^\rho + h_\chi C^A_\chi A^\rho)| H_b; p,r \rangle^* , 
\eea
with $(C^V_\chi V^\alpha + h_\chi C^A_\chi A^\alpha) = \bar c(0) \gamma^\alpha (C^V_\chi + h_\chi C^A_\chi\gamma_5) b(0)$. 
The sum is done over initial (averaged) and final hadron
helicities, and the above tensor should be contracted with  the lepton one 
$L_{\alpha\rho}(k,k';h, h_\chi)$ (Eq.~\eqref{eq:leptensorVA}) to get the contribution
 to $\overline\sum\,  |{\cal M}|^2_{\nu_{\ell\chi}}$, $\chi=L,R$. 
The tensor can be expressed in terms of five SFs as 
\bea
W^{\alpha\rho}_\chi (p,q, C^V_\chi,C^A_\chi)&=&-g^{\alpha\rho}
\widetilde W_{1\chi}+\frac{p^{\alpha}p^{\rho}}{M^2}
\widetilde W_{2\chi}-ih_\chi\epsilon^{\alpha\rho\delta\eta}p_{\delta}q_{\eta}
\frac{\widetilde W_{3\chi}}{2M^2} \nonumber \\
&+&\frac{q^{\alpha}q^{\rho}}{M^2}
\widetilde W_{4\chi}+\dfrac{p^{\alpha}q^{\rho}+p^{\rho}q^{\alpha}}{2M^2}
\widetilde W_{5\chi},
\eea
where all $\widetilde W_{1\chi,2\chi,3\chi,4\chi,5\chi}(q^2, C^V_\chi,C^A_\chi)$ SFs are real. Following the notation  in Ref.~\cite{Penalva:2020xup}, 
\begin{eqnarray}
\widetilde W_{1\chi,2\chi,4\chi,5\chi}(q^2) &=& 
|C^V_\chi|^2 W_{1,2,4,5}^{VV}(q^2)+ |C^A_\chi|^2 W_{1,2,4,5}^{AA}(q^2),\,\nonumber \\
\widetilde W_{3\chi}(q^2) &=& {\rm Re}(C^V_\chi C^{A*}_\chi)W_3^{V\!\!A}(q^2).
\end{eqnarray}

 \item The diagonal contribution of the tensor operator $O_{H\chi}^{\alpha\beta}$ 
 gives rise to 
\bea
W_\chi^{\alpha\beta\rho\lambda}(p,q, C^T_\chi) &=& |C^T_\chi|^2\overline{\sum_{r,r'}} \langle H_c;  p',r' | \bar c(0) \sigma_{\alpha\beta} (1+h_\chi\gamma_5) b(0) | H_b; p,r\rangle \times \nonumber \\
&\times & \langle H_c; p',r' |\bar c(0) \sigma_{\rho\lambda} 
 (1+h_\chi\gamma_5) b(0) | H_b; p,r \rangle^*, 
\eea
which contracted with  the lepton tensor $L_{\alpha\beta\rho\lambda}(k,k';h,h_\chi)$ 
in Eq.~\eqref{eq:leptensor} provides the $L$ or $R$ contributions to the differential decay 
rate.  The total tensor can be expressed in terms of four real SFs,  
\bea 
W^{\alpha\beta\rho\lambda}_\chi &= & |C^T_\chi|^2 \Bigg\{ W_1^T \Big[g^{\alpha\rho}g^{\beta\lambda}-g^{\alpha\lambda}g^{\beta\rho}
+i h_\chi\epsilon^{\rho\lambda\alpha\beta}\Big] + \frac{W_2^T}{M^2} \Big[g^{\alpha\rho}p^\beta p^\lambda-g^{\alpha\lambda}p^\beta p^\rho\nonumber \\ 
&-&g^{\beta\rho}p^\alpha p^\lambda+g^{\beta\lambda}p^\alpha p^\rho+ih_\chi\Big(
\epsilon^{\rho\lambda\alpha\delta}p^\beta p_\delta-
\epsilon^{\rho\lambda\beta\delta}p^\alpha p_\delta 
\Big)\Big] \nonumber \\ 
&+& \frac{W_3^T}{M^2} \Big[g^{\alpha\rho}q^\beta q^\lambda-g^{\alpha\lambda}q^\beta q^\rho-
g^{\beta\rho}q^\alpha q^\lambda+g^{\beta\lambda}q^\alpha q^\rho\nonumber \\
&+&ih_\chi\Big(
\epsilon^{\rho\lambda\alpha\delta}q^\beta q_\delta-
\epsilon^{\rho\lambda\beta\delta}q^\alpha q_\delta\Big)\Big]+ \frac{W_4^T}{M^2} \Big[g^{\alpha\rho}(p^\beta q^\lambda+p^\lambda q^\beta) \nonumber \\
&-& g^{\alpha\lambda}(p^\beta q^\rho+p^\rho q^\beta)-g^{\beta\rho}(p^\alpha q^\lambda+p^\lambda q^\alpha) + g^{\beta\lambda}(p^\alpha q^\rho+p^\rho q^\alpha)\nonumber \\ 
&+& ih_\chi\Big(
\epsilon^{\rho\lambda\alpha\delta}(p^\beta q_\delta+q^\beta p_\delta)-
\epsilon^{\rho\lambda\beta\delta}(p^\alpha q_\delta+q^\alpha p_\delta)
\Big)\Big]\Bigg\}.
\label{eq:WTdecom}
\eea
The $W_{1,2,3,4}^T$ SFs are found from $(W^{\alpha\beta\rho\lambda}_{TT} +  W^{\alpha\beta\rho\lambda}_{pTpT})$~\cite{Penalva:2020xup} and accomplish 
the constraint
\be
2 M^2 W_1^T + p^2 W_2^T + q^2 W_3^T+ 2(p\cdot q) W_4^T =0, \label{eq:relacWTs}
\ee
which can be used to re-write $W_1^T$ in terms  of $W^T_{2,3,4}$. In any case, 
the contraction of the $W_1^T$-part of the tensor    with $L_{\alpha\beta\rho\lambda}(k,k';h,h_\chi)$ is zero, and thus the contribution of 
 $ W^{\alpha\beta\rho\lambda}_\chi$ to $\overline\sum  |{\cal M}|^2_{\nu_{\ell\chi}}$ 
 is given  only in terms of $W_2^T$, $W_3^T$ and $W_4^T$. The common factor $|C^T_\chi|^2$ was absorbed in \cite{Penalva:2020xup} by introducing $\widetilde W^T_{1\chi,2\chi,3\chi,4\chi} = |C^T_\chi|^2 
W^T_{1,2,3,4}$. 

\item  The diagonal contribution of the  operator $O_{H\chi}$ leads to the real scalar SF
\bea
W_\chi(p,q)&=& \widetilde W_{SP\chi}(q^2) = |C^S_\chi|^2 \overline{\sum_{r,r'}} |\langle H_c;  p',r' | \bar c(0)
 b(0) | H_b; p,r\rangle|^2 \nonumber \\
 &+& |C^P_\chi|^2 \overline{\sum_{r,r'}} |\langle H_c;  p',r' | \bar c(0)
 \gamma_5 b(0) | H_b; p,r\rangle|^2,  
\eea
which should be multiplied by the scalar lepton term of Eq.~\eqref{eq:leptensorSP}. 

\item The $O_{H\chi}^\alpha$ and $O_{H\chi}$ interference contribute to  $\overline\sum\,  |{\cal M}|^2_{\nu_{\ell\chi}}$ as 
$2{\rm Re}\Big[L_\alpha(k,k';h,h_\chi)\times$ $W^{\alpha}_\chi(p,q, C^{V,A,S,P}_\chi)\Big]$, with 
the lepton tensor defined in Eq.~\eqref{eq:leptensor-vec} and 
 \bea 
 W^{\alpha}_\chi(p,q, C^{V,A,S,P}_\chi) &=& \overline{\sum_{r,r'}} \langle H_c;  p',r' | (C^V_\chi V^\alpha+h_\chi C^A_\chi A^\alpha) | H_b; p,r\rangle\times \nonumber \\ 
 &\times& \langle H_c; p',r' |\bar c(0) (C^S_\chi+h_\chi C^P_\chi\gamma_5) b(0) | H_b; p,r \rangle^*\nonumber\\
&=& \frac{1}{2M} \left( \widetilde  W_{I1\chi} p^\alpha +\widetilde 
W_{I2\chi} q^\alpha \right),
 \eea
where $\widetilde W_{I1\chi,I2\chi}$ are obtained as
\be
\widetilde W_{I1\chi, I2\chi} (q^2)= C^V_\chi C^{S*}_\chi W^{VS}_{I1,I2}(q^2) + C^A_\chi C^{P*}_\chi
W^{AP}_{I1,I2}(q^2),
\ee
with all four $W^{VS,AP}_{I1,I2}$ real functions of $q^2$~\cite{Penalva:2020xup}.

\item The $O_{H\chi}$ and $O_{H\chi}^{\rho\lambda}$ interference contribute to  $\overline\sum\,  |{\cal M}|^2_{\nu_{\ell\chi}}$ as $2{\rm Re}\Big[L'_{\rho\lambda}(k,k';h,h_\chi)\times$ $W^{\prime\rho\lambda}_\chi(p,q, C^{S,P,T}_\chi)\Big]$,
 with the lepton tensor defined in Eq.~\eqref{eq:leptensor-ST} and 
\bea 
 W_\chi^{\prime\rho\lambda}(p,q, C^{S,P,T}_\chi) &=& C^{T*}_\chi\overline{\sum_{r,r'}} \langle H_c;  p',r' |
  \bar c(0) (C^S_\chi + h_\chi C^P_\chi\gamma_5) b(0) | H_b; p,r\rangle \times \nonumber \\
 &\times& \langle H_c; p',r' |\bar c(0) \sigma^{\rho\lambda}(1+ h_\chi \gamma_5) b(0) | H_b; p,r \rangle^* \nonumber \\
 &=&\frac{\widetilde W_{I3\chi}}{2M^2} \left[ 
  i(p^\rho q^\lambda-p^\lambda q^\rho) -h_\chi\epsilon^{\rho\lambda\delta\eta} p_\delta q_\eta
 \right],
\eea
with~\cite{Penalva:2020xup}
\be
\widetilde W_{I3\chi}(q^2) = C^{T^*}_\chi \left(C^S_\chi W_{I3}^{ST}(q^2)+ C^P_\chi
 W_{I3}^{PpT}(q^2)\right)
\ee
and $W_{I3}^{ST,PpT}$ real scalar functions of $q^2$.

\item The $O^\alpha_{H\chi}$ and $O_{H\chi}^{\rho\lambda}$ interference contribute to the decay width 
as $2{\rm Re}\Big[L_{\alpha\rho\lambda}(k,k';h)\times$ $W^{\alpha\rho\lambda}_\chi(p,q, C^{V,A,T}_\chi)
\Big]$, with the  lepton tensor defined in Eq.~\eqref{eq:lepdif} and 
\bea 
 W^{\alpha\rho\lambda}_\chi(p,q, C^{V,A,T}_\chi) &=& C^{T*}_\chi\overline{\sum_{r,r'}} \langle H_c;  p',r' | 
 (C^V_\chi V^\alpha+h_\chi C^A_\chi A^\alpha) | H_b; p,r\rangle\times \nonumber \\ 
 &\times&\langle H_c; p',r' |\bar c(0) \sigma^{\rho\lambda}(1+h_\chi\gamma_5) b(0) | H_b; p,r \rangle^*\nonumber \\
&=&\frac{p^\alpha \widetilde  W_{I4\chi} + q^\alpha \widetilde W_{I5\chi}}{2M^3} 
 \left[i(p^\rho q^\lambda-p^\lambda q^\rho) -h_\chi\epsilon^{\rho\lambda\delta\eta} p_\delta q_\eta  
 \right] \nonumber \\
&+& \frac{p_\delta \widetilde  W_{I6\chi} + q_\delta \widetilde W_{I7\chi}}{2M} \left[i(g^{\alpha\rho}g^{\lambda\delta}-g^{\alpha\lambda}g^{\rho\delta}) -h_\chi \epsilon^{\rho\lambda\alpha\delta}  
 \right],
 \eea
where the SFs are obtained from~\cite{Penalva:2020xup}
\be
 \widetilde W_{I4\chi,I5\chi,I6\chi,I7\chi}(q^2) = C^{T*}_\chi\left(C^V_\chi W_{I4,I5,I6,I7}^{VT}(q^2)+ C^A_\chi
 W_{I4,I5,I6,I7}^{ApT}(q^2)\right),
\ee
with $W_{I4,I5,I6,I7}^{VT}$ and $W_{I4,I5,I6,I7}^{ApT}$ real scalar functions of $q^2$, which are given in terms of the 
form-factors used to parameterize the hadronic matrix elements.

\end{itemize}

\section{ $\left(|{\cal M}|_{\nu_{\ell L}}^2 + |{\cal M}|_{\nu_{\ell R}}^2\right)$  in terms of the $\widetilde W$ SFs}
\label{app:coeff}

In this appendix we collect the expressions of the  
${\cal A}, {\cal B}$ and ${\cal C}$   and ${\cal A}_H,{\cal B}_H$,
${\cal C}_H, {\cal D}_H, {\cal E}_H, {\cal F}_H$ and ${\cal G}_H$  functions 
introduced in Eq.~\eqref{eq:pol}, and that describe, respectively, the semileptonic decay
for the cases of unpolarized and  polarized outgoing charged leptons. They are combinations of the hadronic $\widetilde W$ SFs 
and receive contributions from both neutrino chiralities (symbolically $\widetilde W_\chi = C_\chi W$) . For ${\cal A}, {\cal B}, {\cal C}$ and the 
CP-violating ${\cal F}_H$ and ${\cal G}_H$, it always appears the combination $(L+R)$, i.e. $(\widetilde W_{iL}+ \widetilde W_{iR})$, while 
 for ${\cal A_H}, {\cal B_H}, {\cal C_H}, {\cal D_H}$ and ${\cal E_H}$ the structure is ($L-R):$ ($\widetilde W_{iL}- \widetilde W_{iR})$.  The explicit expressions, for any semileptonic decay  driven by a $q\to q' \ell \bar\nu_\ell$ transition, read ($M_\omega= M-M'\omega$)
\begin{eqnarray}
 {\cal A}(\omega) &=&\frac{q^2-m_\ell^2}{M^2}\sum_{\chi=L,R}\Bigg\{ 2\widetilde W_{1\chi} -
 \widetilde W_{2\chi}+\frac{M_\omega}{M}\widetilde W_{3\chi}+ \widetilde W_{SP\chi} 
 +8\widetilde W_{2\chi}^T-\frac{8q^2}{M^2}\widetilde W_{3\chi}^T\nonumber \\
&-& \frac{16 M_\omega}{M} \widetilde W_{4\chi}^T  + \frac{m_\ell}{M}{\rm Re}\left[\widetilde W_{I2\chi}+4\,\widetilde W_{I4\chi}+
 \frac{4M_\omega}{M} \widetilde W_{I5\chi}+12\,\widetilde W_{I7\chi}\right]\nonumber \\
 &+&   \frac{4M_\omega}{M} {\rm Re}[\widetilde W_{I3\chi}]
 +\frac{m_\ell^2}{M^2}\left( \widetilde W_{4\chi}-16 \widetilde W_{3\chi}^T\right)
 \Bigg\}, \nonumber \\
 {\cal B}(\omega) & = & \sum_{\chi=L,R} \Bigg\{-\frac{2q^2}{M^2}\left(\widetilde W_{3\chi}+ 
 4{\rm Re}[ \widetilde W_{I3\chi}]\right)+\frac{4M_\omega}{M}
 \Big(\widetilde W_{2\chi} -16\, \widetilde W_{2\chi}^T \Big)\nonumber
\\&+&   \frac{2m_\ell}{M}{\rm Re}\left[\widetilde W_{I1\chi} 
 -\frac{4 M_\omega}{M}\widetilde W_{I4\chi}-\frac{ 4 q^2}{M^2}\widetilde W_{I5\chi}+12\,
 \widetilde W_{I6\chi}\right] \nonumber\\
 &+&\frac{2m_\ell^2}{M^2} \left( \widetilde W_{5\chi} -32\, 
 \widetilde W_{4\chi}^T\right) \Bigg\}, \nonumber \\
 {\cal C}(\omega)&=& -4\sum_{\chi=L,R}\Big(\widetilde W_{2\chi} -16\, \widetilde W_{2\chi}^T \Big),\label{eq:abc}
\end{eqnarray}
\begin{eqnarray}
{\cal A_H}(\omega)&=& \frac{q^2-m^2_\ell}{2M^2}\sum_{\chi=L,R}h_\chi\Bigg\{{\rm Re} \left[\widetilde W_{I1\chi}
+\frac{4 M_{\omega}}{M} \widetilde W_{I4\chi}-4\,\widetilde W_{I6\chi} \right]
 \nonumber \\
&+&\frac{m_\ell}{M}\left( \widetilde W_{3\chi} +\widetilde W_{5\chi} 
-4\,{\rm Re}[\widetilde W_{I3\chi}]+ 32\, \widetilde W_{4\chi}^T   \right)-\frac{4 m_\ell^2}{M^2} {\rm Re} [\widetilde W_{I5\chi}]\Bigg\},
\nonumber\\
\nonumber\\
{\cal B_H}(\omega)&=& -\sum_{\chi=L,R}h_\chi\Bigg\{ \frac{M_\omega}{M}{\rm Re} \left[\widetilde W_{I1\chi}+\frac{4 M_{\omega}}{M} \widetilde W_{I4\chi}-4\,\widetilde W_{I6\chi} \right] 
-\frac{m_\ell}{M}\left(2\, \widetilde W_{1\chi}    \right. \nonumber\\
&-&\left.  \widetilde W_{2\chi} -\frac{M_\omega}{M}\widetilde W_{5\chi}-\widetilde W_{SP\chi}  - 8\, \widetilde W_{2\chi}^T+\frac{8q^2}{M^2}\widetilde W_{3\chi}^T - \frac{16 M_\omega}{M} \widetilde W_{4\chi}^T  \right)\nonumber \\
&+& \frac{m_\ell^2}{M^2} {\rm Re} \left[\widetilde W_{I2\chi}-4\,\widetilde W_{I4\chi}-4\,
\widetilde W_{I7\chi}\right] + \frac{m_\ell^3}{M^3}\left(\widetilde W_{4\chi} + 16\, 
\widetilde W_{3\chi}^T \right)\Bigg\}, \nonumber\\
 %
{\cal C_H}(\omega)&=& -\sum_{\chi=L,R} h_\chi \Bigg\{ \frac{4q^2}{M^2}{\rm Re}[\widetilde W_{I4\chi}]-\frac{2m_\ell}{M}
\left(\widetilde W_{2\chi}+16 \widetilde W_{2\chi}^T\right)  + \frac{4 m_\ell^2}{M^2}{\rm Re}[\widetilde
W_{I4\chi}]\Bigg\}, \nonumber \\\nonumber\\
{\cal D_H}(\omega)&=& \sum_{\chi=L,R}h_\chi\Bigg\{ {\rm Re} \left[\widetilde W_{I1\chi}+\frac{12 M_\omega}{M} 
\widetilde W_{I4\chi}-4\, \widetilde W_{I6\chi}\right] \nonumber \\
&&- \frac{m_\ell}{M}\left(
\widetilde W_{3\chi} - \widetilde W_{5\chi}- 32 \widetilde W_{4\chi}^T  - 4 {\rm Re}
[\widetilde W_{I3\chi}]\right) + 
\frac{4m^2_\ell}{M^2} {\rm Re}[\widetilde W_{I5\chi}]\Bigg\}, \nonumber\\
{\cal E_H}(\omega)&=& -8\, \sum_{\chi=L,R}h_\chi{\rm Re}[\widetilde
  W_{I4\chi}],\\\nonumber\\\nonumber\\
%
{\cal F_H}(\omega)&=& 4 \sum_{\chi=L,R} {\rm Im}\bigg[\frac{\widetilde W_{I1\chi}}{4} +\frac{m_\ell}{M}\widetilde W_{I3\chi}+\frac{M_\omega}{M}\widetilde W_{I4\chi}
 +\frac{m^2_\ell}{M^2}\widetilde W_{I5\chi}- \widetilde W_{I6\chi} \bigg],\nonumber \\
{\cal G_H}(\omega)&=& -8 \sum_{\chi=L,R} {\rm Im}[\widetilde W_{I4\chi}]
.\label{eq:polgral2}
\end{eqnarray}
with $h_{\chi=R}= 1 $ and $h_{\chi=L}=-1$. Finally, expressions for the 
$\widetilde W_{i\chi}$ SFs in terms of the Wilson coefficients and the form-factors, 
used to parameterize the genuine  hadronic responses $W_i$, can be obtained from the 
Appendices E of Ref.~\cite{Penalva:2020xup} and B of \cite{Penalva:2020ftd} for the $\Lambda_b^0\to  \Lambda_c^+\ell^-\bar\nu_\ell$
  and $P_b\to  P_c^{(*)}\ell^-\bar\nu_\ell$ decays, respectively\footnote{Here  $P_b$ and $P_c$ are pseudoscalar mesons 
($\bar B_c$ or $\bar B$ and $\eta_c$ or $D$) and $P_c^{(*)}$ a  pseudoscalar or a vector   meson ($\eta_c$, $D$, $J/\psi$ or $D^*$).}. In fact, replacing
\be
C_{V,A, S,P,T} \to C^{V,A, S,P,T}_{\chi} \label{eq:repl}
\ee
in these last works, all $\widetilde W_{i\chi}$ SFs are obtained. Furthermore, using 
the appropriate  form-factors, the results of Refs.~\cite{Penalva:2020xup} and
 \cite{Penalva:2020ftd} can be used to describe any $1/2^+ \to 1/2^+$, $0^- \to 0^-$ or 
 $0^- \to 1^-$ semileptonic decay, regardless of the involved flavors (see also the last 
 comment in Appendix~\ref{sec:app-antiquark}).  

\section{Antiquark-driven semileptonic decays}
\label{sec:app-antiquark}
The hermitian conjugate terms of the effective Hamiltonian of Eq.~\eqref{eq:hnp}, not explicitly written in that equation, 
can be used to evaluate the semileptonic decay
\be
H_{\bar b}\to H_{\bar c} \ell^+ \nu_\ell
\ee
driven by the antiquark $\bar b \to \bar c  \ell^+ \nu_\ell$ transition, and obviously this reaction can be related to that involving $b$ and $c$ quarks. 
Looking at the ${\cal M}|_{\ell\bar\nu_\ell}$ and ${\cal M}|_{\bar\ell\nu_\ell}$ amplitudes, and using charge-conjugation transformations of the hadron operators and states~\cite{Itzykson:1980rh} we first find,
\be
 \langle H_{\bar c};  p',r' |  \bar b~ 
\widetilde O_{H\chi}^{(\alpha\beta)}c | H_{\bar b}; p,r\rangle = \langle H_{c};  p',r' |  \bar c~ 
O_{H\chi}^{(\alpha\beta)}b | H_{ b}; p,r\rangle\Bigg|_{\begin{array}{c}
C_{\chi}^{V,A,T} \to  -\left(C_{\chi}^{V,A,T}\right)^*\\ C_{\chi}^{S,P} 
\to \left(C_{\chi}^{S,P}\right)^* \\ h_\chi\to -h_\chi  \end{array}}, \label{eq:anti-had-ampl}
\ee
with   $\widetilde O_{H\chi}= \gamma^0 O_{H\chi}^\dagger \gamma^0$.  For the leptonic part
of the amplitude, we use  now the properties of the charge conjugation matrix in 
the Dirac space and its action on Dirac spinors and matrices~\cite{Itzykson:1980rh} to get

\bea
J^{L; \bar \ell\nu_\ell}_{(\alpha\beta)}(k,k';h,h_\chi) &=& \frac{1}{\sqrt{2}} 
\bar u_{\nu_\ell} (k) P_5^{-h_\chi} \gamma^0 \Gamma_{(\alpha\beta)}^\dagger 
\gamma^0    v_{\ell}^S(k';h)\nonumber \\
&=& (-1)^{n_{L}+1}\frac{1}{\sqrt{2}} 
\bar u_\ell^S (k';h) P_h \Gamma_{(\alpha\beta)} P_5^{-h_\chi} v_{\nu_\ell}(k) \nonumber \\
&=&(-1)^{n_{L}+1} J^{L; \ell\bar\nu_\ell}_{(\alpha\beta)}(k,k';h,-h_\chi),
\label{eq:anti-lep-ampl}
 \eea
with $n_{L}=1$ for $\Gamma_{(\alpha\beta)}=(\gamma_\alpha,\sigma_{\alpha\beta})$ and $n_L=0$ for $\Gamma_{(\alpha\beta)}=1$. 
The factor $(-1)^{n_{L}}$ compensates the relative sign between  
$C_{\chi}^{S,P}$ and $C_{\chi}^{V,A,T}$ in Eq.~\eqref{eq:anti-had-ampl}, while
the extra minus sign in Eq.~\eqref{eq:anti-lep-ampl} is of no consequence in
evaluating the amplitude squared. 
Taking the complex-conjugate of the Wilson coefficients has no effects when calculating their 
squared moduli or the real part of the product of two of them, but it does
produce a minus sign when  the imaginary part of the product of two of them is 
considered instead.   All together, from 
Eqs.~\eqref{eq:anti-had-ampl}--\eqref{eq:anti-lep-ampl} and 
\eqref{eq:abc}--\eqref{eq:polgral2}, we conclude that  ${\cal A}, {\cal B}$ and  
${\cal C}$ are identical for both quark and antiquark decays, while  the ${\cal A_H}, 
{\cal B_H}, {\cal C_H}, {\cal D_H}$, ${\cal E_H}$, ${\cal F_H}$ and ${\cal G_H}$ antiquark 
functions get a global sign. The first five  because they are proportional to $h_\chi$, 
while  in the case of ${\cal F_H}$ and ${\cal G_H}$, they are proportional to  the imaginary 
part of the product of two  Wilson coefficients\footnote{Note that  ${\cal A}, {\cal B}$ 
and  ${\cal C}$ and ${\cal A_H}, {\cal B_H}, {\cal C_H}, {\cal D_H}$ and ${\cal E_H}$ 
involve only   squared moduli of Wilson coefficients or
the real part of the product of two of them.}. Hence, we obtain
\begin{eqnarray}
\frac{2\,\overline\sum\, |{\cal M}|^2 }{M^2}\Big|_{H_{\bar b}\to H_{\bar c}}&=&
{\cal N}(\omega, p\cdot k) - h\bigg\{ \frac{(p\cdot S)}{M}\,
{\cal N_{H_{\rm 1}}}(\omega, p\cdot k) \nonumber\\
&+&\frac{(q\cdot S)}{M}\,
{\cal N_{H_{\rm 2}}}(\omega, p\cdot k)+\frac{\epsilon^{ S k' qp}}{M^3}\,{\cal N_{H_{\rm 3}}}(\omega, p\cdot k)
 \ \bigg\},\label{eq:pol-anti}
\end{eqnarray}
with the ${\cal N}$ and  $\cal N_{H_{\rm 123}}$ scalar functions identical to 
those that appear in the $H_{b}\to H_{c}$ decay, and the charged-lepton produced 
in the polarized state $v_\ell^S(k';h)$ that satisfies
\be
\gamma_5\slashed{S}\, v_\ell^S(k';h)=h\,v_\ell^S(k';h).
\ee

Note also that
Eq.~\eqref{eq:anti-lep-ampl} and the results for $\widetilde W$ SFs compiled in Appendix~\ref{sec:appH}, for NP 
operators involving both left- and right-handed neutrino fields, can be 
straightforwardly used to describe  quark charged-current transitions giving rise to a 
final $\ell^+\nu_{\ell}$ lepton pair (e.g. $c\to s \ell^+\nu_{\ell}$). From 
Eq.~\eqref{eq:anti-lep-ampl} it is clear that  left/right leptonic currents for  
$\bar \ell\nu_\ell$ production are related to  right/left leptonic currents for  
$\ell\bar \nu_\ell$ production.

\section{Phase space integrations}
\label{app:phspin}
Here we give several results and formulae used in the integration of the available phase space in the sequential $H_b\to H_c \tau\, (\pi \nu_\tau, \rho \nu_\tau, \mu \bar \nu_\mu \nu_\tau) \bar\nu_\tau$ decays. 

First, we note
\bea
\frac{\delta^4\left(q-k-\tilde p_d\right)}{\left[(q-k)^2-m^2_\tau\right]^2+ (q-k)^2 \Gamma^2_\tau\left[(q-k)^2\right]} &=& \int d^4k' \delta^4(k'-(q-k))\times \nonumber \\
&\times& \int dz^2 \frac{\delta (z^2-k^{\prime\,2})\delta^4(k'-\tilde p_d)}
{\left(z^2-m^2_\tau\right)^2+ z^2 \Gamma^2_\tau (z^2)} ,
\eea
with $\tilde p_d$ the total four-momentum of all decay products of the virtual $\tau$ (e.g. $\tilde p_d= p_\pi+p_{\nu_\tau}$ for the pion mode).
Now using $\delta(z^2-k^{\prime\,2}) = \delta\big[k^{\prime\,0}-\sqrt{z^2+\vec{k}^{\prime\, 2}}\,\big]/(2k^{\prime\,0})$, since  $k^{\prime\,0}= \tilde p_d^0 >0$, 
\bea
\frac{\delta^4\left(q-k-\tilde p_d\right)}{\left[\tilde p_d^2-m^2_\tau\right]^2+ \tilde p_d^2 \Gamma^2_\tau\left[\tilde p_d^2\right]} &=& 
\int \frac{dz^2} {\left(z^2-m^2_\tau\right)^2+ z^2 \Gamma^2_\tau (z^2)} \times \nonumber\\
&\times&\int \frac{d^3k'}{2\sqrt{z^2+\vec{k}^{\prime\, 2}}}
\delta^4(q-k'-k)\delta^4(k'-\tilde p_d), \label{eq:putintau}
\eea
with $k^{\prime\,0}=\sqrt{z^2+\vec{k}^{\prime\, 2}}$. Finally, the narrow width approximation of Eq.~\eqref{eq:narrow} leads to
\bea
\frac{\delta^4\left(q-k-\tilde p_d\right)}{\left[\tilde p_d^2-m^2_\tau\right]^2+ 
\tilde p_d^2 \Gamma^2_\tau\left[\tilde p_d^2\right]} &\simeq&  
\frac{\pi}{m_\tau \Gamma_\tau}\int \frac{d^3k'}
{2\sqrt{m_\tau^2+\vec{k}^{\prime\, 2}}}\delta^4(q-k'-k)\delta^4(k'-\tilde p_d),
\eea
with the $\tau$ on the mass shell.

In the second place,  we find for the $(\nu_\tau,\bar\nu_\mu)$ phase-space integration of the muon polarization sum ${\cal R}_{\mu\bar\nu_\mu}$ (Eq.~\eqref{eq:polsum3})
\bea
\int\frac{d^3p_{\nu_\tau}}{2|\vec{p}_{\nu_\tau}|}\int\frac{d^3p_{\bar\nu_\mu}}{2|\vec{p}_{\bar\nu_\mu}|}\delta^4(Q-p_{\nu_\tau}-p_{\bar\nu_\mu}){\cal R}_{\mu\bar\nu_\mu} &=& \frac{\pi m_\tau^3}{3}H(1+y^2-x) \Big[G_1(x,y)\nonumber \\
&+& \frac{2(p_\mu {\cal P})}{m_\tau}\left(1+3y^2-2x\right)\Big]. \label{eq:inte-muon-mode}
\eea
with $Q=k'-p_\mu$,  $x=2(p_\mu\cdot k')/m^2_\tau$, $y^2=m^2_\mu/m^2_\tau$, 
$G_1(x,y)= x(3-2x)-y^2(4-3x)$ and $H[...]$  the step function. This result 
is obtained by using that for massless neutrinos
\be
\int\frac{d^3p_{\nu_\tau}}{2|\vec{p}_{\nu_\tau}|}\int\frac{d^3p_{\bar\nu_\mu}}
{2|\vec{p}_{\bar\nu_\mu}|}\delta^4(Q-p_{\nu_\tau}-p_{\bar\nu_\mu})
p_{\nu_\tau}^\alpha p_{\bar\nu_\mu}^\beta = \frac{\pi Q^2}{24}\left(g^{\alpha\beta}+
 2\frac{Q^\alpha Q^\beta}{Q^2}\right)H[Q^2].
\ee

Finally, in section~\ref{sec:visible}, and in order to perform the $\cos\theta_\tau$ integrations in the CM frame, we make use of
\be
\frac{1}{\pi} \int_a^b d\alpha \frac{t_0+t_1\alpha+t_2\alpha^2}
{\sqrt{(b-\alpha)(\alpha-a)}} = t_0+\frac{t_2}{3}+t_1\costd\cos\theta_d+ 
\frac{2t_2}{3}P_2(\costd)P_2(\cos\theta_d), \label{eq:integrals}
\ee
with $a=\cos(\theta_d+\theta_{\tau d}^{\rm CM})$, 
$b=\cos(\theta_d-\theta_{\tau d}^{\rm CM})$ and $P_2$ the Legendre polynomial of order 2. 

\section{$C_n, C_{P_L}, C_{A_{FB}}, C_{Z_L},C_{ P_T}, C_{A_Q}, C_{Z_Q}$ 
and $C_{Z_\perp}$ coefficients and their dependence on $(\omega, \xi_d)$}
\label{app:coeff2}
\begin{figure*}[!h]
\centering
\includegraphics[scale=0.5]{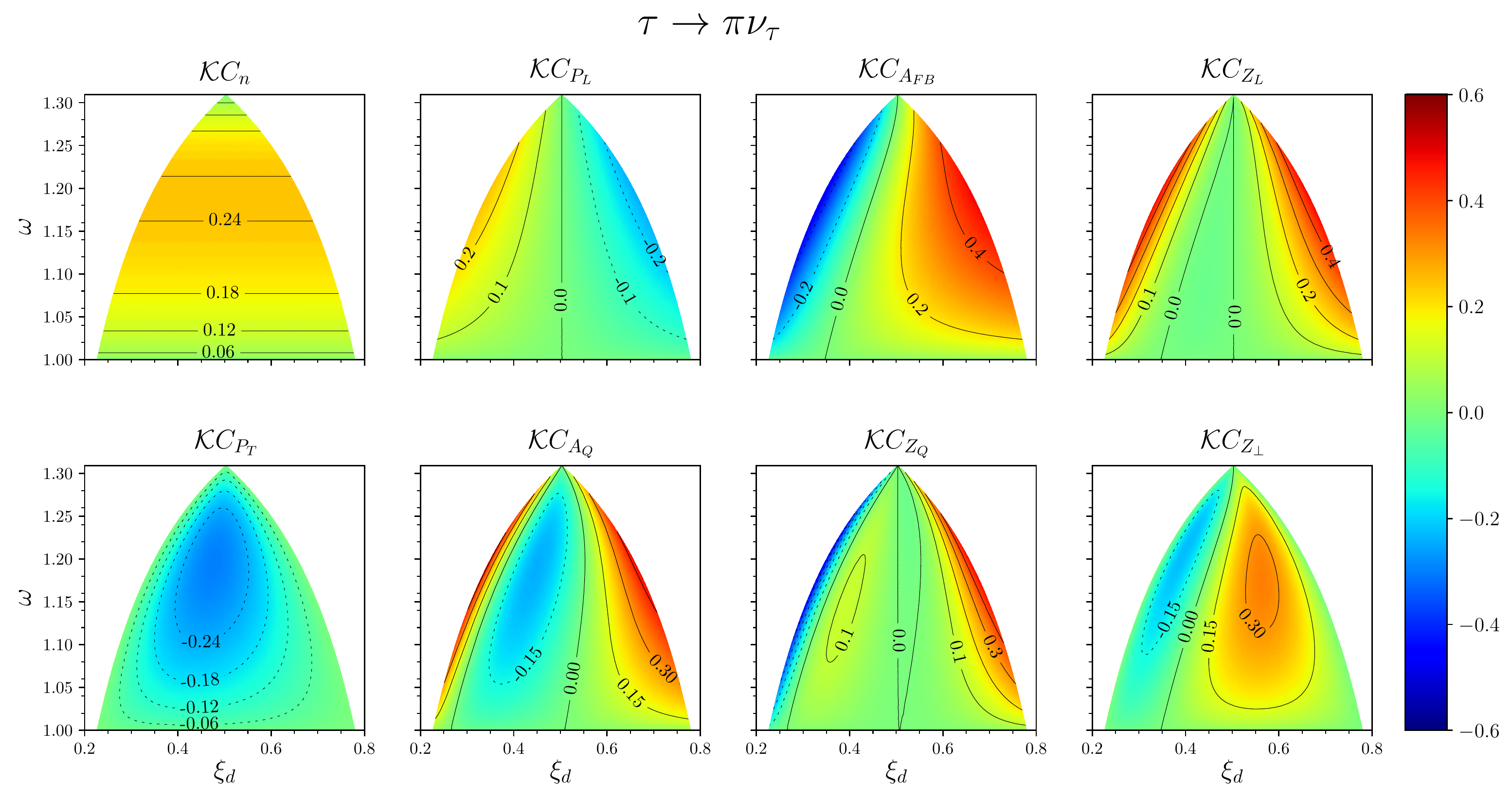}\vspace{.5cm}\\
\includegraphics[scale=0.5]{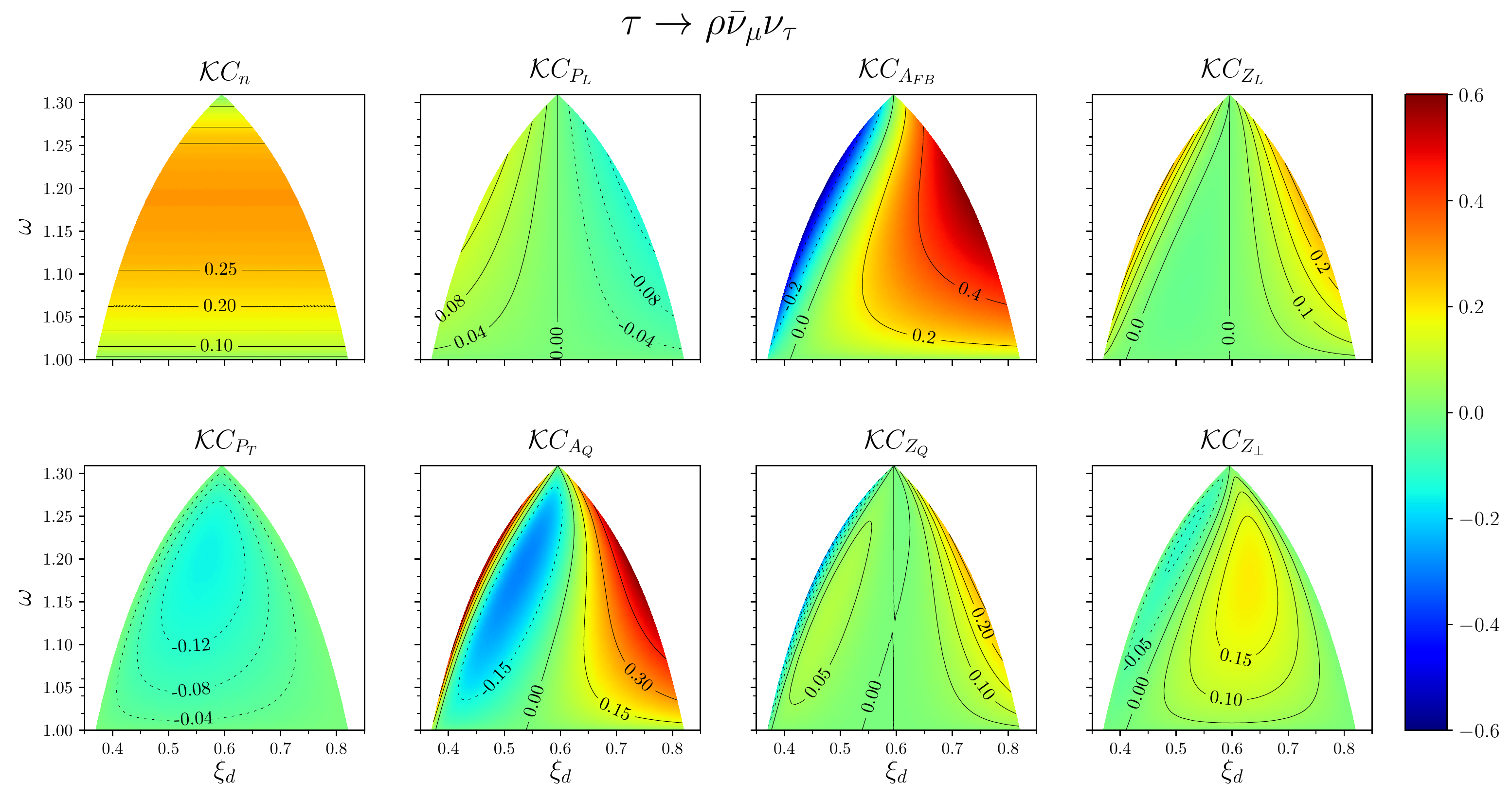}
\caption{ Two dimensional $(\omega, \xi_d)$ dependence of the coefficients 
[multiplied by the kinematical factor ${\cal K}(\omega)=\sqrt{\omega^2-1}\,(1-m_\tau^2/q^2)^2$ ]
 introduced in 
Eq.~\eqref{eq:coeff} for the $\pi$ and $\rho$ hadronic tau-decay modes. Each of the
coefficients  
 affects  one of the  
observables introduced in Eq.~\eqref{eq:observ}, within the corresponding  
$F^d_{0,1,2}(\omega,\xi_d)$ partial wave contribution to  
the $d^3\Gamma_d/(d\omega  d\xi_d d\cos\theta_d)$ differential decay width 
of  Eq.~\eqref {eq:visible-distr} (see text at the end of section~\ref{sec:visible} 
for further details). The used $(\omega, \xi_d)$  ranges correspond to those 
available for the $\Lambda_b\to \Lambda_c\tau (\pi \nu_\tau,
\rho \nu_\tau )\bar\nu_\tau$ sequential decays. Note that dotted contour lines 
correspond to negative values, while solid ones stand for positive (or zero) values.}
\label{fig:coeff1}
\end{figure*}
\begin{figure*}[!h]
\centering
\includegraphics[scale=0.5]{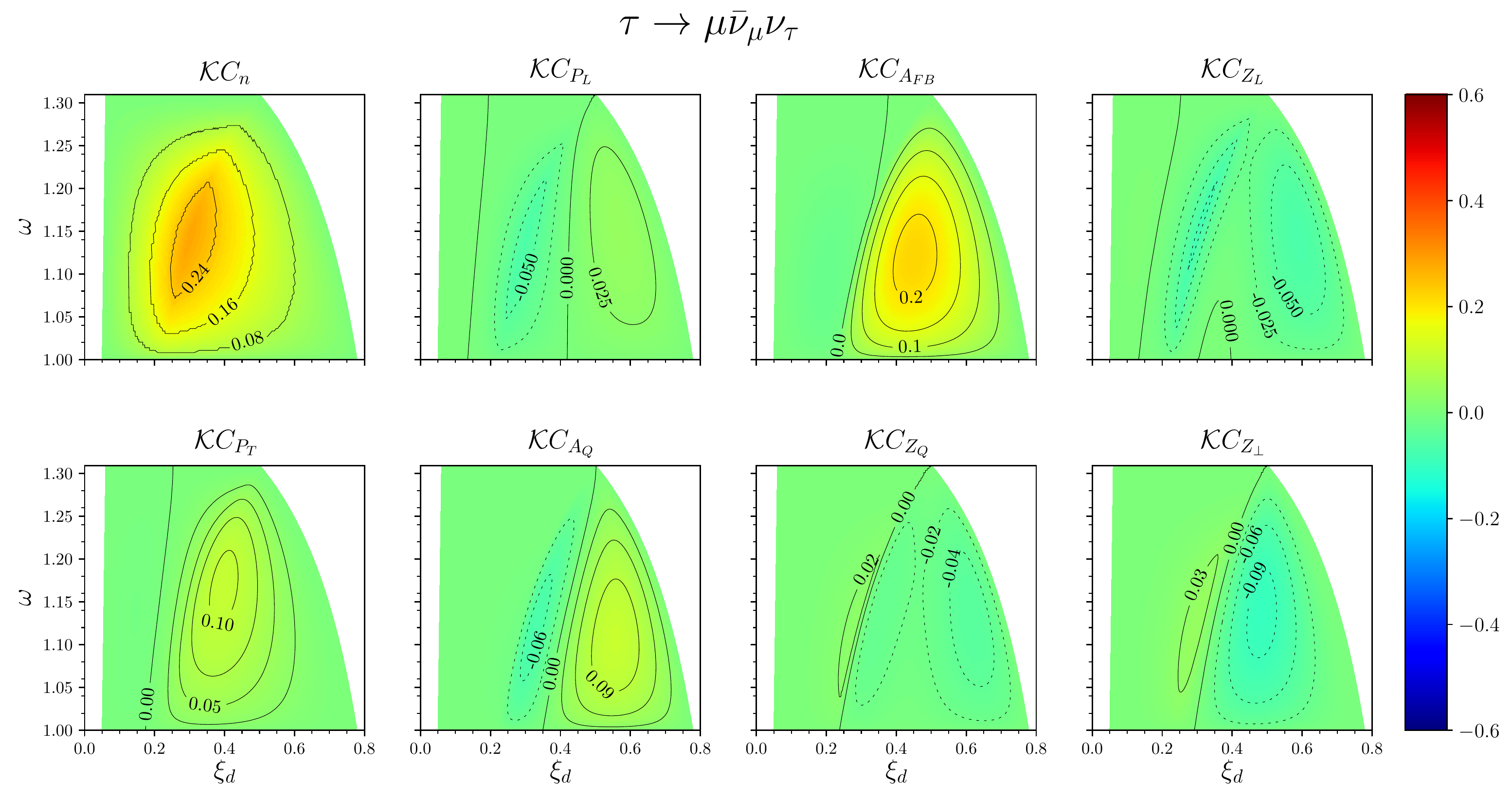}\\
\caption{ Same as Fig.~\ref{fig:coeff1} but for the $\mu\bar\nu_\mu$ tau-decay mode.}
\label{fig:coeff2}
\end{figure*}
The numerical values of the different coefficients introduced in Eq.~\eqref{eq:coeff}  are shown in 
Figs.~\ref{fig:coeff1} (for the
$\pi$ and $\rho$ hadronic tau-decay modes) and \ref{fig:coeff2} (for the $\mu\bar\nu_\mu$ channel). 
We display the coefficients 
multiplied by the kinematical factor
\begin{equation}
 {\cal K}(\omega) = \sqrt{\omega^2-1}\left(1-\frac{m_\tau^2}{q^2}\right)^2, 
\end{equation}
which makes part of the $d\Gamma_{\rm SL}/d\omega$ semileptonic decay width 
(Eq.~\eqref{eq:GammaCM}). The overall
 normalization $n(q^2)=(3a_0(q^2)+a_2(q^2))$, that also multiplies
the 
 asymmetry coefficients, 
provides a smooth $\omega$-dependence (see  the top-left panel of Figure~\ref{fig:ZsyAes}). This latter dynamical factor depends of the effective 
 $b\to c \tau \bar\nu_\tau$ Hamiltonian, but as mentioned,  it should not 
 significantly modify the $\omega$-dependencies displayed in Figs.~\ref{fig:coeff1} 
 and ~\ref{fig:coeff2}.
The whole available $(\omega, \xi_d)$ phase-space  for the sequential tau decay 
from the $\Lambda_b \to \Lambda_c$ transition is explored in the plots. From the 
figures one can easily identify which
zones of the phase-space are more appropriate to extract the different observables
associated with each of the coefficients. One can also  infer that $\tau$-decay hadronic 
modes  are better suited for that purpose than the purely leptonic one.

The coefficients $C_i^d(\omega,\xi_d)$ (Eq.~\eqref{eq:coeff}) for the $\pi$ and $\rho$ hadronic decay modes can be easily read out from Eq.~\eqref{eq:coeffs-1}.
Below, we collect the values for the pure $d=\mu\bar\nu_\mu$ leptonic mode ($y \le \sqrt{(1-\beta)/(1+\beta)}= m_\tau/\sqrt{q^2}$). Their expressions depend on the $\xi_d$ value for which we have to distinguish two different regions. We find (for simplicity, we omit the subindex $d$ in  the variable $\xi_d$):  
\vspace{0.2cm}

For $\underline{y/\gamma \le \xi \le \xi_1}$
\begin{eqnarray}
 C_n^{\mu\bar\nu_\mu}(\omega,\xi) &=& \frac{1}{\beta f(y)}\int_{x_-}^{x_+} dx \,G_1(x,y)=\alpha_n(\omega,\xi), \nonumber \\
 C_{P_L}^{\mu\bar\nu_\mu}(\omega,\xi) &=& \frac{1}{\beta f(y)}\int_{x_-}^{x_+} dx \,G_L(x,y)=\alpha_{P_L}(\omega,\xi),\nonumber \\
 C_{A_{FB}}^{\mu\bar\nu_\mu}(\omega,\xi) &=& \frac{2}{\beta f(y)}\int_{x_-}^{x_+} dx 
 \,G_1(x,y)\costd=\alpha_{A_{FB}}(\omega,\xi),\nonumber\\ 
 C_{Z_L}^{\mu\bar\nu_\mu}(\omega,\xi) &=& -\frac{2}{\beta f(y)}\int_{x_-}^{x_+} dx \,G_L(x,y)\costd=\alpha_{Z_{L}}(\omega,\xi),\nonumber \\
 C_{P_T}^{\mu\bar\nu_\mu}(\omega,\xi) &=& -\frac{8  \sqrt{\gamma ^2 \xi ^2-y^2} }
 {\pi\beta  f(y)}\int_{x_-}^{x_+} dx \,G_T(x,y)\left[\sin\theta_{\tau d}^{\rm CM}\right]^2
 =\alpha_{P_{T}}(\omega,\xi),\nonumber 
\end{eqnarray}
\begin{eqnarray}
 C_{A_Q}^{\mu\bar\nu_\mu}(\omega,\xi) &=& \frac{2}{\beta f(y)}\int_{x_-}^{x_+} dx \,G_1(x,y)P_2(\costd)=\alpha_{A_Q}(\omega,\xi),\nonumber \\
 C_{Z_Q}^{\mu\bar\nu_\mu}(\omega,\xi) &=& -\frac{2}{\beta f(y)}\int_{x_-}^{x_+} dx \,G_L(x,y)P_2(\costd)=\alpha_{Z_Q}(\omega,\xi),\nonumber \\
 C_{Z_\perp}^{\mu\bar\nu_\mu}(\omega,\xi) &=&\frac{6 \sqrt{\gamma ^2 \xi ^2-y^2} }{\beta  f(y)}\int_{x_-}^{x_+} dx \,G_T(x,y)\left[\sin\theta_{\tau d}^{\rm CM}\right]^2\costd=\alpha_{Z_{\perp}}(\omega,\xi),
\end{eqnarray}
where we have introduced 
\begin{eqnarray}
\alpha_n(\omega,\xi) &=& \frac{8\gamma \sqrt{\gamma ^2 \xi ^2-y^2}}{3f(y)}   \left[\gamma ^2   \left(9+\left(4-16
   \gamma ^2\right) \xi \right)\xi+y^2 \left( 9\gamma ^2 \xi +4\gamma ^2-10\right) \right],\nonumber \\
   \nonumber \\
 \alpha_{P_L}(\omega,\xi)  &=& \frac{8 \beta \gamma^3 \sqrt{\gamma ^2 \xi ^2-y^2} }{3  f(y)}\left[ 
   \left(3-16 \gamma ^2 \xi \right)\xi+y^2(9 \xi +4) \right],\nonumber\\
%
   \alpha_{A_{FB}}(\omega,\xi) &=& \frac{16 \beta \gamma ^2(\gamma ^2 \xi ^2-y^2)}{3f(y)} \left[8 \gamma ^2 \xi-3 -3 y^2\right],\nonumber \\
\nonumber \\
\alpha_{Z_{L}}(\omega,\xi) &=& \frac{16 \gamma ^2(\gamma ^2 \xi ^2-y^2)}{3f(y)}\left[\left(4-8 \gamma
   ^2\right) \xi +1 +3 y^2\right],\nonumber \\ \nonumber \\
 \alpha_{P_{T}}(\omega,\xi)  &=&\frac{64\gamma (\gamma ^2 \xi ^2-y^2)}{3\pi f(y)}\left[4 \gamma ^2 \xi-1 -3
   y^2\right],\nonumber\\
    \alpha_{A_Q}(\omega,\xi)&=& -\beta \alpha_{Z_Q}(\omega,\xi)=-\frac{2\gamma\beta}{3} \alpha_{Z_\perp}(\omega,\xi_d)=-\frac{128\beta ^2 \gamma ^3 }{15 f(y)} \left(\gamma ^2 \xi ^2-y^2\right)^{3/2}.
\end{eqnarray}

Finally, for the other region of the phase-space  $\underline{\xi_1 \le \xi \le \xi_2}$
\begin{eqnarray}
 C_n^{\mu\bar\nu_\mu}(\omega,\xi) &=& \frac{1}{\beta f(y)}\int_{x_-}^{1+y^2} dx \,G_1(x,y)=\frac{\alpha_n(\omega,\xi)}{2}+ \frac{1}{6\beta f(y)}\Bigg\{5 y^6+9 \left(4 \gamma ^2-5\right) y^4
 \nonumber \\
 && +y^2 \left[-24 \gamma ^4 \xi  (3 \xi +4)+36 \gamma ^2 (\xi 
   (\xi +4)+1)-45\right]\nonumber \\
   &&+ 4 \gamma ^2 \xi ^2 \left[32 \gamma ^4 \xi -6 \gamma ^2 (4 \xi
   +3)+9\right]+5\Bigg\},\nonumber 
   \end{eqnarray}
   \begin{eqnarray}
 C_{P_L}^{\mu\bar\nu_\mu}(\omega,\xi) &=& \frac{1}{\beta f(y)}\int_{x_-}^{1+y^2} dx \,G_L(x,y)=\frac{\alpha_{P_L}(\omega,\xi)}{2}+ \frac{1}{6\beta^2 f(y)}\Bigg\{5y^6 + 3 y^4 \left(12 \gamma ^2-8 \xi -9\right)\nonumber \\
 &&-3 y^2 \left[4 \gamma ^2 \left(\left(2 \gamma ^2-3\right) \xi  (3
   \xi +4)-1\right)+24 \xi +3\right]\nonumber \\
  && +4 \gamma ^2 \xi ^2 \left[32 \gamma ^4 \xi -6 \gamma ^2 (8 \xi +1)+12 \xi
   +9\right]-1\Bigg\},\nonumber\\
%
 C_{A_{FB}}^{\mu\bar\nu_\mu}(\omega,\xi) &=& \frac{2}{\beta f(y)}\int_{x_-}^{1+y^2} dx \,G_1(x,y)\costd=\frac{\alpha_{A_{FB}}(\omega,\xi)}{2}- \frac{1}{6\gamma\beta^2\sqrt{\gamma ^2 \xi ^2-y^2} f(y)}\Bigg\{ 3 y^8 
 \nonumber \\
 &&-10 \gamma ^2 \xi 
   y^6 +6 y^4
   \left[4 \gamma ^4 (3 \xi +2)-3 \gamma ^2 (3 \xi +8)+15\right]\nonumber \\
   &&-6 \gamma ^2 \xi  y^2 \left[4 \gamma ^2
   \left(\left(2 \gamma ^2-3\right) \xi  (\xi +4)-3\right)+24 \xi +9\right]\nonumber \\
   &&-2 \gamma ^2 \xi  \left[5-4 \gamma ^2 \xi ^2 \left(16 \gamma ^4 \xi -6 \gamma ^2
   (4 \xi +1)+6 \xi +9\right)\right]+3\Bigg\},\nonumber \\  
 C_{Z_L}^{\mu\bar\nu_\mu}(\omega,\xi) &=& -\frac{2}{\beta f(y)}\int_{x_-}^{1+y^2} dx \,G_L(x,y)\costd=\frac{\alpha_{Z_{L}}(\omega,\xi)}{2} + \frac{1}{6\gamma\beta^3\sqrt{\gamma ^2 \xi ^2-y^2} f(y)}\Bigg\{ 3 y^8 
 \nonumber \\
 &&-2 y^6 \left[5 \left(\gamma ^2+1\right) \xi
   -4\right]+6 y^4 \left[4 \gamma ^4 (3 \xi +2)+\gamma ^2 \left(8 \xi
   ^2-27 \xi -16\right)+9 (\xi +1)\right]\nonumber \\
   &&-6 y^2 \left[\xi  \left(8 \gamma ^6 \xi  (\xi +4)-4 \gamma ^4 \left(3
   \xi ^2+16 \xi +1\right)+3 \gamma ^2 \left(4 \xi ^2+8 \xi
   +3\right)-3\right)\right]\nonumber \\
   &&+ 2 \left(\gamma ^2+1\right) \xi +16 \gamma ^4 \left(8 \gamma ^4-16 \gamma
   ^2+9\right) \xi ^4-8 \gamma ^2 \left(2 \gamma ^4-3 \gamma ^2+3\right) \xi
   ^3-1\Bigg\},\nonumber \\  
 C_{P_T}^{\mu\bar\nu_\mu}(\omega,\xi) &=& -\frac{8  \sqrt{\gamma ^2 \xi ^2-y^2} }{\pi\beta  f(y)}\int_{x_-}^{1+y^2} dx \,G_T(x,y)\left[\sin\theta_{\tau d}^{\rm CM}\right]^2=\frac{\alpha_{P_{T}}(\omega,\xi)}{2}
 \nonumber \\
 &&+ \frac{1}{3\pi\gamma^2\beta^3\sqrt{\gamma ^2 \xi ^2-y^2} f(y)}\Bigg\{ 3 y^8 +y^6
   \left[\gamma ^2 (48-20 \xi )-40\right]\nonumber \\
   &&-6 y^4 \left[8
   \gamma ^4 (3 \xi +1)-2 \gamma ^2 \left(4 \xi ^2+9 \xi +12\right)+15\right]\nonumber \\
   &&+12 \gamma ^2 \xi  y^2 \left[8 \gamma ^4 \xi  (\xi
   +2)-4 \gamma ^2 \left(3 \xi ^2+6 \xi +1\right)+12 \xi +3\right]\nonumber \\
   &&-128 \gamma ^8 \xi ^4+32 \gamma ^6 \xi ^3 (6 \xi +1)-48 \gamma ^4 \xi ^3 (\xi
   +1)+4 \gamma ^2 \xi -1\Bigg\}, \nonumber\\
 C_{A_Q}^{\mu\bar\nu_\mu}(\omega,\xi) &=& \frac{2}{\beta f(y)}\int_{x_-}^{1+y^2} dx \,G_1(x,y)P_2(\costd)=\frac{\alpha_{A_Q}(\omega,\xi)}{2}
 \nonumber \\
 &&+ \frac{1}{240\gamma^2\beta^3\left[\gamma ^2 \xi ^2-y^2\right] f(y)}\Bigg\{ 63 y^{10}-5 y^8 \left[8 \gamma ^2 (9 \xi
   -5)+25\right]\nonumber \\
 &&+10 y^6 \left[8 \gamma ^4 \left(5 \xi ^2-9\right)+4
   \gamma ^2 \left(5 \xi ^2+27\right)-45\right] \nonumber \\
 &&+30 y^4 \left[64
   \gamma ^6 \xi -8 \gamma ^4 \left(9 \xi ^2+8 \xi +3\right)+12 \gamma ^2 \left(3 \xi
   ^2+2 \xi +3\right)-15\right] \nonumber \\
   &&-5 y^2
   \left[512 \gamma ^8 \xi ^3-1280 \gamma ^6 \xi ^3+48 \gamma ^4 \xi ^2 \left(3 \xi
   ^2+8 \xi +9\right)-8 \gamma ^2 \left(27 \xi ^2+5\right)+25\right]\nonumber \\
   &&+1024 \gamma ^{10} \xi ^5-2560 \gamma ^8 \xi ^5+1920 \gamma ^6 \xi ^5-80 \gamma
   ^4 \xi ^2 \left(9 \xi ^2-5\right)+40 \gamma ^2 \xi  (5 \xi -9)+63\Bigg\},\nonumber 
 \end{eqnarray}
  \begin{eqnarray}  
 C_{Z_Q}^{\mu\bar\nu_\mu}(\omega,\xi) &=& -\frac{2}{\beta f(y)}\int_{x_-}^{1+y^2} dx \,G_L(x,y)P_2(\costd)=\frac{\alpha_{Z_Q}(\omega,\xi)}{2}
 \nonumber \\
 &&- \frac{1}{240\gamma^2\beta^4\left[\gamma ^2 \xi ^2-y^2\right] f(y)}\Bigg\{ 63 y^{10}-5 y^8 \left[8 \gamma
   ^2 (9 \xi -5)+36 \xi -5\right]\nonumber \\
 &&+10 y^6 \left[8 \gamma ^4 \left(5 \xi ^2-9\right)+4
   \gamma ^2 \left(35 \xi ^2-48 \xi +45\right)+48 \xi -81\right] \nonumber \\
 &&+30 y^4 \left[64 \gamma ^6 \xi -8 \gamma ^4 \left(8
   \xi ^3-9 \xi ^2+20 \xi +1\right)+4 \gamma ^2 \left(-8 \xi ^3+9 \xi ^2+18 \xi
   +5\right)-3 (4 \xi +3)\right] \nonumber \\
   &&+5 y^2
   \left[-512 \gamma ^8 \xi ^3+1280 \gamma ^6 \xi
   ^3+48 \gamma ^4 \xi ^2 \left(9 \xi ^2-32 \xi +3\right)+8 \gamma ^2 \left(24 \xi
   ^3+9 \xi ^2-1\right)-1\right]\nonumber \\
   &&+720 \gamma ^4 \xi ^4-40 \gamma ^2 \left(2 \gamma ^2+7\right) \xi ^2+60 \left(2
   \gamma ^2+1\right) \xi \nonumber \\
   &&+64 \gamma ^4 \left(16 \gamma ^6-40 \gamma ^4+30 \gamma
   ^2-15\right) \xi ^5-27\Bigg\},\nonumber \\
 C_{Z_\perp}^{\mu\bar\nu_\mu}(\omega,\xi) &=&\frac{6 \sqrt{\gamma ^2 \xi ^2-y^2} }{\beta  f(y)}\int_{x_-}^{1+y^2} dx \,G_T(x,y)\left[\sin\theta_{\tau d}^{\rm CM}\right]^2\costd=\frac{\alpha_{Z_{\perp}}(\omega,\xi)}{2}\nonumber \\
 &&+ \frac{1}{80\gamma^3\beta^4\left[\gamma ^2 \xi ^2-y^2\right] f(y)}\Bigg\{ 21 y^{10}-5 y^8 \left[4
   \gamma ^2 (9 \xi -10)+25\right]\nonumber \\
 &&+10 y^6 \left[8 \gamma ^4 \left(5 \xi ^2-12
   \xi +9\right)+4 \gamma ^2 \left(5 \xi ^2+12 \xi -27\right)+45\right] \nonumber \\
 &&-30 y^4 \left[32
   \gamma ^6 \xi +8 \gamma ^4 \left(4 \xi ^3-9 \xi ^2-4 \xi -1\right)+12 \gamma ^2
   \left(3 \xi ^2+\xi +1\right)-5\right] \nonumber \\
   &&+5 y^2
   \left[256 \gamma ^8 \xi ^3-640 \gamma ^6 \xi ^3+48 \gamma ^4 \xi ^2 \left(3 \xi
   ^2+4 \xi +3\right)-8 \gamma ^2 \left(9 \xi ^2+1\right)+5\right]\nonumber \\
   &&-512 \gamma ^{10} \xi ^5+1280 \gamma ^8 \xi ^5-960 \gamma ^6 \xi ^5+80 \gamma ^4
   \xi ^2 \left(3 \xi ^2-1\right)+20 \gamma ^2 (3-2 \xi ) \xi -9\Bigg\}.
\end{eqnarray}
This is the first calculation which includes $m_\mu/m_\tau$ contributions, though neglecting the muon mass ($y=0$) should be a good approximation for the above coefficients.

\bibliographystyle{jhep}
\bibliography{B2Dbib}

\end{document}